\pdfoutput=1
\documentclass[cup6b]{cupbook}

\usepackage{graphicx} 
\usepackage{color}
\usepackage{amsmath}
\usepackage{amssymb}

\def\thechapter{}

\newcommand{\msun}{\mbox{$M_{\odot}$}}
\newcommand{\lsun}{\mbox{$L_{\odot}$}}
\newcommand{\degree}{\ensuremath{^\circ}}

\author[Nick Z. Scoville]{Nick Z. Scoville\\                              
        California Institute of Technology\\
        1200 East California Boulevard, Pasadena, CA 91125, USA\\
        NZS@astro.caltech.edu}   

\begin{document}

% Dummy out the chapter number in the preprint, in part because it does not get centered.

\chapter{Evolution of star formation and gas}

%
%%%%%%%%%%%%%%%%%%%%%%%%%%%%%%%%%%%%%%%%%%%%%%%%%%%%%%%%%%%%%%%%%%%%%%%%%%%%%%%%
%

\abstract{
In these lectures I review observations of star-forming molecular clouds in our
Galaxy and nearby galaxies to develop a physical intuition for understanding
star formation in the local and high-redshift Universe. A lot of this material
is drawn from early work in the field since much of the work was done two
decades ago and this background is not generally available in the present
literature. I also attempt to synthesise our well-developed understanding of
star formation in low-redshift galaxies with constraints from theory and
observations at high redshift to develop an intuitive model for the evolution of
galaxy mass and luminosity functions in the early Universe. 

The overall goal of this contribution is to provide students with background
helpful for analysis of far-infrared (FIR) observations from \textit{Herschel}
and millimetre/submillimetre (mm/submm) imaging with ALMA (the Atacama
Large Millimetre/submillimetre Array). These two instruments will revolutionise
our understanding of the interstellar medium (ISM) and associated star formation
and galaxy evolution, both locally and in the distant Universe. To facilitate
interpreting the FIR spectra of Galactic star-forming regions and high-redshift
sources, I develop a model for the dust heating and radiative transfer in order
to elucidate the observed infrared (IR) emissions. I do this because I am not
aware of a similar coherent discussion in the literature.}

\def\thechapter{8}

\section{Star-forming molecular clouds}
\label{gmc} 

Here, I provide an overview of the observations and physics of star-forming
molecular clouds in the Galaxy and nearby normal galaxies. The motivation is to
develop the intuitive background for analysis of observations at high redshift,
hopefully avoiding naive and simplistic assumptions. The physics of the
molecular ISM is not at all what one would have guessed before the advent of
mm-line astronomy and the understanding developed over the last 30 years is
vital to galactic evolution studies. 

\subsection{Background}

All known star formation in low-redshift galaxies occurs in molecular gas clouds
and inclusion of atomic H{\sc i} clouds in discussions of star formation is a
`red herring', except in as much as the atomic gas may feed the buildup of
molecular clouds. In fact, there is no instance of star formation demonstrably
observed to occur in atomic gas in nearby galaxies (as distinct from atomic gas
dissociated by recently formed stars). Even in the extended ultraviolet (UV)
disks of nearby galaxies, the young stars producing the UV could have been
formed from molecular clouds simply too low in mass or surface density to be
detected in present observations. The dominant process for formation of hydrogen
molecules (H$_2$) at low redshift is believed to be on the surface of dust
grains where two H atoms remain in proximity long enough for a radiative decay
to a bound state. The hydrogen molecule is dissociated from the ground state by
a two-step process involving UV photon absorption in the Lyman and Werner bands
longward of the H{\sc i} Lyman limit, followed by radiative decay to unbound
vibrational states in the ground electronic state (Hollenbach \textit{et al.}
1971). Thus, the destruction rate for H$_2$ depends sensitively on the ambient
UV radiation field and hence on the amount of dust extinction and H$_2$ column
density to the cloud surface -- the latter since the foreground H$_2$ may absorb
all the incoming UV when the H$_2$ Lyman and Werner bands become optically thick
(i.e., self-shielding). Thus, whether a gas parcel is atomic or molecular will
depend complexly on a number of factors: the volume density (since the formation
rate depends on the densities of dust and atoms), the clumping of the ISM  and
the column density to the cloud surface and velocity dispersion which determine
the dust continuum and H$_2$ line opacities. For typical cloud densities of a
few hundred per cm$^{-3}$, the phase transition from H{\sc i}- to
H$_2$-dominated will occur at a column density $N_{{\rm H+2H}_2} \sim
2\times10^{21}$ cm$^{-2}$, corresponding to an extinction $A_V$\,=\,1\,mag for a
standard dust to gas abundance ratio (Hollenbach \textit{et al.} 1971). 

\begin{figure}
\vspace{2mm}
\includegraphics[width=0.8\linewidth]{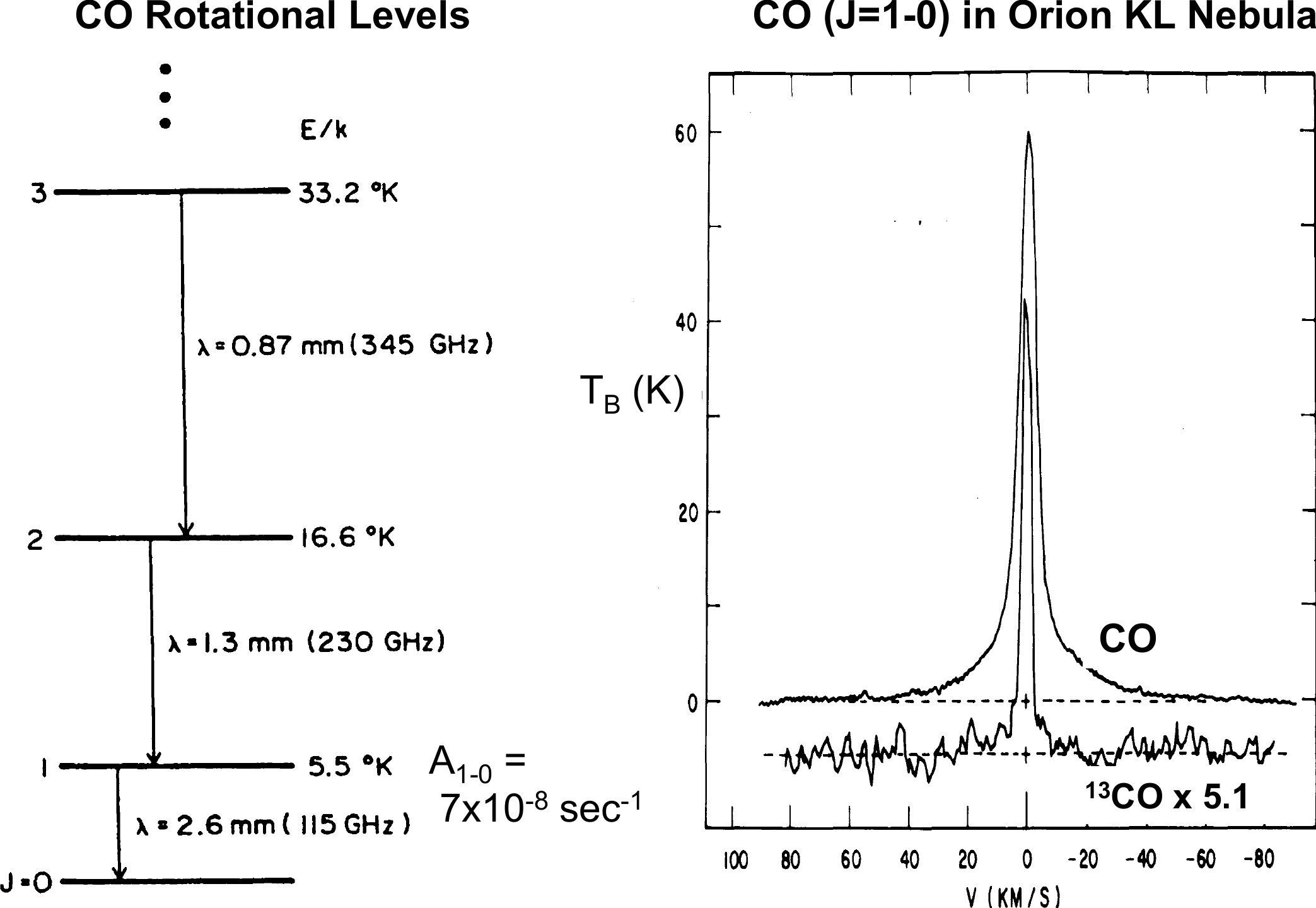}
\caption{The low-lying rotational levels of CO are shown on the left; the
spontaneous decay rate for the $J$\,=\,1--0 transition is
$7\times10^{-8}$\,sec$^{-1}$ and the critical density for pure collisional
excitation by H$_2$ is $\sim$3000\,cm$^{-3}$. With line photon trapping and a
typical optical depth of $\tau\sim$10, the critical density for thermalisation
is reduced to $\sim$300\,cm$^{-3}$. Thus, the CO line brightness temperature
provides a measure of the gas kinetic temperature. On the right, the CO and
$^{13}$CO $J$\,=\,1--0 emission lines are shown in the direction of the Orion
Kleinmann-Low (KL) nebula.}
\label{co}
\end{figure}

The H$_2$ molecule is homonuclear and therefore has no permanent dipole moment
-- its pure rotational transitions are weak quadrupole transitions with low
emissivity. In addition, H$_2$ has a small moment of inertia and its first
rotational transition ($J$\,=\,2--0) at 28.2\,$\mu$m is therefore at relatively
high energy ($h\nu/k\simeq508$\,K), requiring a minimum gas temperature of
$T_{\rm k}\sim150$\,K for appreciable collisional excitation. Thus trace
molecules such as CO and HCN have become the standard surrogate probes of
molecular clouds. CO and HCN have dipole moments of 0.1 and 2.3 Debye,
respectively, and their larger moments of inertia (due to having heavier atoms
than H) yield lowest rotational transitions at 115.2 and 88.6\,GHz (2.6 and
3\,mm wavelength), corresponding to $h\nu/k$\,$\simeq$\,5.5 and $4.2$\,K (see
Fig.~\ref{co}). These molecules also have rare isotopic forms (e.g., $^{13}$CO
and C$^{18}$O) with slightly different moments of inertia, and hence shifted
rotational lines which can be used to probe regions of high column density where
the abundant isotope lines are very optically thick. 

%-------------------------------------------------------------------------------
%-------------------------------------------------------------------------------

\subsection{Molecular excitation}

Excitation for the rotational levels of molecules like CO observed at mm and
submm wavelengths is provided by collisions with H$_2$. If these collisions are
sufficiently frequent (compared to the spontaneous decay rate) the levels will
come into thermal equilibrium with the H$_2$ and their excitation temperature
will approach the gas kinetic temperature. For optically thin transitions, this
occurs at the critical density $n_{{\rm H}_{2},{\rm crit}}=A_{\rm
u-l}/\langle\sigma\nu_{\rm u-l}\rangle$ where $A_{\rm u-l}$ is the Einstein
spontaneous decay rate ($7\times10^{-8}$\,sec$^{-1}$ for CO, $J$\,=\,1--0), and
$\langle\sigma\nu_{\rm u-l}\rangle$ is the collisional de-excitation rate averaged
over the Maxwellian distribution of H$_2$ thermal velocities. For CO
($J$\,=\,1--0) this critical density is $\sim$3000\,cm$^{-3}$ and for HCN (and
most other molecules) it is $\sim$10$^{4-5}$\,cm$^{-3}$ due to their higher
dipole moments and therefore higher $A$ coefficients. Hence, the non-CO
molecules are generally taken as probes of high-density molecular gas as
compared with that traced by CO. 

Measurements of the rare $^{13}$C isotopes of both CO and HCN indicate that the
$^{12}$CO emission is optically thick. The rare isotopes are typically seen at
an intensity approximately 0.1--0.5 of the abundant species whereas the ISM
abundance ratio is $^{13}$C/C\,=\,l/89--1/40, implying that the emission from the
abundant species is optically thick and therefore saturated (see Fig.~\ref{co},
right). For optically thick transitions, the upper-level population can be
enhanced due to absorption of line photons, leading to excitation temperatures
higher than those expected simply due to H$_2$ collisions, since the line
photons emitted upon spontaneous decay cannot easily escape the cloud. This
so-called radiative trapping of the line photons builds up the radiation field
at the frequency of the line, leading to enhanced excitation of the upper state
via photon absorption. 

The escape probability formalism used to treat this optically thick situation
was first applied to molecular clouds by Scoville \& Solomon (1974) and
Goldreich \& Kwan (1974) and is now routinely used to analyse interstellar
molecule excitation (often called large velocity gradient [LVG], non-thermal
equilibrium [non-LTE] analysis; see, e.g., van der Tak \textit{et al.} (2007) and
their publicly available RADEX
code)\footnote{{\tt http://www.strw.leidenuniv.nl/$\sim$moldata/radex.html}}. \emph{One
of the biggest advantages of the LVG formalism is that it permits treatment of
the coupled radiative transfer and molecular excitation as a local problem} -- a
fact probably not fully appreciated by current routine users of these codes.
This formalism is applicable to situations in which systematic velocity
gradients are large compared to the small-scale thermal motions. The line
photons from one region of the cloud are then incoherent with other regions due
to the Doppler shift; they can then only interact with molecules in the local
region near where they were emitted.  

In the photon trapping regime, the spontaneous decay rates (\textit{A}) used in
analysing the equilibrium molecular excitation are reduced by a factor $\beta$
equal to the effective probability for escape of line photons from the emission
region (Scoville \& Solomon 1974; Goldreich \& Kwan 1974). Thus, the\linebreak critical
density for thermalisation of the levels is reduced by the same factor $\beta$.
For a spherically symmetric cloud with radial velocity field ($\nu\propto r$),
the escape probability is given by 
\begin{equation}
\beta = {1\over{\tau}}(1 - {\rm e}^{-\tau}),
\end{equation}
where
\begin{equation}
\tau = {A_{\rm u-l}\lambda^3 g_{\rm u}\over{8\pi{\rm d}\nu/{\rm d}r}}(n_{\rm l}/g_{\rm l}-n_{\rm u}/g_{\rm u}),
\end{equation}
for $\tau\ll1$, $\beta=1$ and for $\tau\gg1$, $\beta=1/\tau$. Thus the critical
density is reduced by a factor of $\tau$ in the optically thick regime. For most
giant molecular clouds (GMCs -- see below), $\tau_{\rm CO}\gtrsim10$ (based on
the relative strength of $^{13}$CO), so the critical density of CO (1--0) is
reduced from 3000 to 300\,cm$^{-3}$. \emph{Thus, the CO excitation temperature
will be approximately thermalised at the gas kinetic temperature ($T_{\rm k}$)
in most clouds. And since the CO line is optically thick, the brightness
temperature at the emission line peak will indicate the H$_2$ kinetic
temperature.}

Similar considerations apply for the optically thick lines of molecules such as
HCN and CS which have higher dipole moments -- i.e., their critical densities
are reduced by a factor $1/\tau$ but since those critical densities are higher
than that of CO, their excitation temperatures and line brightness temperatures
will be subthermal (except in the densest cloud core regions). In fact, since
the critical density for optically thick transitions is $n_{{\rm H}_{2},{\rm crit}}=(A_{\rm
u-l}/\tau)/\langle\sigma\nu_{\rm u-l}\rangle$, the radiative line strength
(appearing in the $A$-coefficient in the numerator and in $\tau$ in the
denominator) cancels out -- i.e., \emph{the critical density then just depends
inversely on the molecular abundance}. 

In the regime of photon trapping, one should \textit{not} assume that the
emission of a rare isotope like $^{13}$CO is optically thin simply because it is
observed to have lower intensity than the abundant isotope emission. The rare
isotope may simply have a lower excitation temperature since it has less photon
trapping. This fact is under-appreciated by `casual' observers. Typically, the
$^{13}$CO emission is 1/5 to 1/10 of CO -- does this mean the optical depth is
always in the small range 1/5 to 1/10? -- of course not!

Molecular excitation can also be provided by absorption of continuum photons --
either the IR continuum of dust or the cosmic microwave background
radiation (CMB). In the escape probability formalism it is easy to include such
continuum excitation. If $\beta$ is the probability of a line photon escaping
from the local region, then $\beta$ is also the probability of an externally
produced background photon making it into the local volume. The low-lying
molecular levels will maintain a base excitation temperature at the cosmic
background level, $2.7\times(1+$redshift)\,K (even in the absence of H$_2$
collisions). However, since the line observations measure only the excess
emission above the local continuum background, this CMB excitation does not
result in detectable emission. The continuum from dust emission is not generally
important for exciting low-lying molecular levels at mm wavelengths (since the
continuum is optically thin and weak); but for higher-energy transitions at
$\lambda_{\rm rest}<500$\,$\mu$m the IR continuum should be
included.

\subsubsection{Summary}

In summary, we can expect that:
\begin{enumerate}[(a)]\listsize
\renewcommand{\theenumi}{(\alph{enumi})}

\item The low CO levels will be thermalised and the brightness temperatures (as
observed in spatially resolved clouds) will approximate the gas kinetic
temperature, i.e., $T_{\rm B}({\rm CO})\simeq T_{\rm k}({\rm H_2})$.

\item Photon trapping in optically thick lines can greatly enhance the molecular
excitation above the excitation due to collisions with H$_2$ molecules. 

\item The CO emission will trace the overall distribution of H$_2$ for densities
$n_{{\rm H}_2}\sim10^{2-3}$\,cm$^{-3}$ -- this turns out to be well tuned to
characteristic densities of the majority of the molecular gas mass.

\item The emissions from higher dipole moment molecules (HCN, CS, etc.) will
trace higher density regions with $n_{{\rm H}_2}\geq10^{4-5}$\,cm$^{-3}$.
\end{enumerate}

\subsection{Observed properties of molecular gas}
\label{prop}

%-------------------------------------------------------------------------------
\begin{figure}
\vspace{2mm}
\includegraphics[height=12cm,width=\linewidth]{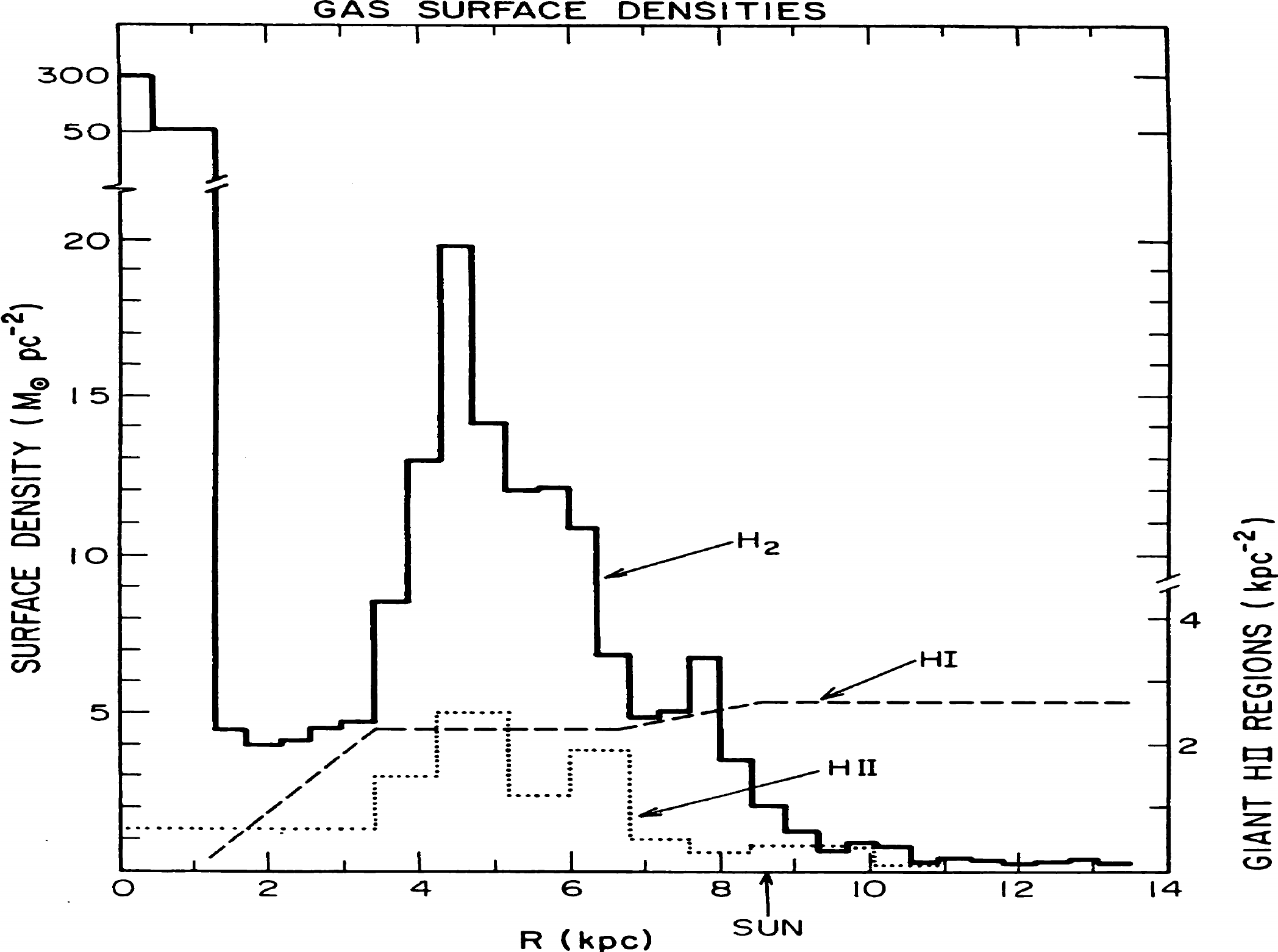}
\caption{The gas surface densities in the Milky Way disk are shown for H$_2$
(Clemens \textit{et al.} 1988), H{\sc i} (Burton \& Gordon 1978) and giant H{\sc
ii} regions ($\geq$M\,42). Both the H$_2$ and H{\sc i} values include a 1.36
correction factor for He.}
\label{gas_sigma}
\end{figure}
%-------------------------------------------------------------------------------

Extensive surveys of the CO and $^{13}$CO ($J$=1--0 at $\lambda$\,=\,2.6\,mm and
$J$=2--1 at $\lambda$\,=\,1.3\,mm) emissions in the Galaxy have been used to
determine the overall distribution of star-forming H$_2$ gas and its properties
(Scoville \& Solomon 1975; Sanders \textit{et al.} 1985; Dame \& Thaddeus 1985;
Scoville \textit{et al.} 1987; Clemens \textit{et al.} 1988; Matsunaga
\textit{et al.} 2001; Dame \textit{et al.} 2001; Jackson \textit{et al.} 2006).
CO emission is ubiquitous at all Galactic longitudes in the inner Galaxy
($|l|<90$\degree) and within 1$\degree$ latitude of the Galactic plane. The CO
emission is particularly strong in the central 1$\degree$ around the Galactic
centre and within an annulus/ring from $|l|$\,=\,20--50 degrees -- the so-called
molecular cloud ring at 3--7\,kpc radius. This concentration of the H$_2$ to the
interior of the Galaxy is remarkably different from that of H{\sc i}; the 21-cm
H{\sc i} distribution is much smoother and the overall H{\sc i} distribution is
fairly constant with Galactic radius out to $R$\,$\sim$\,12--15\,kpc with mean column
density perpendicular to the plane of $\langle N_{\rm
H}\rangle\simeq10^{21}$\,cm$^{-2}$ (see Fig.~\ref{gas_sigma}). 

Within the Galaxy the overall mass contents of the molecular and atomic gas are
approximately equal ($\sim$\,2\,$\times$\,10$^9$\,\msun\ in each) but most of the
molecular gas is inside the Solar-circle whereas most of the H{\sc i} is outside
8.5\,kpc. Since the total Galactic star formation rate is $\sim$3\msun\ per yr,
the characteristic cycling time for H$_2$ into new stars is
$\sim$\,2\,$\times$\,10$^9$\msun/3\msun/yr\,$\sim$10$^9$\,yr. Thus, the H$_2$ clouds are
forming stars at a rate much slower than their free-fall collapse time within
the clouds ($(G\rho)^{-1/2}$\,$\sim$\,3--5\,Myr) but on a timescale much shorter
than the Galactic lifetime. \emph{Star formation is inefficient but a resupply
of gas is required if star formation is not to die out in the next Gyr} (which
seems unlikely).

%-------------------------------------------------------------------------------
\begin{figure} 
\includegraphics[width=\linewidth]{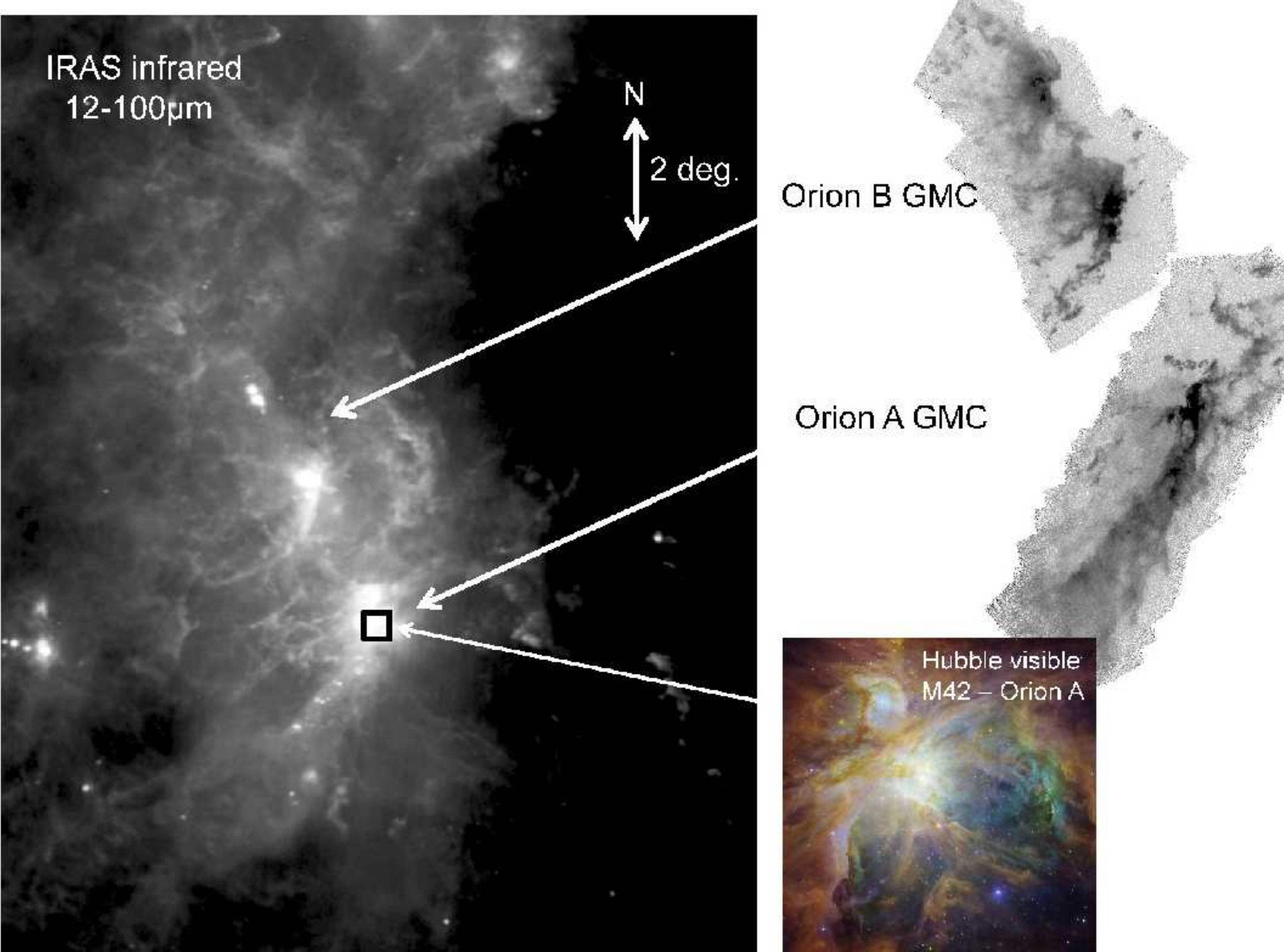}
\caption{The star-forming complex in Orion provides a close view of the
complexity of both the ISM and the star formation activity. The left panel shows
a very large-scale image of the FIR emission as imaged by the IR astronomical
satellite (\textit{IRAS}) (Courtesy NASA/JPL-Caltech). The upper-right panel
shows the integrated CO (1--0) line emission from the two Orion GMCs (Ripple
\textit{et al.} 2012). A \textit{Hubble} Heritage image of the visible H{\sc ii}
region M\,42 is shown at the lower right -- the visible H{\sc ii} region
occupies a very small area of the Orion complex, as outlined in the
\textit{IRAS} image.}
\label{oriongmc}
\end{figure}
%-------------------------------------------------------------------------------

The smaller-scale distribution of CO emission indicates that the H$_2$ gas
resides in discrete clouds, rather than a diffuse, continuous medium. Along a
typical line of sight in the Galactic plane there will be three to six
kinematically discrete CO emission features, occupying a small fraction of the
overall permitted velocity range. Spatial mapping of the individual emission
features indicates extends perpendicular to the line of sight typically 10 to
80\,pc -- hence the name, Giant Molecular Clouds (GMCs). 

The Orion GMC (actually two clouds) is the nearest GMC with an overall extent of
40\,pc (see Fig.~\ref{oriongmc}) -- in striking contrast to the optically
visible Orion Nebula (M\,42) for which the extent is $\lesssim$1\,pc! In the
vicinity of M\,42, the CO brightness temperature increases to a peak of
$\sim$60\,K (see Fig.~\ref{co}, right), compared to 5--20\,K in most of the rest
of the cloud. These elevated temperatures are largely due to heating by the
dust-embedded cluster of young stars in the Kleinmann-Low (KL) IR source,
located behind the visible Trapezium/M\,42 OB star cluster. 

The gas heating associated with the KL nebula (and other IR sources associated
with active star formation) occurs in two steps: heating of the embedding dust
by absorption of photons emitted by the young stars, followed by H$_2$ heating
through collisions with the heated dust grains. Thermal equilibrium between the
dust and the H$_2$ is theoretically expected to occur at gas densities 
$n_{\rm H_2}$\,$>$\,$10^4$\,cm$^{-3}$ (Goldreich \& Kwan 1974). Indeed, the dust colour
temperature in KL is $\sim$60\,K, i.e., similar to the gas kinetic temperature
determined from the CO line brightness temperature. 

%-------------------------------------------------------------------------------
\begin{figure}
\vspace{2mm}
\includegraphics[width=0.7\linewidth]{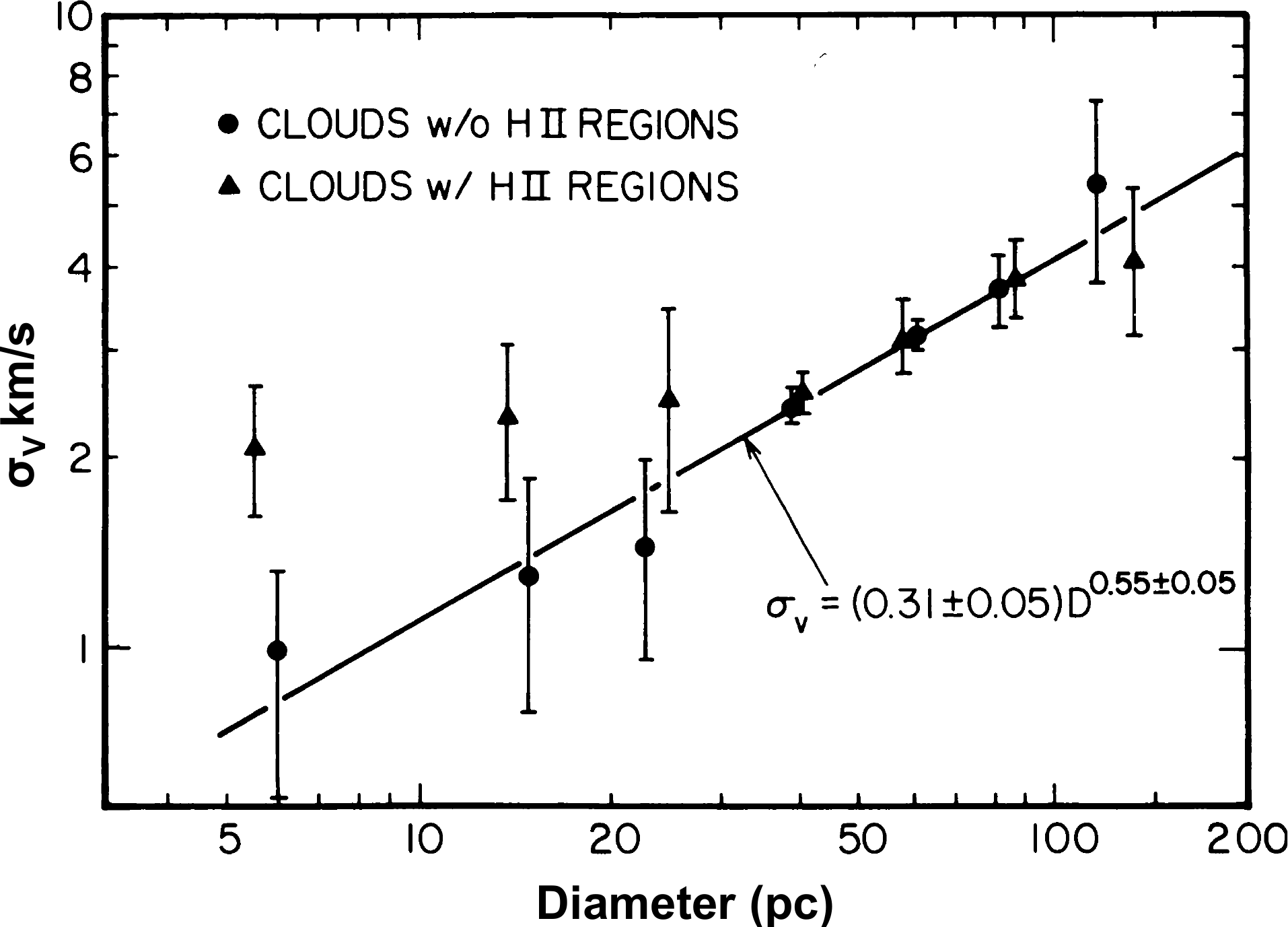}
\caption{The internal velocity dispersions of GMCs are shown as a function of
diameter for clouds with and without giant H{\sc ii} regions, i.e., H{\sc ii}
regions more luminous than M\,42 (Scoville \textit{et al.} 1987). This
illustrates the empirical correlation found between cloud size and linewidth
(the so-called size-linewidth correlation). The H{\sc ii} region clouds depart
from this size-linewidth at low masses/sizes presumably due to feedback effects
from massive star formation.}
\label{size}
\end{figure}
%-------------------------------------------------------------------------------
      
\subsubsection{Self-gravitating GMCs}
    
One of the most perplexing and still not understood features of the GMCs is
their large velocity dispersions, as determined from the width of the molecular
emission lines. At gas temperatures of 10--80\,K, the thermal velocity
dispersion (sound speed) should be $\sim$0.1--0.3\,km/s, yet the observed
velocity dispersions are typically 1--4\,km/s. The linewidths are correlated
with cloud diameter, $D$. The observed relation is $\sigma_{\rm
V}=0.31D_{\rm pc}^{0.55}$\,km/s (see Fig.~\ref{size}). Thus the kinetic
energy in large-scale supersonic motions is $\sim$100 times the expected thermal
energy. The containment of these supersonic turbulent motions is provided by the
self-gravity of the molecular gas within each cloud. Independent estimates of
the cloud masses using both dust extinction measurements and non-LTE analysis of
the molecular lines, indicate\linebreak approximately enough gas mass for the GMCs to be
self-gravitating and in virial equilibrium with the observed internal motions.
Nevertheless, it should be pointed out that although the kinetic energy and
gravitational potential energy approximately balance, the GMCs do not have the
expected spherical shapes; they are often elongated and have internal
filamentary or sheet-like structures.

In any case, it is clear that the GMCs are not in pressure equilibrium with the
external diffuse H{\sc i} and H{\sc ii} phases of the ISM. Their pressures are 
generally taken to be $nT=3000$\,cm$^{-3}$ K, i.e., a factor of 100 lower than 
the turbulence pressures within the GMCs as reflected in their internal velocity
dispersions. 

Given that the GMCs are self-gravitating, one can obtain their mean densities as
a function of cloud diameter using the above empirical relation between the 
velocity dispersion and cloud diameters (i.e., the observed size-linewidth 
relation):
\begin{equation}
\langle n_{{\rm H}_2}\rangle=180\left(\frac{D}{40 {\rm pc}}\right)^{-0.9}{\rm cm}^{-3},
\end{equation}
where the scaling is to 40\,pc -- the size at which half of the overall Galactic
H$_2$ mass is in clouds larger and half is in clouds smaller. The cloud mass
distribution function is $N(M)$\,$\propto$\,$M^{-1.6}$. For a cloud of diameter
40\,pc, the above relation implies a mass of 4\,$\times$\,$10^5$\,\msun\
(including the 36\% mass contribution of He).

\subsubsection{Molecular masses from $\tau\gg1$ CO emission}
    
%-------------------------------------------------------------------------------
\begin{figure}
\vspace{1mm}
\includegraphics[width=0.9\linewidth]{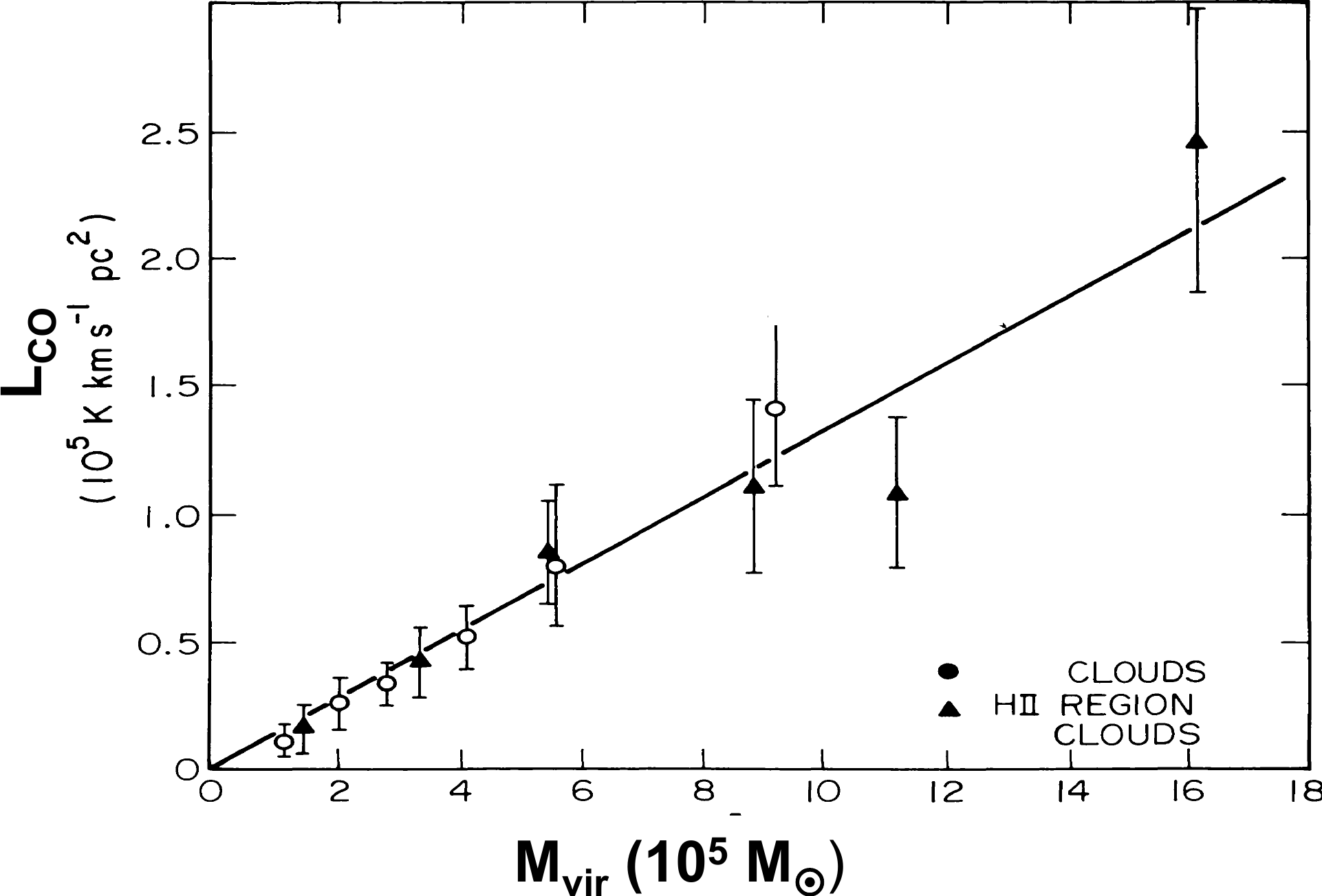}
\caption{The CO luminosities of GMCs are shown as a function of their virial
masses for clouds with and without giant H{\sc ii} regions (Scoville \textit{et
al.} 1987).}
\label{col}
\end{figure}
%-------------------------------------------------------------------------------

How is it that the emission in \emph{optically thick} CO lines can be used to
estimate the total mass of H$_2$ in galaxies? Observers of extragalactic
CO commonly make use of a so-called $X$-factor or $\alpha_{\rm CO}$ to translate 
measurements of CO emission into estimates of molecular gas masses -- let us see 
how this is physically justifiable. 

For a resolved cloud the integrated line brightness temperature, $L_{\rm
CO}=\int T_{\rm B}{\rm d}v$, can be integrated over the projected area of the
cloud to yield a `luminosity' $L_{\rm CO}=d^2\int I_{\rm CO}{\rm d}\Omega$
where $d$ is the distance to the cloud. Thus, 
\begin{equation}
\label{ldef}
L_{\rm CO}=T_{\rm B}({\rm CO})\Delta V\pi R^2\quad{\rm K~km/s ~pc^2},
\end{equation}
where $R$ is the radius of an assumed spherical cloud. For clouds in virial
equilibrium, $\Delta V = (GM/R)^{1/2}$ and therefore,
\begin{equation}
\label{lco}
L_{\rm CO}=(3\pi G/4\rho)^{1/2}T_{\rm B}({\rm CO}) M_{\rm GMC},
\end{equation} 
where $\rho$ is the mass density. Equation~\ref{lco} predicts a linear scaling
between cloud mass and the CO luminosity provided the clouds have approximately 
similar mean density and temperature. Equation~\ref{lco} indicates that 
\emph{the constant of proportionality between L$_{\rm CO}$ and M will vary as 
$T/\sqrt{\rho}$}. Physically, what is going on here? Although the CO line is 
optically thick (and often thermalised), the increased mass of larger clouds 
gets reflected in an increased surface area emitting CO photons and an increased 
linewidth over which they are emitted.

Figure~\ref{col} illustrates the extremely good empirical correlation between
measured CO luminosities and their virial masses (determined from resolved
measurements of the size and linewidth). In fact, if the temperatures are
determined by radiative heating of dust within the cloud (plus a low-level
background heating by cosmic rays heating the gas to $\sim$10--15\,K), the
overall range of mean temperatures is expected to be rather small. The dust
radiatively cools as $\propto T_{\rm D}^{5-6}$ and it is hard to get the dust
much above 40\,K except in very localised regions of active star formation.
Typical FIR colour temperatures of GMCs are in the range 20--40\,K and only a
few ultra-luminous IR galaxy (ULIRG) nuclei get up to 60\,K. Regions with
elevated gas temperatures and hence high $T_{\rm B}({\rm CO})$ are also likely
to be denser (since their high temperatures probably reflect more concentrated
star formation due to higher gas density). The effects of increased temperature
and density will therefore partially compensate each other (see
Equation~\ref{lco}). 

For extragalactic observations where individual GMCs are not resolved,
Equation~\ref{lco} can still be used since the total CO luminosity is just the
sum of that of the individual clouds, provided the GMCs do not overlap both on
the sky and in velocity (unlikely given their small-volume filling factor in
normal galaxies). In the extragalactic observations, the cloud diameters are not
usually measurable but the unresolved apparent brightness temperatures vary as
distance $d^{-2}$ so that the distance to the galaxy may replace $R$ in
Equation~\ref{ldef}. For the external galaxies, Equation~\ref{lco} is translated
to  
\begin{equation}
M_{{\rm H}_2}(\msun)=\alpha_{\rm CO}L_{\rm CO}\quad{\rm K ~km/s ~pc^2},
\end{equation} 
where $\alpha_{\rm CO}\simeq4.9$ (Solomon \& Barrett 1991). 

We should also note that for modest changes in metallicity (affecting the CO
abundance relative to H), $\alpha_{\rm CO}$ will change slowly (until CO becomes
optically thin in large areas) -- this is dramatically demonstrated by the fact
that $^{13}$CO luminosities of clouds are typically only a factor of 4--10 times
lower than those of CO despite the much lower abundance ratio of $^{13}$C/C
(1/40$\rightarrow$1/90). For the optically thick regime where photon trapping is
important for the CO excitation, one can show analytically that the $\alpha_{\rm
CO}$ should scale as CO abundance or metallicity $Z^{-0.4}$
(see Scoville \& Solomon 1974). 

One instance where the above analysis must be modified is in a galactic nucleus
if the gas is smoothly distributed in the central galactic potential. Here, the
CO linewidth will be determined not by the self-gravity of individual clouds but
instead by that of the stars plus the gas, i.e., the linewidth will be larger
than that associated with just the gas mass. There will then  be more CO photons
emitted per unit gas mass. For parameters associated with the most extreme cases
in ULIRGs, the CO conversion factor may be reduced by factors of 2--5 (see
Downes \& Solomon 1998 and Bryant \& Scoville 1999). In the photon trapping
regime, the CO excitation temperature (and hence the brightness temperature) of
the line will vary as the molecular abundance to the $\sim$0.4 power (Scoville
\& Solomon 1974). Thus, one expects that the CO line emissivity to mass
conversion factor will vary with metallicity as $Z^{-0.4}$ at high redshift
where the metallicities are generally lower.\looseness-1

\subsubsection{Lifetimes of GMCs}

Although much current theoretical analysis of star formation adopts the point of
view that GMCs are relatively short-lived (10--30 Myr), I would like to provide
several arguments why this can't be the case. These arguments were well known in
the early days of molecular line astronomy but seem to have been overlooked in
current discussions.

The first argument is simply based on the required conservation or
\emph{continuity of ISM mass between the various ISM phases} (H{\sc i}, H{\sc
ii} and H$_2$). If the GMCs, or more specifically the H$_2$ molecules, are
short-lived then there must be cycling of gas between these phases. The mass
flux from one phase to another  and back must then be in equality (see Scoville
\& Hersh 1979; Koda \textit{et al.} 2009). This mass flux equality is given
simply by
\begin{equation}
\label{mci}
\dot M_{{\rm H_2}\rightarrow{\rm HI+HII}}~(\equiv M_{\rm H_2}/\tau_{\rm H_2})=\dot M_{\rm HI+HII\rightarrow H_2}~(\equiv M_{\rm HI+HII}/\tau_{\rm HI+HII}),
\end{equation}
where we assume relatively little cycling to young stars (since that timescale
was found to be $\sim$1\,Gyr above). In the interior of our Galaxy and most
nearby spirals, the H$_2$ dominates in mass compared to H{\sc i} and H{\sc ii}
by a substantial factor (e.g., more than a factor of five in the Galactic
molecular cloud ring). Thus if cycling between the phases is to be plausible at
all (without disobeying mass conservation!) at these interior radii, the
timescale for a molecule to remain a molecule must be correspondingly longer
than the timescale for a particle in the diffuse phases (H{\sc i} or H{\sc ii})
to remain atomic or ionic. The latter might be expected to be given by the
dynamical time at density a few H cm$^{-3}$ ($\sim$10$^8$\,yr) or by the time
needed to pass between the compressive spiral arms (also
$\sim$\,2\,$\times$\,10$^8$\,yr). Inverting the mass continuity equation
(Equation~\ref{mci}), one finds:
\begin{equation}
\tau_{\rm H_2}={M_{\rm H_2}\over{M_{\rm HI+HI}}}\times\tau_{\rm HI+HII}\sim5\times(2\times10^8)\sim10^9\,{\rm yr}.
\end{equation} 
Thus, for cycling between the phases to work, the characteristic lifetime of a
H$_2$ molecule should be at least several times $10^8$\,yr. And if there is no
such cycling, then the timescale is even longer. Although this calculation was
done for the interior of the Galaxy where H$_2$ dominates H{\sc i}, the estimate 
can be expected to hold elsewhere since it is in the interior of the Galaxy where
there is the most disruptive feedback from star formation and other processes.
In addition, the internal structure of the GMCs in the outer Galaxy is probably
similar to that in the inner Galaxy (i.e., they are equally difficult to
disrupt). 

%-------------------------------------------------------------------------------
\begin{figure} 
\includegraphics[width=1\linewidth]{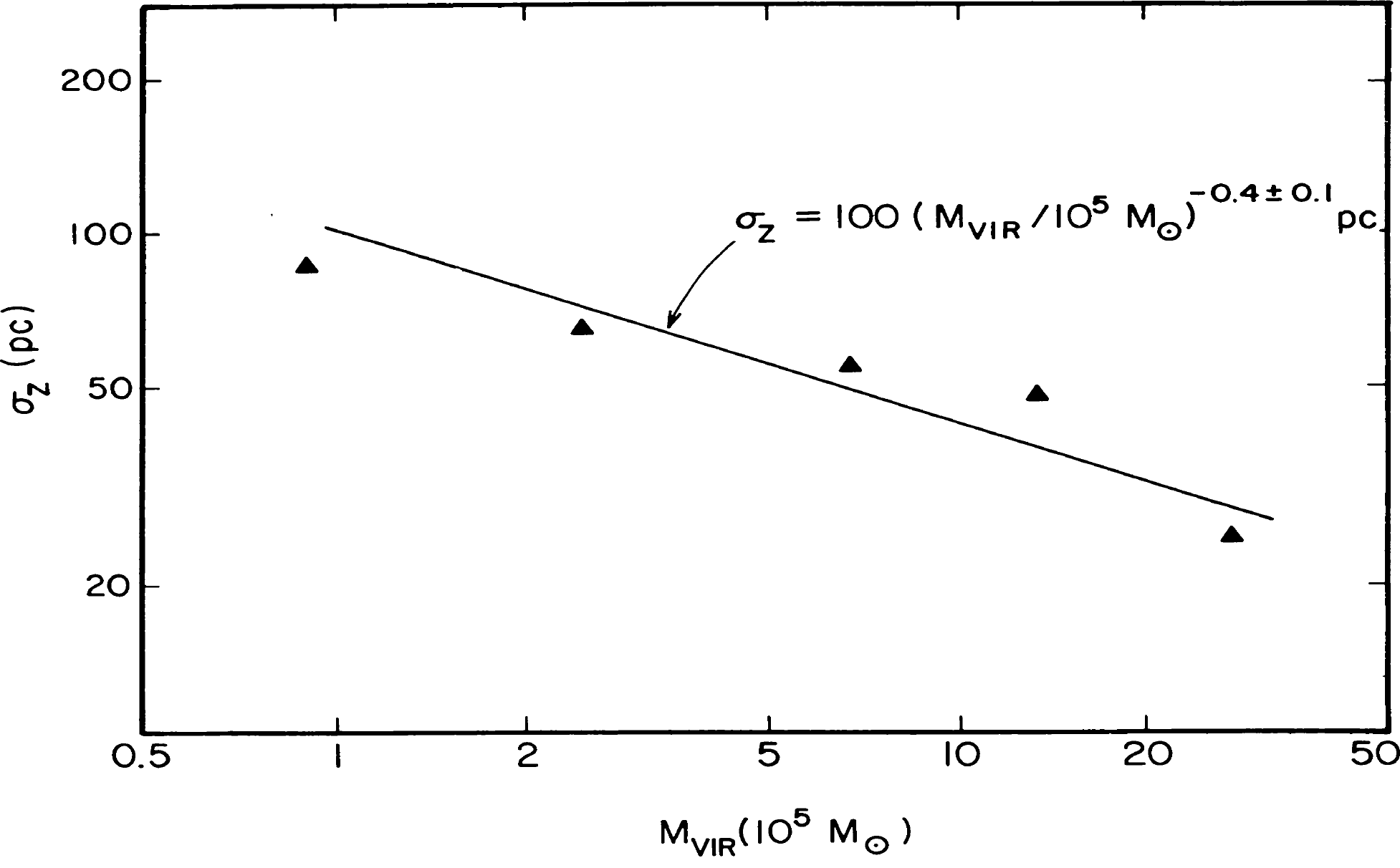}
\caption{The scaleheights of the GMC distribution perpendicular to the Galactic
disk are shown as a function of cloud virial mass (Scoville \textit{et al.}
1987). The nearly square-root dependence on cloud mass suggests an equipartition
of the cloud kinetic energy distribution, requiring that the GMCs last longer
than several $10^8$\,yr.}
\label{scaleht}
\end{figure}
%-------------------------------------------------------------------------------

One must note that this is the timescale for a molecule to remain a molecule and
not necessarily the timescale for a given GMC to retain its physical structure.
Clouds may and probably do grow within the spiral arms and fragment into smaller
clouds outside the arms -- but the molecules cannot cycle all the way back to
the diffuse H{\sc i} or H{\sc ii} phases on a short timescale.

Another line of reasoning suggesting long GMC lifetimes is the fact that, in
general, \emph{feedback processes from star formation within the GMCs are
inefficient} at disrupting the massive clouds (Scoville \& Hersh 1979; Scoville
2003; Murray \textit{et al.} 2010). Most of the stellar feedback will be
radiated and dissipated in shock fronts or leak out the cloud boundaries (as is
surely the case in M\,42). The total momentum needed to unbind the clouds is
enormous, typically $\sim$\,5--10\,$\times$\,10$^5$\,\msun\,km/s. Lastly, I point out that
the scaleheights (perpendicular to the Galactic disk) of the GMCs as a function
of their mass suggest that they have achieved \emph{approximate equipartition}.
In Fig.~\ref{scaleht}, the $z$-scaleheights of clouds indicate that
$\sigma_z\propto M_{\rm vir}^{-0.4}$, i.e., approximately the 1/2 power expected
for equipartition. Equipartition requires that the clouds survive at least
several cloud-cloud scattering times. The latter varies with the local Galactic
space density of clouds but is almost always $\geq10^8$\,yr.

\subsubsection{GMC supersonic internal motions}

I mentioned earlier that the GMCs have highly supersonic internal motions as
determined from the molecular line Doppler widths and that to date there is no
satisfactory explanation. The long cloud lifetimes highlight this problem. The
maps of molecular emission within individual clouds show that the molecular
emissions and their large velocity spread are fairly smooth across the projected
area of the clouds, implying that the area covering factor is $\gtrsim$\,1. If
this is the case then the supersonic turbulent gas parcels will necessarily
collide and dissipate their energy with the time it takes to cross the cloud
diameter. This timescale is 40\,pc/3\,km/s~$\sim$\,10$^7$\,yr. Thus the turbulent
energy required to support the GMCs against gravitational collapse needs to be
replenished within a similar timescale. (Reduction of the dissipation by
postulating that magnetic fields are driving these motions requires very strong
fields and does not change this estimate significantly since the collisional
dissipation can still occur along the field direction.) 

Although the feedback energy released from embedded young stars is of the right
order of magnitude to replenish the turbulent energy, most of the energy
released from the protostellar outflows is radiated away in high-velocity shock
fronts and is also deposited on small length scales rather than scales
comparable to the cloud size needed for the large scale support. Turbulent
cascades generally transport energy from large scale to small scale (except in
2D turbulence, see Robertson \& Goldreich 2012) so it is difficult to see how
the stellar feedback can maintain the turbulent energy.

A possible source of the turbulent energy might be large-scale `corrugations'
in the galactic plane mass distribution. It should be pointed out that despite
the apparent virial equilibrium of the GMCs (in the sense that their
gravitational potential energy is approximately twice their internal kinetic
energy), the morphologies of the clouds do not appear virialised and spherical
-- this provides a strong argument that they are constantly being distorted by
the external force gradients. Those associated with normal density wave spiral
structure pass by only every $\sim$\,2\,$\times$\,10$^8$\,yr and are therefore not
sufficient. Jog \& Ostriker (1988) propose that GMC-GMC scattering is viable to
maintain the internal motions but such scattering occurs also on a typical
timescale of only every $\sim$\,10$^8$\,yr, which seems too long compared to the
dissipation timescale. 

D'Onghia \textit{et al.} (2011) have recently suggested that many observed spiral
structures may be stochastic in nature, where mass seeds corresponding to GMC
masses (very likely the GMCs themselves) will induce a mass overdensity in the
stellar disk via the swing-amplification process. Is it viable that these
associated stellar disk mass enhancements actually feed back gravitational
perturbation energy into the internal supersonic motions of the GMC gas?\looseness-2

\subsubsection{Summary}

In summary, the observations of Galactic GMCs and local galaxies indicate that :
\begin{enumerate}[(a)]\listsize
\renewcommand{\theenumi}{(\alph{enumi})}

\item The CO emission arises from discrete, self-gravitating clouds (GMCs).

\item These GMCs have high internal supersonic motions such that their overall
kinetic energy content is dominated by turbulence with effective pressure 100
times the gas thermal pressure and 100 times the pressure of the external
diffuse ISM.
 
\item The GMCs have a mass spectrum $N(M)\propto M^{-1.6}$ with the midpoint in
the mass contributions at cloud diameter $\sim40$\,pc and mass
$4\times10^5$\,\msun. For this size cloud, the mean density is 
$n_{\rm H_2}\sim180$\,cm$^{-3}$ -- larger clouds have lower density and smaller 
clouds higher mean density.

\item For such self-gravitating clouds with optically thick CO lines, the
emitted CO luminosity is approximately proportional to the cloud mass with a
constant of proportionality scaling as $T/(\rho)^{1/2}$ (i.e., $\alpha_{\rm CO}$
varying as $\rho^{1/2}/T$) and having much lower than linear dependence on
metallicity.

\item The lifetimes of the H$_2$ and probably the GMCs are apparently larger than 
$10^8$\,yr and possibly $10^9$\,yr based on simple mass-conservation
arguments in the ISM between the diffuse and dense phases. This is consistent
with what one might expect based on the great inertia of the GMCs (i.e.,
resistance to disruption/dissociation) and the fact that their extraordinarily
high effective internal pressures (compared to the external diffuse phases)
makes it very difficult for pressure disturbances in the external medium to
significantly influence their internal structure. 

\item The maintenance of the supersonic motions within GMCs, providing support
against collapse, remains unsolved since this energy must be replenished on the
timescale of a few $10^7$\,yr, yet the clouds last $>10^8$\,yr. Internal
feedback from star formation deposits energy on length scales which are too
small. The non-spherical shapes of the GMCs suggest large-scale external force
gradients may be responsible, such as those associated with GMC-GMC scattering,
clouds motions perpendicular to the galactic disk potential or the recently
proposed stochastic spiral structure.
\end{enumerate}

%
%%%%%%%%%%%%%%%%%%%%%%%%%%%%%%%%%%%%%%%%%%%%%%%%%%%%%%%%%%%%%%%%%%%%%%%%%%%%%%%%
%

\section{Star formation} 
 
Here I describe the various probes of star formation in galaxies and their
correlation with the molecular gas contents. I devote considerable effort to
developing an interpretive framework for the FIR emission from
optically thick dust clouds.
 
\subsection{Probes of star formation}
 
There are a number of observational tracers which have been developed to measure
the star formation rates (SFRs) in galaxies (see Calzetti, this volume). The
H{\sc i} recombination lines in the visible and near-IR (e.g., H$\alpha$
and P$\alpha$) have the fluxes proportional to the H{\sc ii} region emission
measures, hence the OB star formation rate over the last 10$^7$\,yr. The
restframe (far-)UV continuum ([F]UV), at $\lambda<2000$\,\AA, arising
from  hot, early-type stars has been used to infer the SFRs for large samples of
galaxies. Both the emission lines and the UV continuum can be severely
attenuated by dust extinction in star-forming regions. Even for the galaxies
with detected UV continuum, the extinction corrections are often factors of
5--10! For the dust-obscured star formation, the FIR luminosity ($\lambda_{\rm
rest}=8-1000\,\mu$m) provides a much more reliable measure of the SFR. These FIR
SFRs can now be obtained for large samples of galaxies using observations from
the \textit{Spitzer} and \textit{Herschel} space telescopes, albeit with
relatively low angular resolution and sensitivity to SFR (compared to the UV). A
summary of all these techniques, including relevant SFR equations, appears in
the recent paper by Murphy \textit{et al.} (2011) so I will not detail all of
them here -- instead I will focus on developing a physical understanding of the
IR emission.

\subsection{Infrared emission}
\label{ire}
 
The FIR emission from both star-forming and active galactic nucleus (AGN) sources
arises from dust surrounding these sources which has been radiatively heated by
absorption of the outflowing photons. Here I develop the logical steps for
interpreting the IR emission since I have not seen this done systematically
elsewhere.
 
\begin{figure} 
\includegraphics[width=\linewidth]{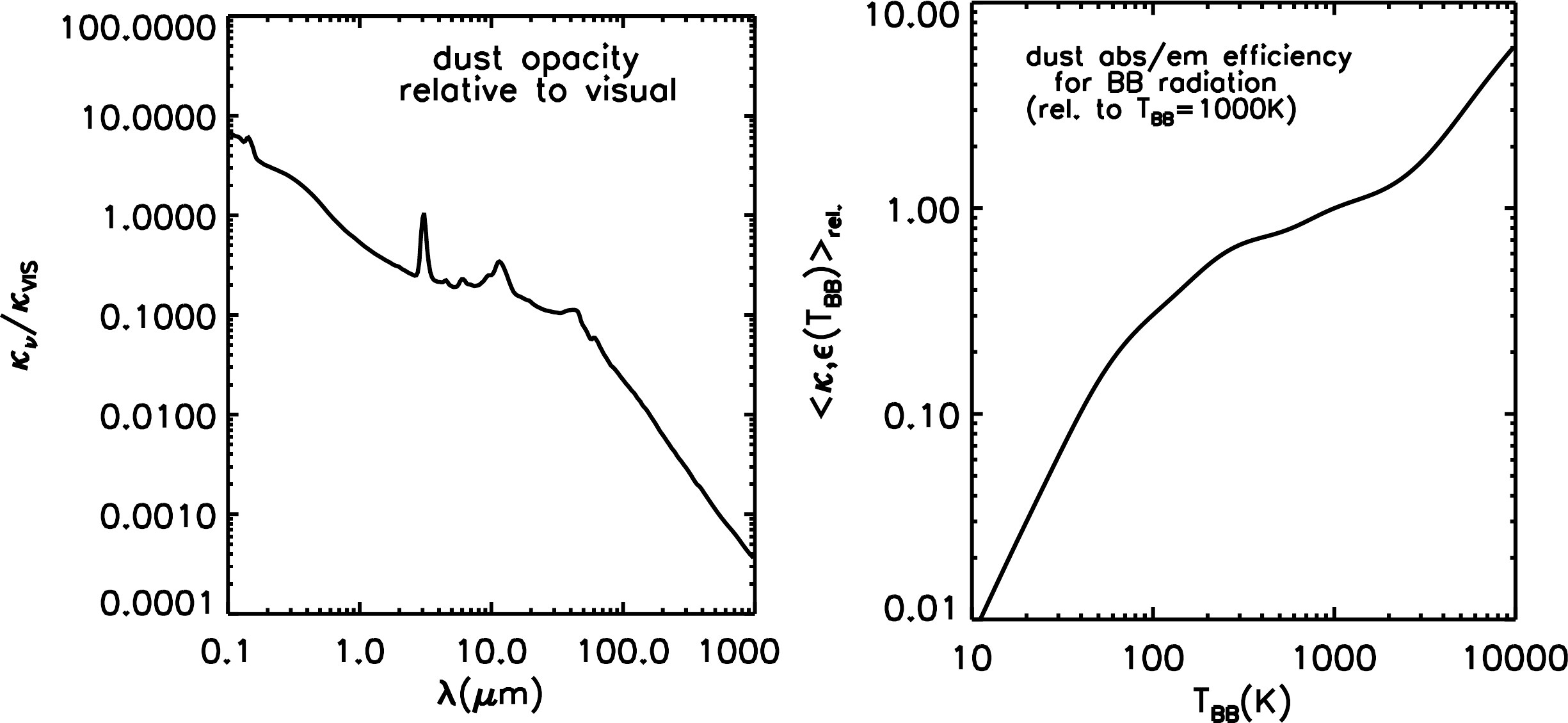}
\caption{Left panel: the dust absorption opacity as a function of wavelength
(Isella \textit{et al.} 2010) for standard interstellar grain composition (12\%
silicates, 27\% organics, and 61\% water ice; Pollack \textit{et al.} 1994), and
size distribution $n(a)\propto a^{-3.5}$ from 0.01 to 1\,$\mu$m radius. The
absorption coefficient is normalised relative to that at visual wavelength
(6060\,\AA, $\kappa_{V}=2.3\times10^4$\,gr$^{-1}$) -- for a standard gas-to-dust
abundance ($\sim$100 in mass) with $N_{\rm H+2H_2} = 2\times10^{21}$\,cm$^{-2}$
per mag of visual extinction ($A_V$). An important feature of the extinction
curve is the relatively flat broadband absorption coefficient over the range
$\lambda$\,=\,2--50\,$\mu$m. In the right panel, I show the Planck integrated
absorption/emission coefficient as a function of blackbody temperature. Due to
the `flat' absorption coefficient in the near- and mid-IR (NIR/MIR), the
Planck-integrated absorption and emission efficiency is quite independent of
temperature from 1000\,K down to 100\,K. The dust opacity curve was calculated
by Andrea Isella and is available from him or myself.}
\label{grains}
\end{figure}

The dust temperatures are determined by radiative equilibrium at distance $R$ 
from a central source of luminosity $L$, with 
\begin{equation}
4\pi a_{\rm d}^2\langle\epsilon_{\nu}\rangle\sigma T_{\rm d}^4=\pi a_{\rm d}^2\langle\kappa_{\nu}\rangle L/(4\pi R^2),
\end{equation} 
where $a_{\rm d}$ and $T_{\rm d}$ are the dust grain radius and temperature, and 
$\langle\epsilon_{\nu}\rangle$ and $\langle\kappa_{\nu}\rangle$ are the dust
emission and absorption efficiencies (shown in Fig.~\ref{grains}, left panel),
weighted, respectively, by the Planck spectrum at the local dust temperature and
that of the luminosity source heating the dust. If the emission efficiency
varies as $1/\lambda$, i.e., $\epsilon_{\nu}\propto T_{\rm d}$, then 
%
%-------------------------------------------------------------------------------
%-------------------------------------------------------------------------------
%
\begin{equation}
T_{\rm d}\propto{L^{1/5}\over{R^{2/5}}} 
\end{equation} 
(Goldreich \& Kwan 1974). More generally, 
\begin{equation}\label{td}
T_{\rm d}\propto{L^{1/4}\over{R^{1/2}}}\left\{{\langle\kappa_{\nu}(T_{L})\rangle\over{\langle\epsilon_{\nu}(T_{\rm d})\rangle}}\right\}^{1/4},
\end{equation} 
where $\langle\kappa_{\nu}(T)\rangle$ and $\langle\epsilon_{\nu}(T)\rangle$
are shown in the right panel of Fig.~\ref{grains}. Due to the flatness of the
broadband absorption coefficient from 2 to 50\,$\mu$m, the grain absorption and
emission efficiencies integrated over black-body spectra are decreasing only
modestly from $T_{\rm BB}$\,=1000--100\,K (Fig.~\ref{grains}, right). 

\begin{figure} 
\includegraphics[width=\linewidth]{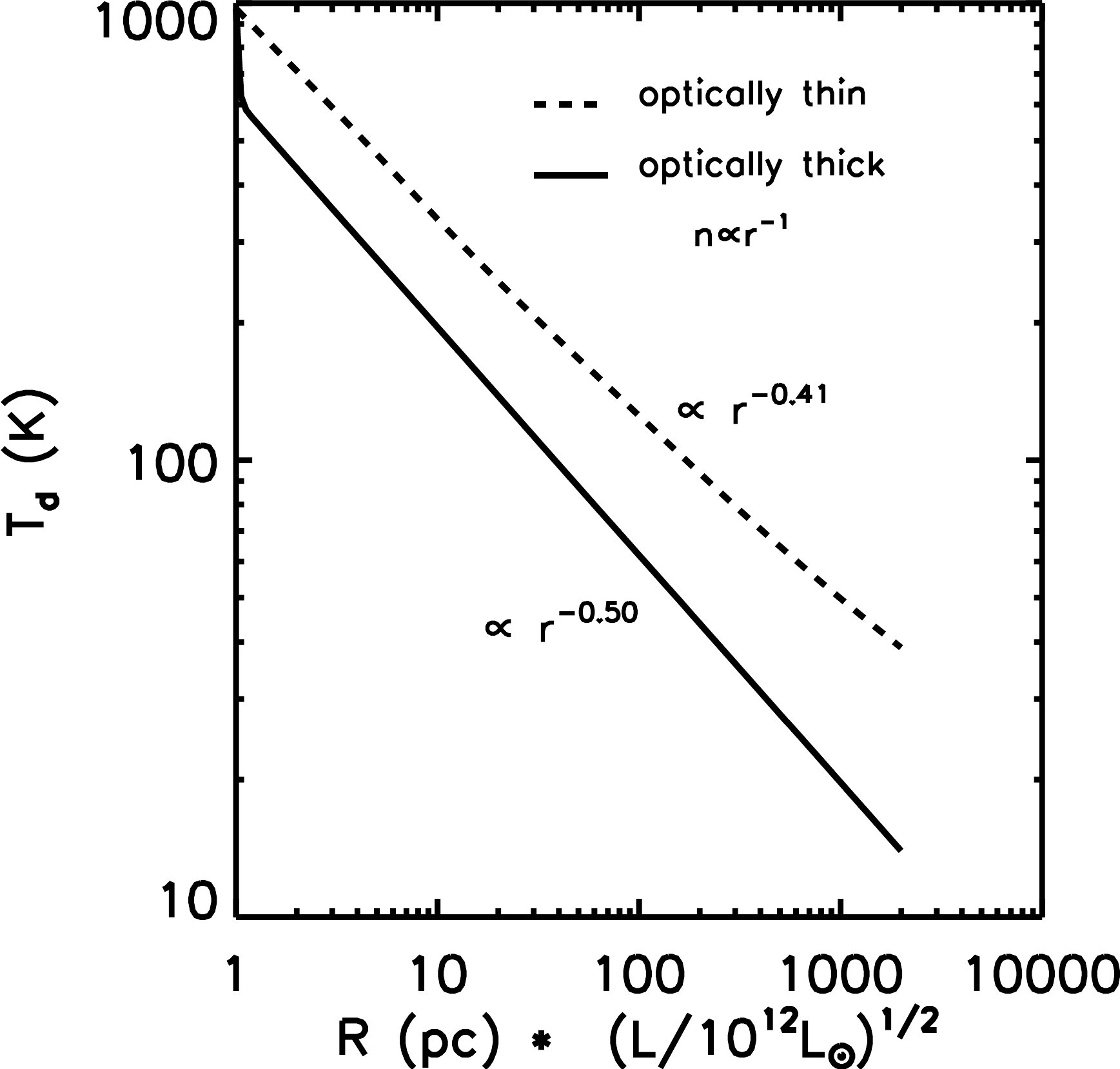}
\caption{The temperature of dust heated by a central luminosity source is shown
as a function of radius for optically thin dust (with all the heating provided
by a $\sim$\,10$^4$\,K blackbody) and for very optically thick dust (where after the
innermost radius, the heating is due to dust at the same radius and
temperature). This resembles closely the very optically-thick case since the
temperature gradients are generally quite shallow. It is important to appreciate
that the radial physical scalelengths will simply stretch homologously with
changing source luminosity. Thus the same curves can be used for sources with
very much higher or lower luminosity. For optically-thick FIR dust emission the
dust temperature varies as $T_{\rm d}\propto r^{-1/2}$.  In real emission
sources, the dust is optically thin right at the inner boundary, optically thick
at intermediate radii and then thin at the outer radii so the temperature
profiles must be pieced together at the appropriate radii. (These equilibrium
dust temperatures were calculated using the grain absorption coefficients shown
in Fig.~\ref{grains} and the inner boundary is taken to be where the grain
temperature is $\sim$\,1000\,K, i.e., somewhat below the expected dust sublimation
temperatures.)}
\label{tds}
\end{figure}

For an optically thin dust distribution surrounding a luminous source, the dust
heating is mainly due to the central short-wavelength source, but for an
optically thick dust envelope, the interior dust does not see the central source
and is instead heated by secondary radiation. This secondary radiation, having
longer wavelength than the central stellar or AGN source, is less efficiently
absorbed and the dust in the optically thick case will therefore be colder (than
it would be if exposed to the shorter-wavelength primary photons of the central
source). In very optically-thick cases, the $\langle\kappa_{\nu}(T_{L})\rangle$ in
Equation~\ref{tds} can be evaluated approximately with $T_{L}(R)\sim T_{\rm
d}(R)$, appropriately weighted over nearby radii. 

%-------------------------------------------------------------------------------
%-------------------------------------------------------------------------------

Figure~\ref{tds} shows the computed dust temperatures for the optically thin and
optically thick cases, evaluated from Equation~\ref{td}. In the optically-thick
regime, the fall-off in dust temperature is $\propto r^{-1/2}$ -- an extremely
simple form due to the fact that the photons heating the grains and those
emitted by the grains have similar wavelength distributions. One will therefore
have  $\langle\kappa\rangle\sim\langle\epsilon\rangle$ in the very optically-thick 
dust clouds.  

The radial scale in Fig.~\ref{tds} is for a central source luminosity of
$10^{12}$\,\lsun, appropriate to ULIRGs and submm galaxies (SMGs), but the
radial distances can be scaled as $L^{1/2}$ for other luminosities (see
Equation~\ref{td}). Thus, these temperature profiles can be equally well applied
to a dust cloud surrounding a luminous protostellar cluster of luminosity
$10^3$--$10^6$\,\lsun. The temperature profiles in Fig.~\ref{tds} start at
$\sim$1000\,K which is a little below the dust sublimation temperatures
($\leq$\,1500\,K). Inside this radius, the dust will not survive. For a less
luminous source, this inner radius will scale inwards; but the physical scales
will all change by the same $L^{1/2}$ and the modelling remains homologous.

\subsection{Dust optical depth: \textbf{$\tau<1$} or \textbf{$\tau>1$}?}

It is quite trivial to observationally distinguish the optically thick and thin
sources since the latter will have a power-law flux distribution on the 
short-wavelength side of the peak whereas the former will appear more exponential (see
Scoville \& Kwan 1976). (If the dust distribution is clumpy, photons can emerge
from the inner regions with hot dust -- thus, in optically-thick dust clouds, a
non-exponential spectral energy distribution [SED] at short wavelengths might
also occur; see Fig.~\ref{clumping}.) 

Virtually all IR sources associated with active star-forming regions (Galactic
GMCs and starburst nuclei) are optically thick into the MIR (based on their
sharp short wavelength fall-off). Since the dust opacity is fairly flat across
the wavelength range (2--50\,$\mu$m), one then expects that the source will
become optically thin only at $\lambda>50\,\mu$m, as long as it is optically
thick at \hbox{3--10\,$\mu$m}. Thus, it is most appropriate to employ the optically
thick dust temperature distribution shown as a solid line in Fig.~\ref{tds} and
fit numerically by
\begin{equation}
\label{tdthick}
T_{\rm d}=630{\left(L/10^{12}\lsun\right)^{1/4}\over{R_{\rm pc}^{1/2}}}\quad{\rm K}.
\end{equation}
One can invert Equation~\ref{tdthick} to find the characteristic size of the 
emitting region:
\begin{equation}
\label{rthick}
R_{\rm thick}=100\left(63/T_{\rm d}\right)^2\left(L/10^{12}\lsun\right)^{1/2}\quad{\rm pc},
\end{equation}
if the total IR luminosity and dust temperature are known. The latter might be
derived by fitting the MIR SED, or more crudely, from the wavelength of the
peak. If the dust is opaque, then the standard Planck expression yields $T_{\rm
d}=100\times(51\,\mu$m$/\lambda_{\rm peak})$\,K but if the dust is optically
thin, then the peak wavelength is reduced by a factor $\sim(3/(3+\alpha))$ where
$\alpha$ is the power-law index for the opacity,
$\kappa_{\nu}\propto\nu^{\alpha}$, near the IR peak.
 
In practice, the opacity can never be much greater than unity at the peak. This
is due to the simple fact that the luminosity cannot escape from the inner
regions which have high opacity to the cloud surface. And once the outward
luminosity flux has shifted to wavelengths where the opacity becomes less than
unity, it escapes. This can be seen analytically using the fact that the
emergent emission for a given grain is proportional to
$B_{\nu}\epsilon_{\nu}{\rm e}^{(-\tau_{\nu})}$. At $\lambda>70\,\mu$m,
$\epsilon_{\nu}$\,$\propto$\,$\nu^{\sim1.6}$. If
$\tau_{\nu}$\,=\,$(\nu/\nu_0)^{1.6}$ (i.e., unity at $\nu$\,=\,$\nu_0$), then
the emission of the grain will peak at $\nu_{\rm
m}\simeq(3+1.6)\nu_{0}/(1.6+h\nu_0/kT_{\rm d})$, where
$\tau=(\nu_m/\nu_0)^{1.6}$. Thus, for example, with $T_{\rm d}$\,=\,100\,K, the
blackbody peak at 51\,$\mu$m is shifted to 78 and 395\,$\mu$m for
$\lambda_0$\,=\,102 and 787\,$\mu$m, respectively. At the peaks the optical
depths are $\tau_{\rm peak}$\,=\,1.5 and $3.0$ in the two cases. This
illustration was for an isothermal dust distribution with $\tau$ simplistically
representing the foreground optical depth. This also provides a strong note of
caution -- if one is deriving the dust temperature from the wavelength of peak
emission, the peak will usually be shifted to a longer wavelength where the dust
starts to become transparent. 

\subsection{Dust temperature of the emergent luminosity}

So how can the characteristic dust temperatures be derived? Probably the best
approach, if the dust is believed to be optically thick on the short wavelength
side of the peak (based on an `exponential' rise), is to simply assume that
$\tau\sim1$ at the peak and estimate the temperature of the dust emitting the
bulk of the emergent luminosity from $T_{\rm
d}\gtrsim100\times(80\,\mu$m$/\lambda_{\rm peak})$\,K. To improve on this {\it
ad hoc} correction requires an assumption of the dust density distribution with
radius and modelling the emergent SED (as is done below in
Section~\ref{sed_model}). 

For an optically-thick source, the emergent radiation at each wavelength will be
from a depth where $\tau_{\lambda}$\,$\simeq$\,1. Since the dust opacity falls off
at longer wavelengths and the temperature is falling at larger radii, this
implies,\linebreak somewhat counter-intuitively, that \emph{at wavelengths short of the
peak wavelength, one will sample dust at increasingly larger radii, i.e., lower
and lower temperatures!}
      
\subsection{Star formation rate from \textbf{$L_{\rm IR}$}}
 
Derivation of SFRs from the IR luminosities is fairly straightforward and
robust. This was first done by Scoville \& Young (1983) using the fact that the
bulk of the luminosity from a stellar population at early times is generated
largely by the OBA stars. For those stars an approximately fixed percentage
(13\%) of their initial mass gets processed through the CNO cycle and one knows
the energy produced per CNO process. This derivation is analogous to the `fuel
consumption theorem' of Renzini \& Buzzoni (1986).

%-------------------------------------------------------------------------------
\begin{figure} 
\includegraphics[width=0.5\linewidth]{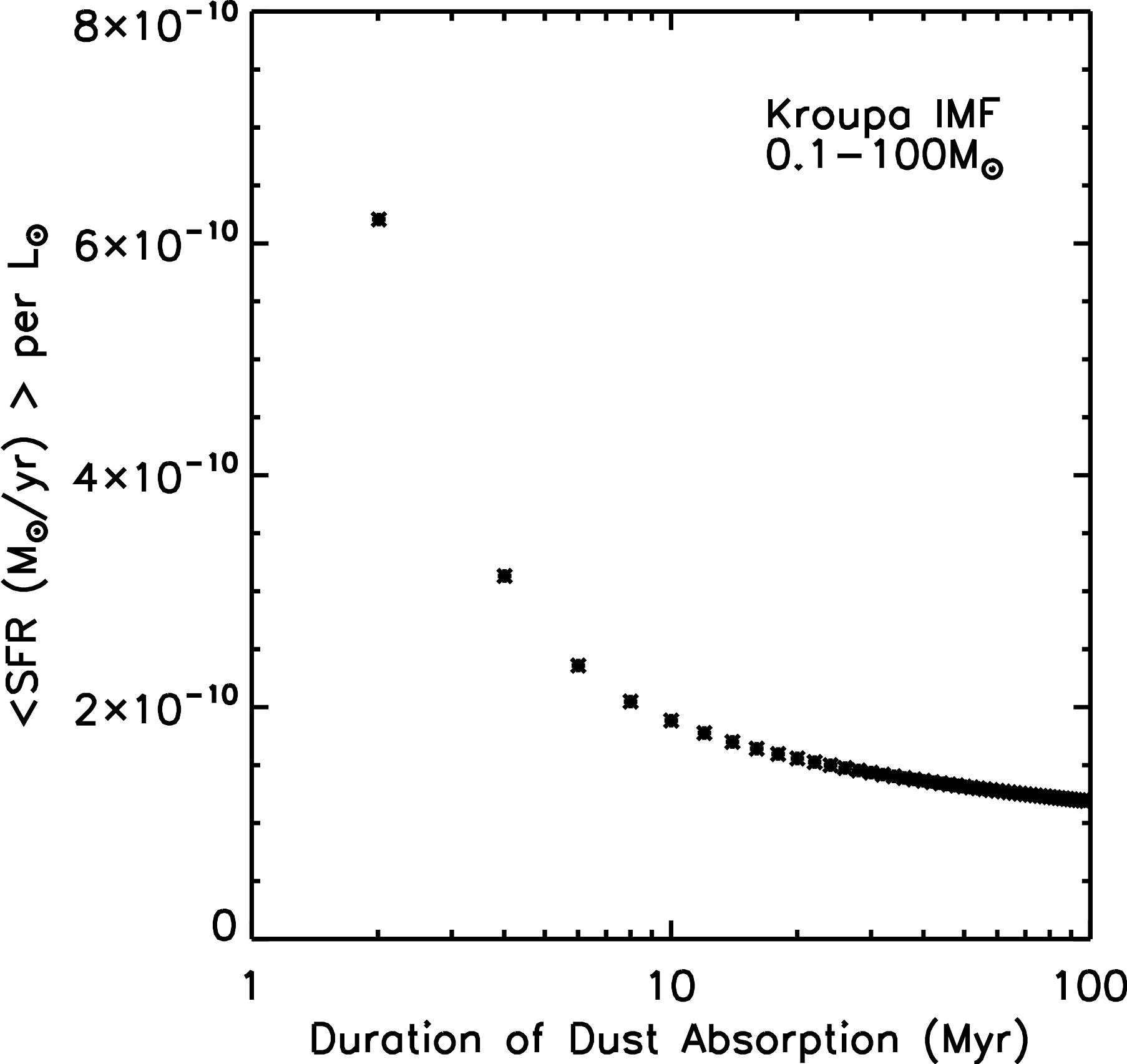}
\caption{The conversion of observed IR luminosities into estimates of the SFR
depends on the duration time of the dust absorption. Here I have used
Starburst99 models with a Kroupa IMF and assumed that all stellar and nebular
photons longward of the Lyman limit are absorbed for the dust duration time (in
Myr) and then none are absorbed after that. For normal star-forming regions this
timescale is $\sim$10\,Myr and for ULIRGs 50--100\,Myr. If the dust envelope
lasts less than 10\,Myr and or is only partially covering, then the
conversion factor is significantly higher -- both are certainly true for an
exposed H{\sc ii} region like the Orion nebula.}
\label{sfr_ir}
\end{figure}
%-------------------------------------------------------------------------------
 
A more precise, modern approach is to run starburst models (e.g., Starburst99;
Leitherer \textit{et al.} 1999) and integrate up the luminosity as a function
of time. One must assume a stellar initial mass function (IMF), and its mass
range, and decide which photons will be absorbed by dust and for how long the
dust absorption persists. In Fig.~\ref{sfr_ir} I show results obtained using
Starburst99 with a Kroupa IMF (0.1 to 100\,\msun, Kroupa 2001) as a\linebreak
function of the duration of the dust absorption. The latter quantity is probably
$\sim10^7$\,yr for Galactic star-forming regions but could be closer to
$10^8$\,yr for merging starburst galaxies where the dust is more widely
distributed. For this calculation, I assume all stellar and nebular photons
longward of the Lyman limit (912\,\AA) are absorbed by the surrounding dust. 
From Fig.~\ref{sfr_ir}, one could reasonably compute a SFR given by
\begin{equation}\label{SFR_i}
{\rm SFR_{IR}}=2-1.2\times10^{-10}~(L_{\rm IR}/\lsun)\quad{\rm \msun/yr},
\end{equation}
with the lower value being appropriate to the ULIRGs which have longer duration
for the dust shrouding. The standard relation given by Murphy \textit{et al.}
(2011) corresponds to $1.5\times10^{-10}$\,\msun~yr$^{-1}$ ($L_{\rm IR}/\lsun$).
The simpler derivation outlined above (based on the CNO cycle energy production)
is quite similar to Equation~\ref{SFR_i} after one corrects for the mass going
into non-OBA stars for a Kroupa IMF. 
 
\subsection{Dust and ISM mass estimates}
\label{rjtail}  
  
On the long-wavelength Rayleigh-Jeans (R-J) tail of the FIR emission, the dust
will be optically thin and the observed continuum fluxes provide an excellent
means of determining the overall mass of dust. If the dust-to-gas abundance is
normal, this dust mass can then be scaled to estimate the overall mass of ISM
within a star-forming region or a distant galaxy.
  
On the optically-thin R-J tail of the IR emission, the observed flux density is 
given by
\begin{equation}
F_{\nu}=\kappa_{\nu}T_{\rm dust}\nu^2 M_{\rm dust}/(4\pi d_{\rm l}^2),
\end{equation}
or in terms of the dust opacity per unit gas mass, $\kappa_{\nu}(\rm
ISM)$\,=\,$\kappa_{\nu}\times M_{\rm dust}/M_{\rm ISM}$,
\begin{equation}
\label{f850eq}
F_{\nu}=\kappa_{\nu}({\rm ISM})~T_{\rm dust}\nu^{2}M_{\rm ISM}/(4\pi d_{\rm l}^2),
\end{equation}
where $d_{\rm l}$ is the source luminosity distance. In normal star-forming
galaxies, the majority of the dust is at $\sim$20--25\,K, and even in the most
vigorous starbursts like Arp\,220 the FIR/submm emission is dominated by dust at
temperatures $\leq$\,45\,K. Thus the expected variations in $T_{\rm dust}$ have 
less than a factor two effect on the observed flux. 

The dust opacity per unit mass of total ISM gas, $\kappa_{\nu}({\rm ISM})$ in
Equation~\ref{f850eq}, can be calibrated from the extensive submm observations
of nearby galaxies. Seventeen of the nearby SINGS survey (\textit{Spitzer}
Infrared Nearby Galaxies Survey; Kennicutt \textit{et al.} 2003) galaxies have
good total submm flux measurements obtained with the SCUBA instrument, mounted
at the James Clerk Maxwell Telescope, at 850\,$\mu$m (see Draine \textit{et al.}
2007), as well as good measurements of the total molecular (H$_2$) and atomic
(H{\sc i}) gas masses. 

%-------------------------------------------------------------------------------
\begin{figure}
\vspace{1mm}
\includegraphics[width=\linewidth]{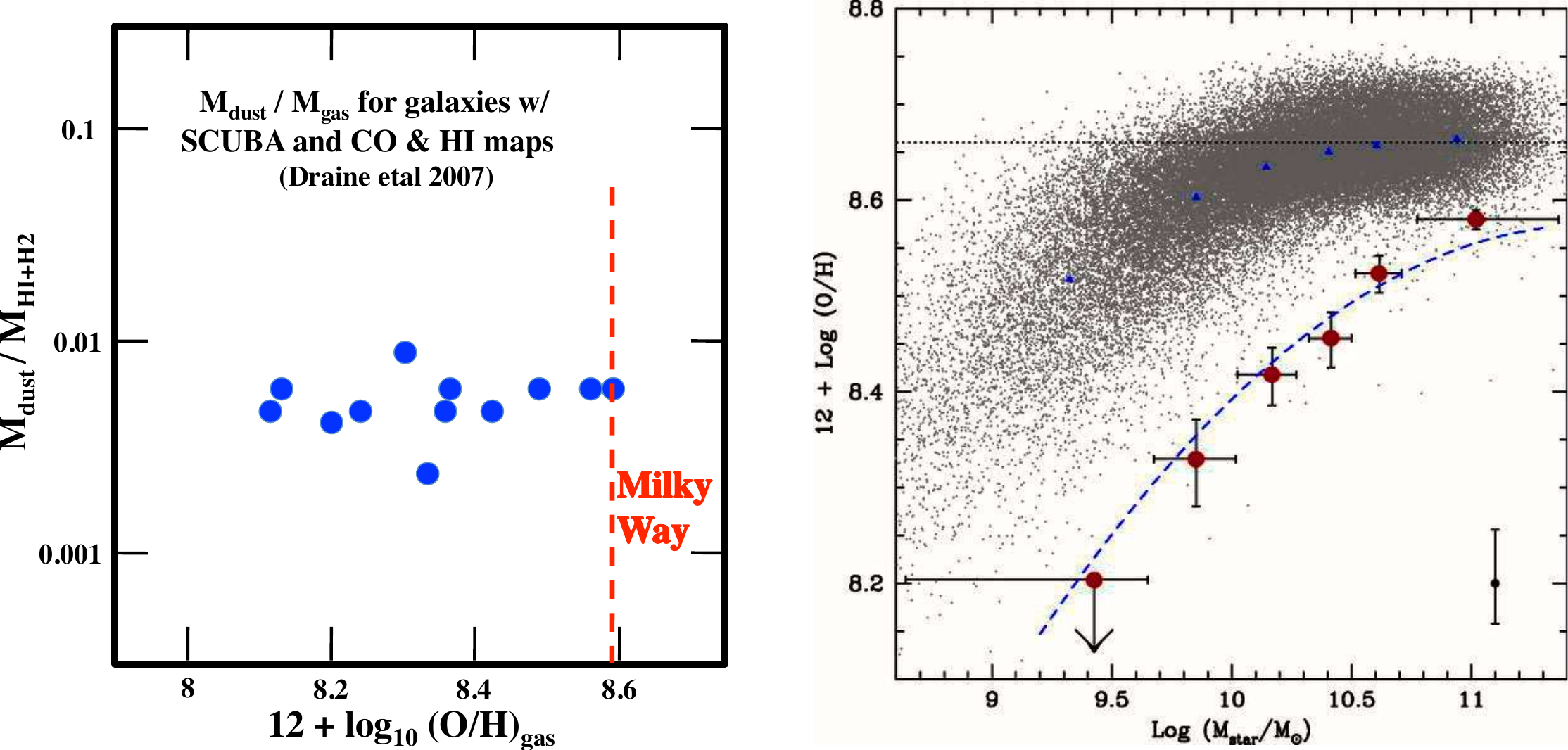}
\caption{Left panel shows the dust-to-gas mass ratios derived by Draine
\textit{et al.} (2007) for galaxies from the SINGS nearby galaxy survey,
selecting only those galaxies with both SCUBA 850\,$\mu$m fluxes and complete
maps of the H$_2$ and H{\sc i} gas; right panel shows the mass-metallicity
relation for low-$z$ galaxies (grey points) and binned values for $z$\,$\sim$\,2
galaxies (red dots) from Erb \textit{et al.} (2006). Over a range of 0.5\,dex in
metallicity below that of the Milky Way there is little variation in the
dust-to-gas ratios. Since the galaxies selected here are massive, their
metallicities even at $z$\,=\,2 are expected to be within this range based on
emission line ratios. Lower-metallicity irregular galaxies probably do show a
decrease in the dust abundance (see Draine \textit{et al.} 2007).}
\label{sings} 
\end{figure}
%-------------------------------------------------------------------------------

In Fig.~\ref{sings}, the derived dust-to-gas (H$_2$ + H{\sc i}) mass ratios from
Draine \textit{et al.} (2007) are shown for the galaxies having SCUBA
850\,$\mu$m measurements, for a range of spiral type (Sa to Sd) and as a
function of mean metallicity. (Equivalent data for low-redshift elliptical
galaxies are not available from Draine \textit{et al.} 2007 due to lack of gas
mass measurements in the early type galaxies.)
 
Figure~\ref{sings} shows that over a range of $\sim$0.5\,dex in metallicity,
there is little empirical evidence of variation in the dust-to-gas mass ratios
and hence the submm flux. If one includes even lower-metallicity galaxies that 
do not have SCUBA 850\,$\mu$m fluxes (hence the submm fluxes must be
extrapolated from an overall SED fit using shorter-wavelength observations),
there is evidence for a metallicity dependence in the dust-to-gas ratio. 
Lastly, it should be emphasised that although some variations in the ratio of
submm flux to ISM mass may be expected, ISM mass estimates at $\sim$30\%
accuracy (see Fig.~\ref{sings}) are still likely to be at least as accurate as
those from CO line measures and much quicker, enabling large samples to be
analysed.
 
In view of the large uncertainty in the submm absorption coefficient $\kappa$
and its scaling with frequency, we adopt an empirical approach based on submm
observations of local galaxies where H{\sc i} and H$_2$ masses have been
estimated. For the local galaxies on which to base this empirical approach, it
is vital that both the submm fluxes and ISM masses are global values.  In
addition as a check on the reliability of the submm measurements we require two 
long wavelength flux measurements so one can check if there is reasonable
consistency with expected values of the spectral index $\beta$.

\begin{table}
\begin{center}
\caption{Low-$z$ galaxies with submm \& ISM data\label{tab:low_z}}
\begin{tabular}{cccccc}
\hline\hline
Galaxy & Distance & S$_{\nu}(450\,\mu)$ & S$_{\nu}(850\,\mu)$ & $\log M_{\rm HI}$ & $\log M_{\rm H_2}$ \\
~ & (Mpc) & (Jy) & (Jy) & (\msun) & (\msun)\\
\hline
NGC\,4631    &   9.0  & 30.7  & 5.73  &  9.2  &  9.5 \\
NGC\,7331    &  15.7  & 18.5  & 2.98  &  9.4  &  9.7 \\
NGC\,7552    &  22.3  & 20.6  & 2.11  &  9.7  & 10.0 \\
NGC\,598     &  76.0  &  2.3  & 0.26  &  9.8  & 10.1 \\
NGC\,1614    &  62.0  &  1.0  & 0.14  &  9.7  & 10.0 \\
NGC\,1667    &  59.0  &  1.2  & 0.16  &  9.3  &  9.6 \\
Arp\,148     & 143.0  &  0.6  & 0.09  &  9.9  & 10.2 \\
1ZW107       & 170.0  &  0.4  & 0.06  & 10.0  & 10.3 \\
Arp\,220     &  79.0  &  6.3  & 0.83  & 10.0  & 10.3 \\
12112+0305   & 293.0  &  0.5  & 0.05  & 10.3  & 10.6 \\
Mrk\,231     & 174.0  &  0.5  & 0.10  &  9.8  & 10.1 \\
Mrk\,273     & 153.0  &  0.7  & 0.08  &  9.9  & 10.2 \\
\hline
\end{tabular}
\end{center}
\end{table}

For the spectral index $\beta$ of the R-J tail (S$_{\nu}$ varying as
$\nu^{\beta}$), the observed flux ratios of submm galaxies can vary between 3
and 4. For most dust models the spectral index of the opacity is typically 1.5
to 2, implying $\beta$\,=\,3.5 to 4. Empirical fits to the observed long
wavelength SEDs give $\beta$\,=\,3.5 to 4 (Dunne \& Eales 2001; Clements
\textit{et al.} 2010) for local galaxies. For high-$z$ submm galaxies, the
spectral index can be between 3.2 and 3.8, but in most cases the
shorter-wavelength point is getting close the IR peak in the restframe and
therefore not strictly on the R-J tail. In the following, we adopt
$\beta$\,=3.8.

Table~\ref{tab:low_z} lists local spiral or star-forming galaxies for which both
450 and 850\,$\mu$m measurements exist with good signal-to-noise and for which
global fluxes were estimated.

The 850\,$\mu$m fluxes were then converted to specific luminosity $L_{\nu(850)}$, 
using
\begin{equation}
L_{\nu}=10^{-23}~4\pi d_{L}^2~S_{\nu}({\rm Jy})=1.19\times10^{27}~S_{\nu}({\rm Jy})~d_{L}^2 ({\rm Mpc}).
\end{equation}
%
%-------------------------------------------------------------------------------
\begin{figure} 
\includegraphics[width=0.75\linewidth]{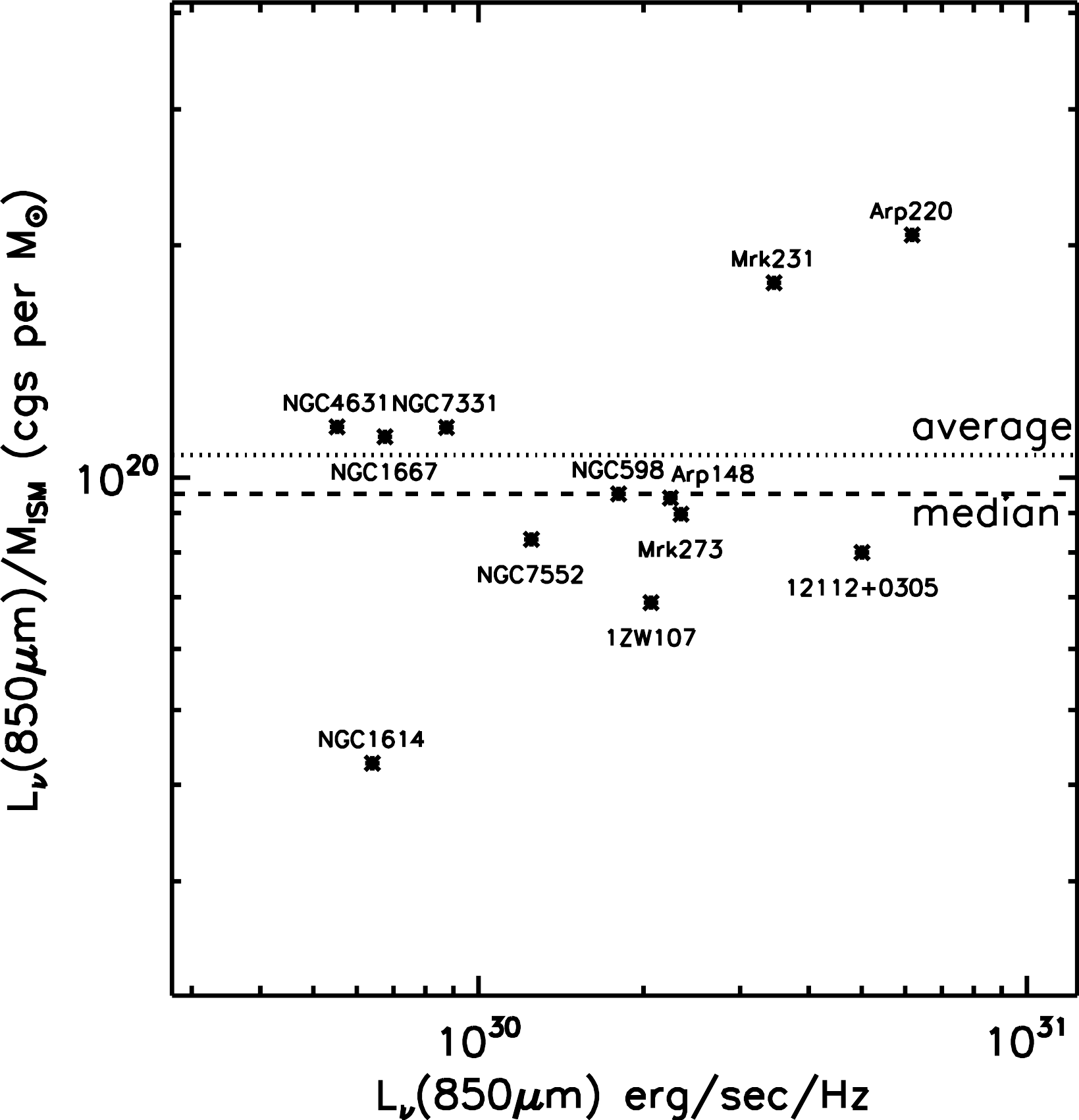}
\caption{The ratio of $ L_{\nu}$ at 850\,$\mu$m to $M_{\rm ISM}$ is shown for a
sample of low-$z$ spiral and starburst galaxies from Dale \textit{et al.} (2005)
and Clements \textit{et al.} (2010). The average and median values for the
sample are shown by horizontal lines.}
\label{sings_ratio} 
\end{figure}
%-------------------------------------------------------------------------------
%
Figure~\ref{sings_ratio} shows the ratio $L_{\nu_{850}}/M_{\rm ISM}$ as a 
function of $L_{\nu_{850}}$ where $M_{\rm ISM}=M_{\rm HI}+M_{H_2}$. Based 
on this plot, we then adopt as a working value
\begin{equation}
{L_{\nu_{850}}\over M_{\rm ISM}}=1\times10^{20}\quad{\rm erg/s/Hz/\msun}. 
\end{equation}
Define this mean value as
\begin{equation}
\alpha_{850} =1\times10^{20}\quad{\rm erg/s/Hz/\msun}.
\end{equation}
For high-redshift observations,
\begin{equation}
{\rm observed:}~(\nu_{obs}~S_{\nu_{obs}})~=~{\rm restframe:}~\left(\nu_{\rm rest}{L_{\nu_{\rm rest}}\over4\pi d_{L}^2}\right),
\end{equation}
with
\begin{equation}
\nu_{\rm rest}=\nu_{\rm obs}\times(1+z).
\end{equation}
Therefore,
\begin{eqnarray}
S_{\nu_{\rm obs}}&=&{\nu_{\rm rest}\over\nu_{\rm obs}}~{L_{\nu_{\rm rest}}\over4\pi d_{L}^2} \nonumber\\
                 &=&(1+z)~{L_{(1+z)\nu_{\rm obs}}\over4 \pi d_{L}^2},
\end{eqnarray}
with
\begin{equation}
%L_{\nu_{\rm rest}}=\alpha_{850}~((1+z)\nu_{\rm obs} 350\,{\rm GHz})^{\beta}~M_{\rm ISM}.
L_{\nu_{\rm rest}}=\alpha_{850}~\left((1+z)~{\nu_{\rm obs}\over350\,{\rm GHz}}\right)^{\beta}~M_{\rm ISM}.
\end{equation}
Then, 
\begin{equation} 
S_{\nu_{\rm o}}={1+z\over4\pi d_{L}^2}~\alpha_{850}~\left( (1+z)\nu_{\rm obs}\over350\,{\rm GHz} \right)^{\beta}~M_{\rm ISM},
\end{equation}
where the 350\,GHz is the frequency corresponding to 850\,$\mu$m.

Normalising to $M_{\rm ISM}$\,=\,2$\times$10$^{10}$\,\msun\ with $\beta=3.8$, 
\begin{equation} 
S_{\nu_{\rm obs}}=\alpha_{850}~2\times10^{10}\times{M_{\rm ISM}\over2\times10^{10}}~(1+z)^{4.8}~\left({\nu_{\rm obs}\over350\,{\rm GHz}}\right)^{3.8} {1\over4\pi d_{L}^2},
\end{equation} 
\begin{equation}
\label{rj_dust} 
S_{\nu_{\rm obs}} ({\rm mJy})=1.67~{M_{\rm ISM}\over2\times10^{10}}~(1+z)^{4.8}~\left({\nu_{\rm obs}\over350\,{\rm GHz}}\right)^{3.8}~{1\over d_{L}^2({\rm Gpc})},
\end{equation}

at $z=0.3, 1, 2$ and 3, $d_L({\rm Gpc})=1.5, 6.6, 15.5$ and 25.4\,Gpc.

\begin{figure} 
\includegraphics[width=0.85\linewidth]{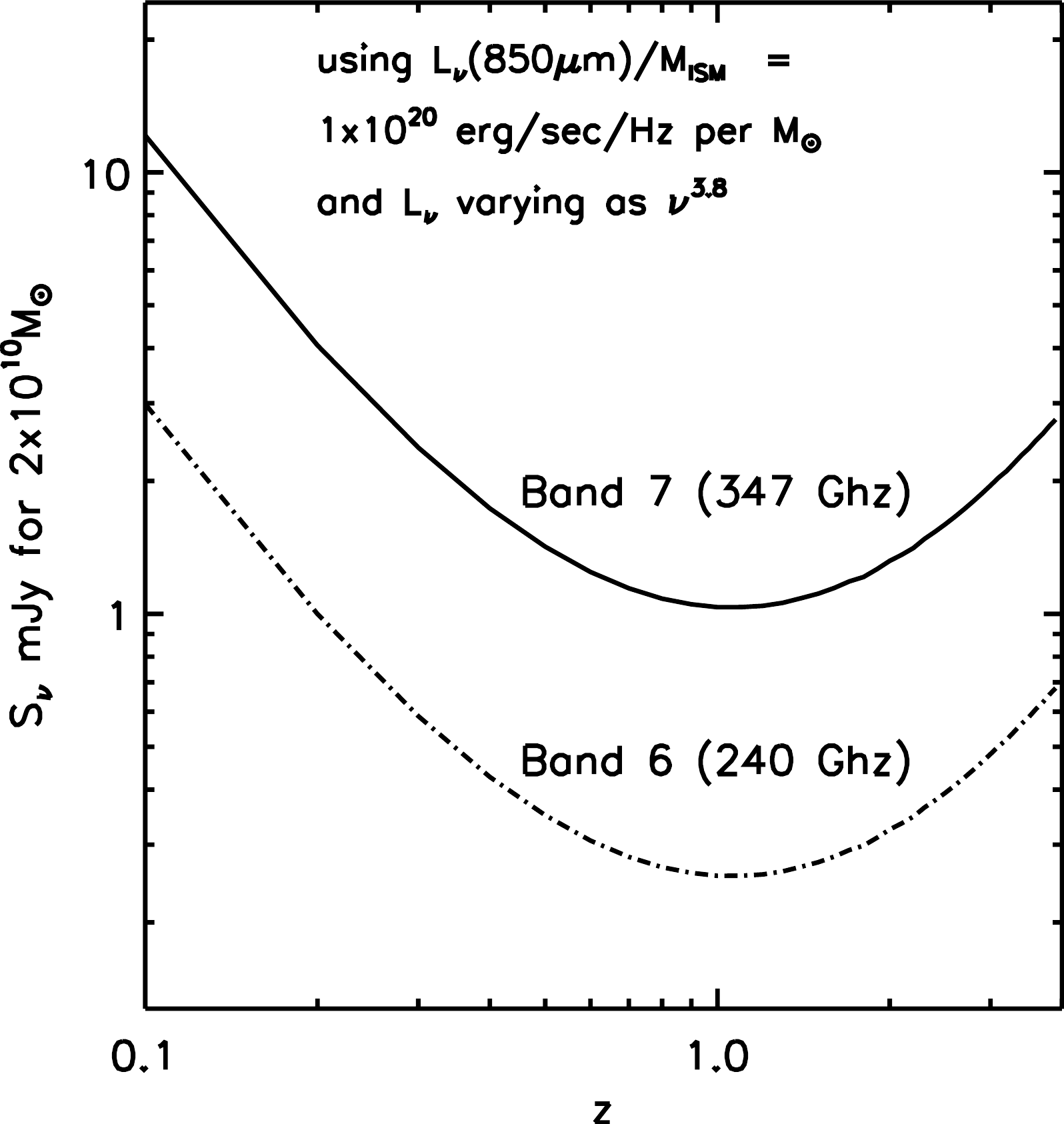}
\caption{The expected ALMA Band~6 (240\,GHz) and Band~7 (345\,GHz) flux 
densities are shown for $M_{\rm ISM}=2\times10^{10}$\,\msun.}
\label{expected}
\end{figure}

Figure~\ref{expected} shows the new predicted fluxes as a function of redshift 
for both Band~6 (240\,GHz) and Band~7 (347\,GHz). 

%-------------------------------------------------------------------------------
%-------------------------------------------------------------------------------

For reference, looking at the ALMA exposure time calculator, with
7.5\,GHz bandwidth in each polarisation and 10.2\,min of integration at
Cycle~1 yields 1\,$\sigma$\,=\,0.075\,mJy at 345\,GHz,
1\,$\sigma$\,=\,0.042\,mJy at 240\,GHz and 0.029\,mJy at 100\,GHz. The expected
fluxes on Fig.~\ref{expected} for $z$\,=\,2 are $\sim$\,1.7 and 0.3\,mJy
respectively for 2\,$\times$\,10$^{10}$\,\msun. Thus Band~7 is optimal since the
expected flux ratio is $\sim$\,5:1. Band~3 (100\,GHz) is not plotted in the
figure since its expected flux density is 28 times below that of Band~6.

To compare with the ability to detect CO, we might use the source BX\,691 from
Tacconi \textit{et al.} (2010) which has a $M_*$\,=\,7.6\,$\times$\,10$^{10}$\,\msun\ and
$M_{{\rm H}_2}$\,=\,3.5\,$\times$\,10$^{10}$\,\msun\ at $z$\,=\,2.19 and
CO(3--2)\,=\,0.15\,Jy~kms$^{-1}$. For a width of 300\,kms$^{-1}$, this has an
average line flux of 0.5\,mJy. If the mass is scaled to
2\,$\times$\,10$^{10}$\,\msun, then the average line flux is 0.28\,mJy. To get
5$\sigma$ or 0.056\,mJy sensitivity in a single 300\,kms$^{-1}$, requires 5
hours!

In summary, {\it measurement of the R-J tail of the emission (and using
Equation~\ref{rj_dust}) thus provides an excellent and fast means of determining
dust and ISM masses in high-redshift galaxies using ALMA}. (The coefficient in
Equation~\ref{rj_dust} was derived empirically from submm observations and
therefore may be slightly different than that obtained from the dust opacity
shown in Fig.~\ref{grains}.)

% {\color{red} \medskip\hrule\medskip\hrule\medskip}
\subsection{Effective source size}   
     
For optically thick IR sources, one can estimate the effective size of the
emitting region from 
\begin{equation}
R=\left({L\over{4\pi\sigma T_{\rm dust}^4}}\right)^{1/2},
\end{equation}
and scaling to ULIRG luminosities, we obtain
\begin{equation}
\label{radius}
R_{\rm kpc}=1.09\times\left({L/(10^{12}\,\lsun)\over{(T_{\rm dust}/35\,{\rm K})^4}}\right)^{1/2} {\rm kpc}.
\end{equation}
This effective radius is the overall size of the optically-thick region emitting
the FIR luminosity. In the event that the emission is optically thin, then the
estimates from Equation~\ref{radius} are of course lower limits. (In
Section~\ref{sed_model} optically-thick, radiative transfer modelling of the
dust emission for a $r^{-1}$ dust density distribution is presented.
Figure~\ref{rmax} shows the effective radius and dust temperature for the
emitting region producing the majority of the emergent flux -- for comparison
with Equation~\ref{radius}.)
  
For local ULIRGs the typical FIR colour temperatures are $\sim$50\,K so the IR
emission radius is $\sim$500\,pc for $10^{12}$\,\lsun. This estimate is similar
to the overall size of the central concentration in Arp\,220 (see below),
indicating that the optically thick assumption is not unreasonable. The most
luminous SMGs observed at high redshift can have $L_{\rm
IR}$\,$>$\,$10^{13}$\,\lsun; for these sources, the emission must come from
galactic-scale regions, not just a compact nucleus.
  
For the Milky Way the FIR luminosity is $\sim$10$^{10}$\,\lsun\ and the mean
dust temperature $\sim$30--35\,K; the effective emitting radius from
Equation~\ref{radius} is $\sim$100\,pc. However, this emission clearly
originates from a large number of separate clouds, and the mean size of each
must be $n_{\rm cloud}^{1/2}$ times smaller. For example, if the Galactic
emission is assumed to be contributed by $\sim$400 IR-luminous GMCs, then the
effective size of the IR-dominant region in each would be $\sim$5\,pc.

\subsection{Luminosity and SFR estimates from submm continuum}     
   
Estimating the FIR luminosity (and hence the SFR) from measurements on the submm
R-J tail is an extremely questionable procedure -- this hasn't stopped observers
from routinely doing it! As noted above the R-J flux provides a measure of the
dust mass weighed linearly by $T_{\rm d}$, but inferring a total bolometric
luminosity requires knowing where the FIR peaks, and the fluxes near the peak.
Observations near the SED peak can now be done using \textit{Herschel} PACS
(Photodetector Array Camera and Spectrometer) and SPIRE (Spectral and
Photometric Imaging REceiver) but many of the SMGs are subject to source
confusion at the SPIRE resolution. In the absence of direct observations at the
SED peak, one must assume a dust temperature (30--50\,K) and optical depth,
perhaps based on the submm flux. For many of the SMGs, FIR luminosities in the
range $10^{13-14}$\,\lsun\ have been estimated, implying SFRs of several
$\times10^3$\,\msun~per~yr (typically assuming $T_{\rm d}$\,$\sim$\,30--50\,K)
but such estimates must be extremely uncertain since the derived luminosity will
vary approximately as $T^{4-5}$. Typical ISM masses of the SMGs derived from CO
measurements or the submm continuum are $\sim$\,1--3\,$\times$\,10$^{10}$\,\msun,\linebreak
implying that the ISM will be used up in star formation in an implausibly short
time of $\sim$\,$10^7$\,yr.
   
Blain \textit{et al.} (2003) attempted to derive empirically a scaling between
the 850\,$\mu$m flux and $L_{\rm IR}$ based on the apparent $T_{\rm d}$ from
fitting the SEDs of local galaxies. Unfortunately, there is large scatter in the
correlation. In fact, as we saw earlier, the notion of a single $T_{\rm d}$ is
extremely shaky -- both because there clearly is a range of temperatures and
more importantly, the apparent $T_{\rm d}$ (derived from fitting near the FIR
peak) is somewhat degenerate with the dust opacity (which also can cause an
exponential fall-off to short wavelengths). 
   
To summarise -- it should be clear that one cannot constrain $T_{\rm d}$ without
observations near the SED peak; and if one has such observations, it would be
best to simply use them directly to estimate the luminosity. (Of course, if the
observed source is at redshift $z>5$, then the submm flux measurements are in
fact probing near the restframe SED peak; they can then provide a decent
luminosity estimate.) In the next section, we model the optically-thick dust
sources in order to appreciate some of the systematics and the range of
uncertainty.

\subsection{Modelling optically-thick dust clouds}
\label{sed_model}  

For internally-heated FIR sources, it is vital to appreciate that the sources
have essentially two totally independent parameters: \textit{the luminosity L of
the central heating source and the total mass of dust in the surrounding
envelope} -- not much else matters! The character of the spectrum of the source
(be it young stars or an AGN) makes little difference since the primary photons
at short wavelength are absorbed in the innermost boundary layer of the dust
envelope. The radius of this inner boundary is set by the radius at which the
dust is heated to sublimation (see Section~\ref{ire}). The overall mass of dust
of course determines the opacity and therefore the radius at which the IR
radiation can escape, and thus the `effective dust temperature' which is
observed. We will see below that the model SEDs of the FIR sources can be
characterised by a single parameter -- the luminosity-to-mass ratio, $L/M$, and
thus the problem has, in essence, really just one independent variable as long
as the source structure is simple (e.g., a single source with radial fall-off in
density and no clumping). The compactness of the dust cloud is parametrised by a
radial power law density distribution which can be varied but for the discussion
below I adopt $R^{-1}$.
  
%-------------------------------------------------------------------------------
\begin{figure}
\includegraphics[width=0.8\linewidth]{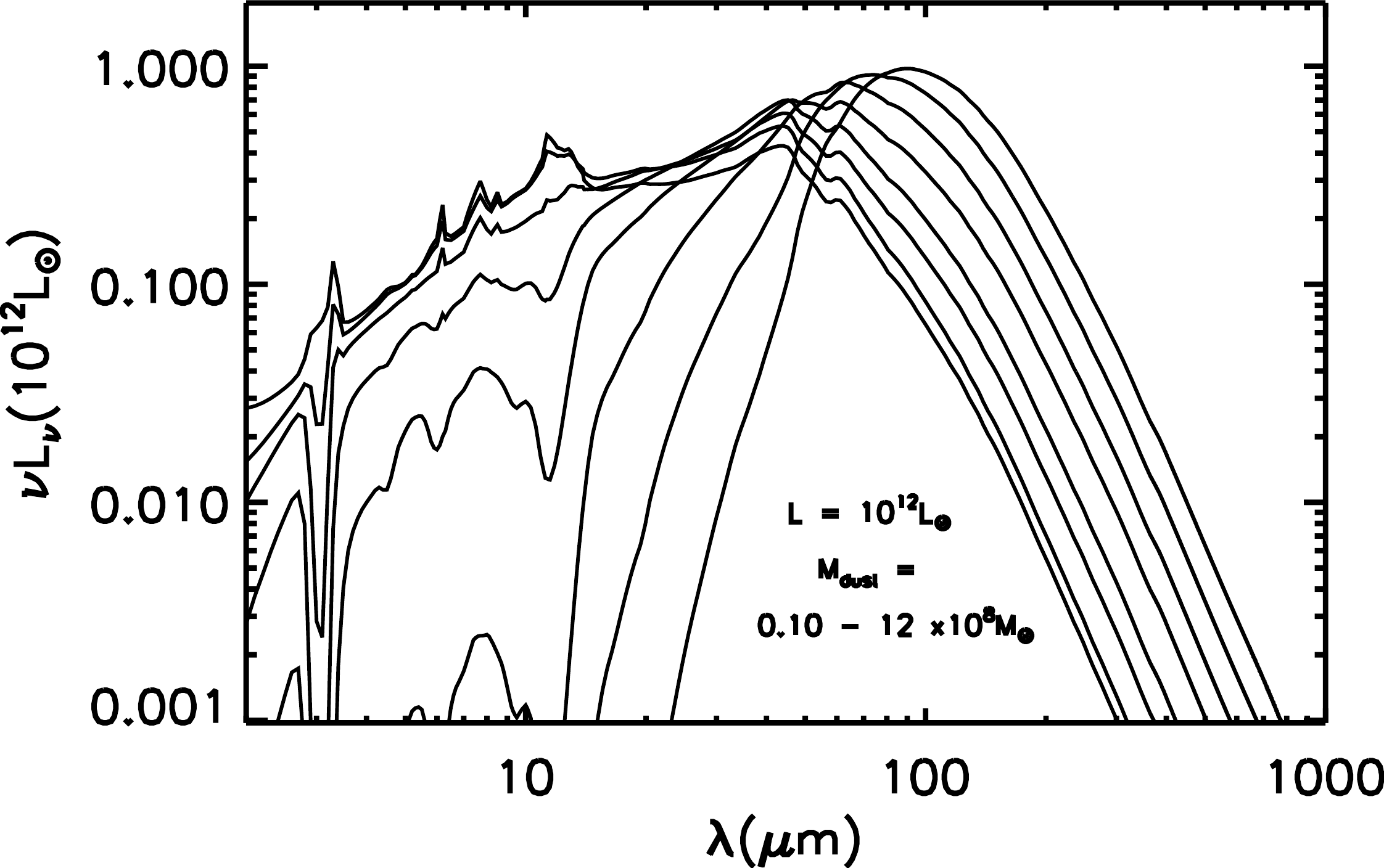}
\caption{The IR SEDs of dust cloud models with $10^{12}$\,\lsun\ are shown for
increasing masses of overlying dust and an $r^{-1}$ density distribution. The
dust heating is provided by both the central source and secondary photons from
warm dust (see text). The peak shifts to longer wavelength as the dust opacity
increases and the R-J tail rises linearly with dust mass.}
\label{irseds} 
\end{figure}
%-------------------------------------------------------------------------------

Using the temperature profiles  derived above, I have computed emergent spectra
for a source of central luminosity $10^{12}$\,\lsun\ with overlying dust masses
ranging from $10^{7-9}$\,\msun\ (i.e., total ISM masses $\sim$100 times greater
or $10^{9-11}$\,\msun). These parameters are directed towards ULIRGs and SMGs
but the results can easily be scaled to lower- or higher-luminosity objects. As
mentioned above, the critical model characteristic is the luminosity-to-mass
ratio. For this modelling, the dust distribution is taken to vary as $R^{-1}$
but similar results are found with other reasonable power laws. The inner radius
is taken at $T_{\rm d}$\,=\,1000\,K, but this is, of course, not critical since
the hot dust is covered by the overlying colder dust unless the cloud is
optically thin at short wavelengths. The outer radius was taken at 2\,kpc --
this also is not critical since the dust is cold and optically thin to the
secondary radiation at the largest radii. 
     
Figure~\ref{irseds} shows the emergent specific luminosities ($\nu L_{\nu}$) for
the models with different enveloping dust masses but constant overall
luminosity. Here one clearly sees the effect of varying dust mass and opacity.
The clouds of lower dust mass have non-exponential short-wavelength SEDs due to
the lack of high dust extinction on the short-wavelength side of the SED peak,
and hence the hotter dust in the interior is exposed to our view. This contrasts
with the higher-mass clouds which show a peak shifted to relatively longer
wavelength -- due to the fact that dust extinction short of the peak precludes
photons escaping from the hotter interior radii. Figure~\ref{irseds} also
clearly demonstrates the anticipated correlation (see Section~\ref{rjtail})
between the flux on the long-wavelength R-J tail and the dust mass. 

\begin{figure}
\includegraphics[width=\linewidth]{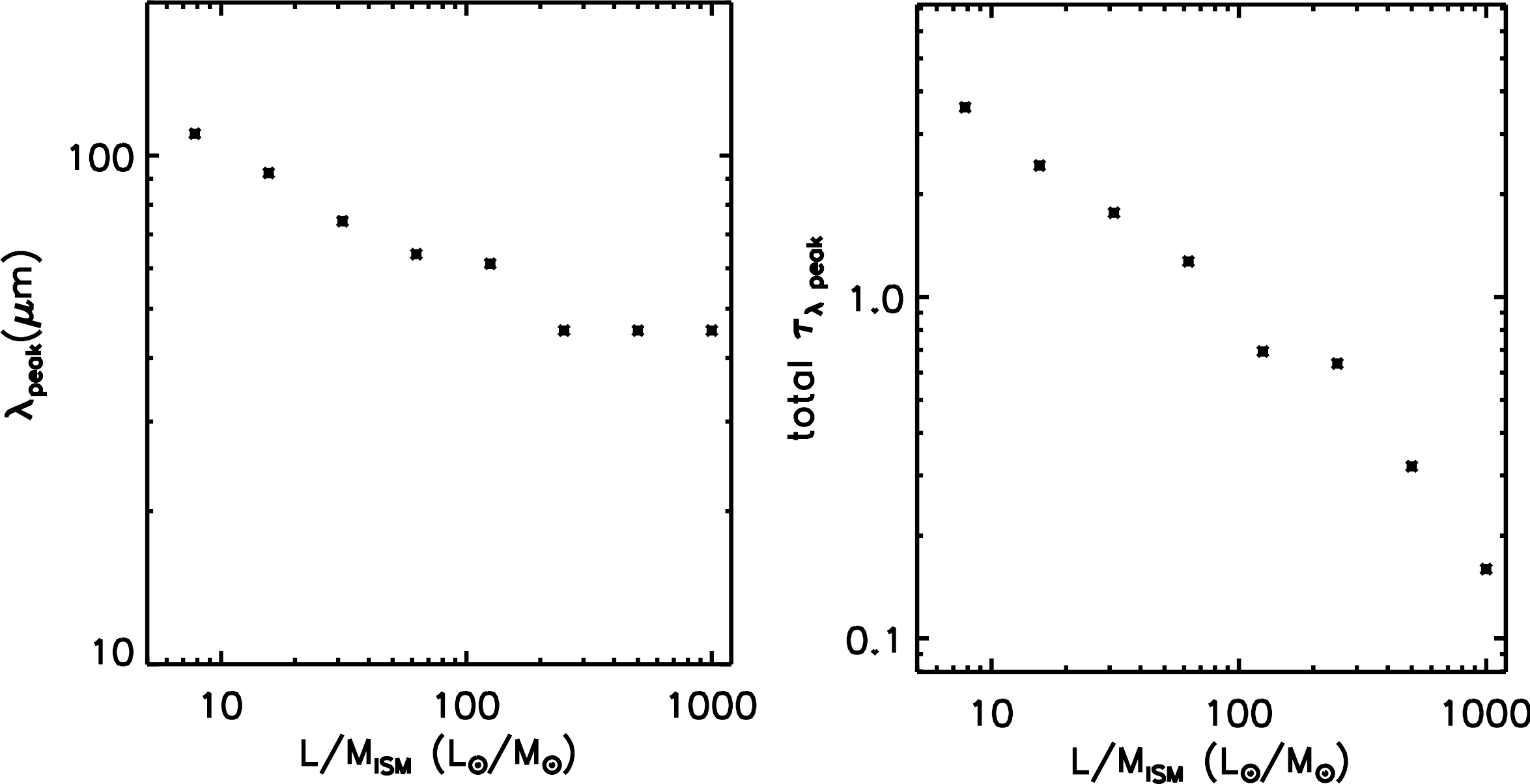}
\caption{The variation in the peak wavelength for $L_{\nu}$ and the optical
depth through the cloud at the peak wavelength are shown for varying $L/M_{\rm
ISM}$ values with a density distribution of dust $\propto R^{-1}$. (Models run
for power laws of 0, $-1$ and $-2$ exhibit similar behaviour as long as the clouds
are optically thick.)}
\label{lam_pk} 
\end{figure}

To quantify some of the spectral characteristics, Fig.~\ref{lam_pk} shows the
shift in the wavelength of peak emission ($\lambda_{\rm peak}$, left panel) and
the optical depth at $\lambda_{\rm peak}$ to the cloud surface (right panel) for
the radial shell contributing most to the emergent luminosity. The right panel
illustrates what was said earlier based on analytics -- typically the opacity at
the peak will be $\sim$1 for a large range of overlying dust masses. Also, since
the opacity must be $\sim$1 at the peak, the wavelength of peak emission will be
determined by \textit{both} opacity and dust temperature in realistic models. 

%-------------------------------------------------------------------------------
%-------------------------------------------------------------------------------

%-------------------------------------------------------------------------------
\begin{figure}
\includegraphics[width=\linewidth]{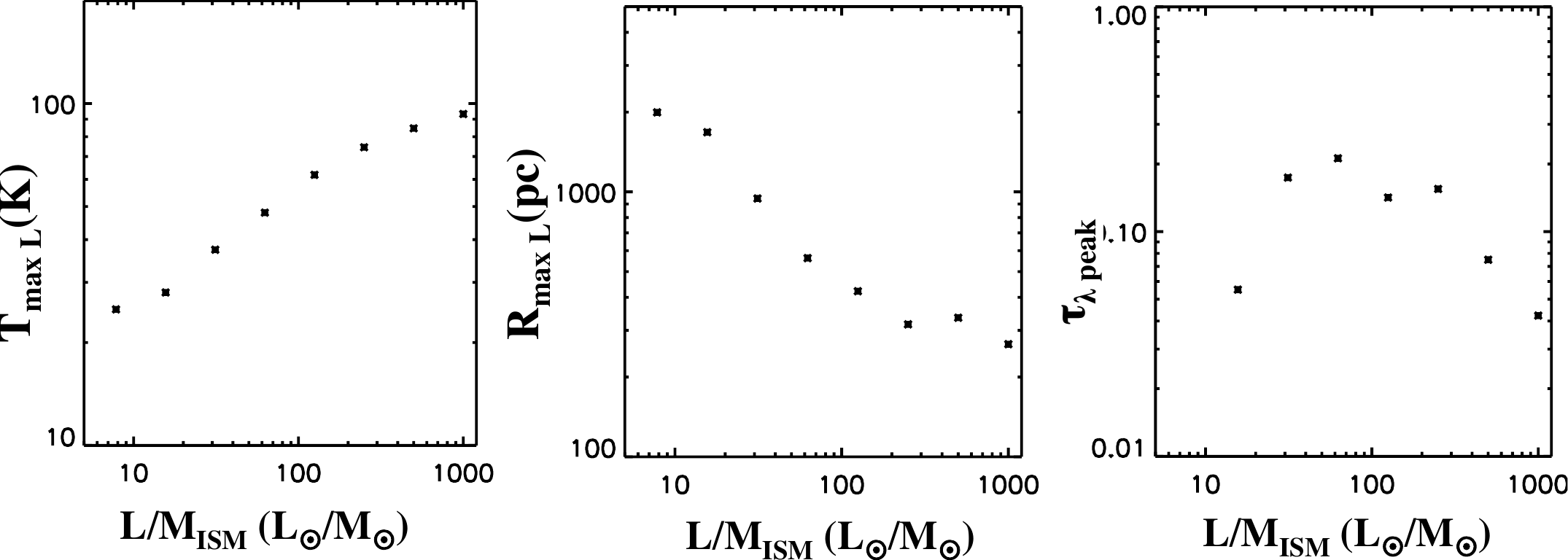}
\caption{Left and middle panels show the dust temperature and radius of the shell in the
model producing the largest fraction of the overall luminosity for varying
$L/M_{\rm ISM}$ values. The right panel shows the optical depth at the peak
wavelength from this shell to the outer cloud surface -- illustrating the fact
that the emergent luminosity is produced at $\tau\lesssim1$. In most cases,
the output luminosity is in fact spread over a fairly broad range of radii.}
\label{rmax} 
\end{figure}
%-------------------------------------------------------------------------------

Figure~\ref{rmax} shows the radius and dust temperature at which the largest
fraction of the overall luminosity escapes as a function of the luminosity-to-mass 
ratio. The total luminosity is of course spread over a broad range of radii
except in the most optically thick models (low $L/M_{\rm ISM}$ values).

\begin{figure}
\includegraphics[width=0.6\linewidth]{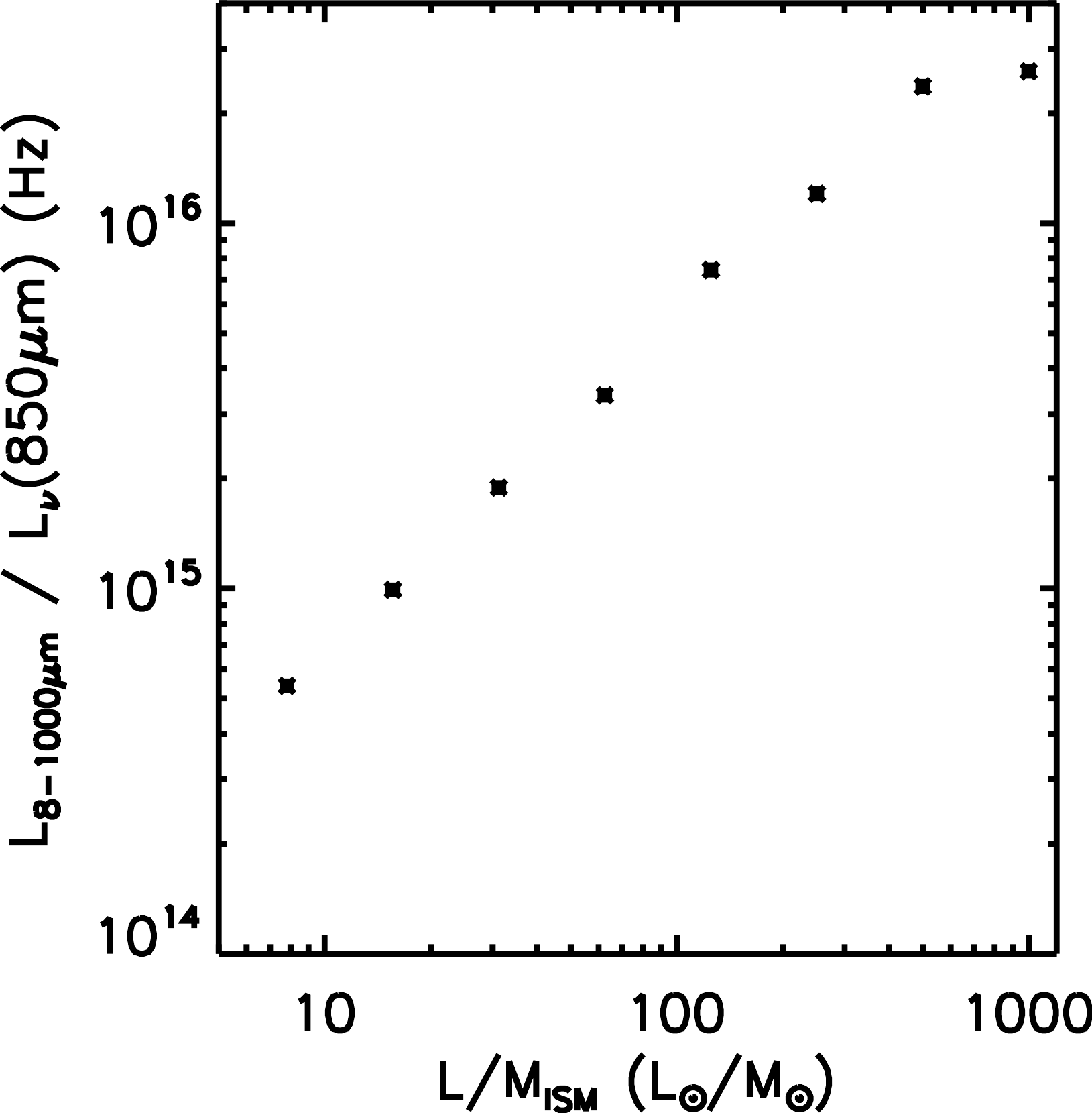}
\caption{The variation in the total IR luminosity to restframe 850\,$\mu$m flux
is shown for a range of $L/M_{\rm ISM}$ values, illustrating the impossibility
of estimating the total IR luminosity from a single long-wavelength flux
measurement.}
\label{f850} 
\end{figure}

Figure~\ref{f850} shows the ratio of total IR luminosity (at
$\lambda$\,=\,8--1000\,$\mu$m) to the specific luminosity at
$\lambda$\,=\,850\,$\mu$m (i.e., a point on the power-law R-J tail). It should
be obvious from this figure that there is really no good single value for the
conversion factor from submm flux density ($\lambda_{\rm rest}\gtrsim150\,\mu$m)
to total IR luminosity -- these models all had the same total luminosity but
varying dust masses.  Thus it is impossible to reliably estimate the total FIR
luminosity from R-J flux measurements unless there are additional constraints,
for example on the luminosity-to-mass ratio or the ISM mass (e.g., from CO, or
dynamical mass estimates). One simply must have measurements close to the IR SED
peak. 

%-------------------------------------------------------------------------------
%-------------------------------------------------------------------------------

\begin{figure}
\includegraphics[width=0.9\linewidth]{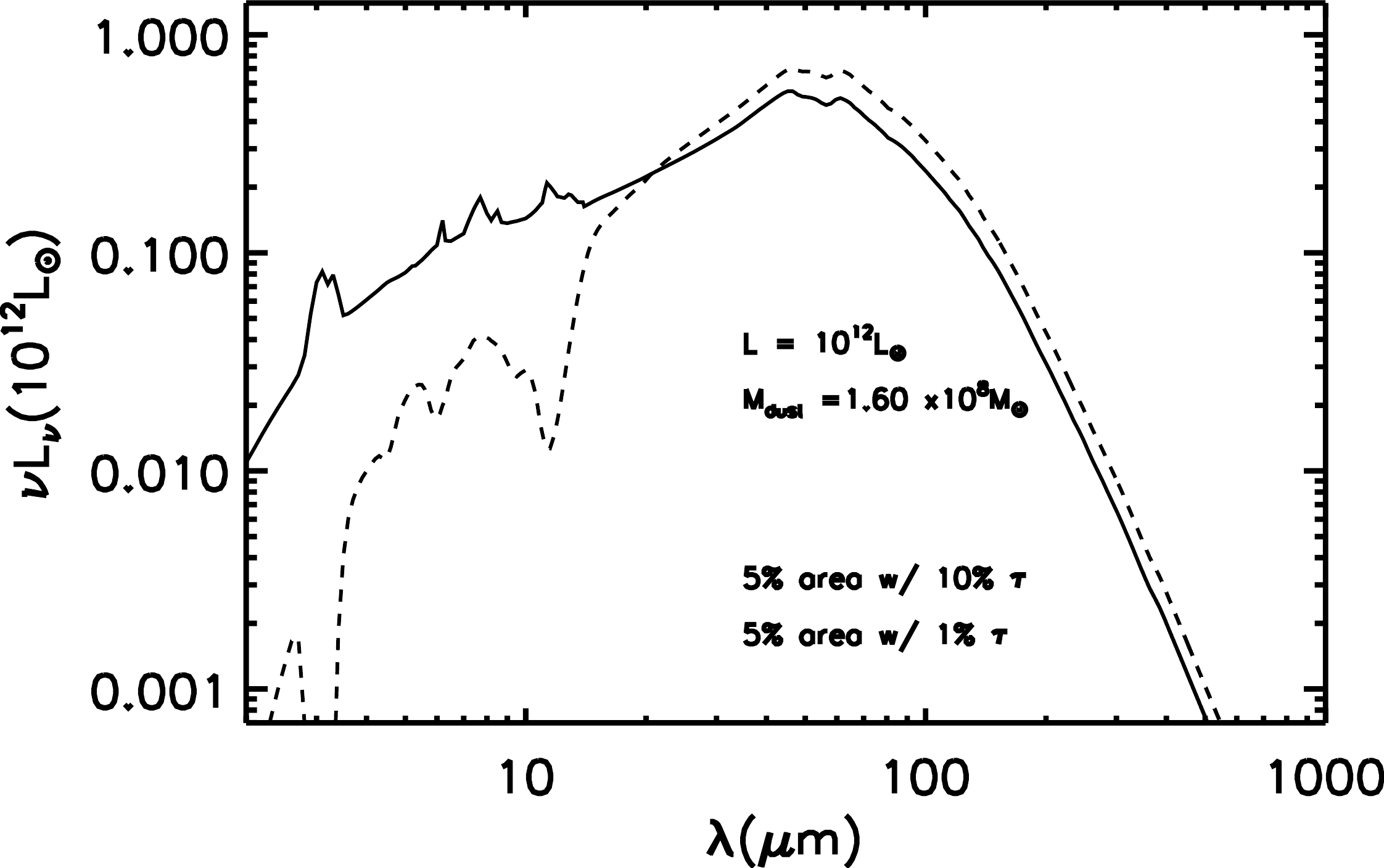}
\caption{Models with complete covering (dashed line) and with 5\% of the area
having extinction reduced to 10\% and 1\%  of the complete covering model (solid
line) are shown to illustrate that a slightly clumpy dust distribution is
required to model the observed short-wavelength hot dust emission. Transient
small grain heating is  included by calculating the UV--visual radiation energy
density at each radius and inserting the PAH emission following Draine \& Li
(2001).}
\label{clumping} 
\end{figure}

There are also some limitations or lessons from this modelling. Most noteworthy
is the polycyclic aromatic hydrocarbon (PAH) emission which arises from
transiently heated small dust grains. This small grain component was included by
calculating the UV--visual radiation energy density at each radius and inserting
the PAH emission following Draine \& Li (2001). These short-wavelength features
require the emergence of flux from regions where the grains are hot. In sources
showing such features, the dust obscuration must be clumpy, or multiple
optically-thin and -thick sources must be contributing. To illustrate this
possibility, Fig.~\ref{clumping} shows the effect of reducing the dust column in
5\% of the surface area to a value 10\% of the normal column density -- the
NIR/MIR becomes much stronger and the silicate feature appears in absorption.

The presence or absence of detectable PAH or silicates features depends on the
covering fraction of the overlying optically-thick dust -- their detection
should not be taken as a reliable indicator of one mode of star formation (as
done by Elbaz \textit{et al.} 2011) or alteration of the grain abundances since
their presence can depend simply on geometry and source non-uniformity. 

%-------------------------------------------------------------------------------
%-------------------------------------------------------------------------------

\subsection{Summary}

We have developed a very simple model for the FIR emission which can be
implemented for fast radiative transfer computations relevant to dust-embedded
luminosity sources. There are several important conclusions:

\begin{enumerate}[(a)]\listsize
\renewcommand{\theenumi}{(\alph{enumi})}

\item For a central luminosity source, the dust temperature profile with radius
can be modelled with very simply power laws: $T_{\rm dust}\propto r^{-0.42}$ for
optically thin dust and $T_{\rm dust}\propto r^{-0.5}$ for optically thick dust.
In a realistic source, it will be optically thin at the very innermost radius,
optically thick at intermediate radii and optically thin at the outer radii; the
temperature profile is then a piecewise fitting together of these two power
laws.

\item To model the FIR sources, it is vital to recognise that there are two
entirely independent parameters: the total luminosity and the mass of dust (and
to a much lesser extent the radial profile of the dust distribution). Since the
temperature profile scales with $L$, then most sources may be uniquely
characterised by the ratio $L/M_{\rm dust}$.

\item Using these temperature profiles, it is then straightforward and quick to
calculate the emergent SED in the IR as a function of the ratio $L/M_{\rm dust}$.

\item At the FIR peak, the emission is often optically thick (based on the steep
fall-off in the SED to shorter wavelengths) but the optical depth will not be
very large (i.e., $\tau_{\rm peak}$\,$\sim$\,2--4) since otherwise the radiant
luminosity would not be able to escape.

\item The effective dust temperature for the emergent luminosity can be
determined approximately from the wavelength of the peak and then used to
estimate an effective radius for the source.
 
\item Flux measurements on the R-J tail of the SED are uniquely capable for
determination of the total dust mass (with a factor of two reliability) since the
dust is optically thin. And if one can assume a dust-to-gas abundance ratio,
such measures lead to quick and reliable estimates of a distant galaxy's total
ISM content. 

\item Determination of the total luminosity and hence the SFR requires
observations near the peak at $\lambda_{\rm rest}$\,$\sim$\,100\,$\mu$m; it can
not be reliably estimated using a single long wavelength R-J flux measurement
(as is often done for SMGs).

\item In optically-thick sources (as judged by the steep fall-off on the short
side of the SED peak), the MIR emission at $\lambda<20\,\mu$m requires that
the dust be somewhat clumped with incomplete covering in order to see the hot
dust and the silicate and PAH features. 
\end{enumerate}
   
%
%%%%%%%%%%%%%%%%%%%%%%%%%%%%%%%%%%%%%%%%%%%%%%%%%%%%%%%%%%%%%%%%%%%%%%%%%%%%%%%%
%

\section{Star formation in galaxies -- two modes}

\subsection{Quiescent or normal mode of star formation}

As discussed in Section~\ref{gmc}, the star-forming GMCs are self-gravitating
with internal velocity dispersions implying an effective internal turbulent
pressure typically 100 times the external diffuse ISM pressures. One's physical
intuition should then suggest that disturbances in the external, diffuse ISM
will have little influence in general on the rate of formation of stars within
GMCs. With this in mind, one might expect that normally SFRs will simply depend
linearly on the overall mass of H$_2$ and/or the internal density of the cloud
and this logic should hold in other galaxies as long as the GMC properties are
roughly similar. Galaxy type or the location within a galaxy should have little
influence on what happens within the self-gravitating clouds. 

\begin{figure} 
\includegraphics[width=\linewidth]{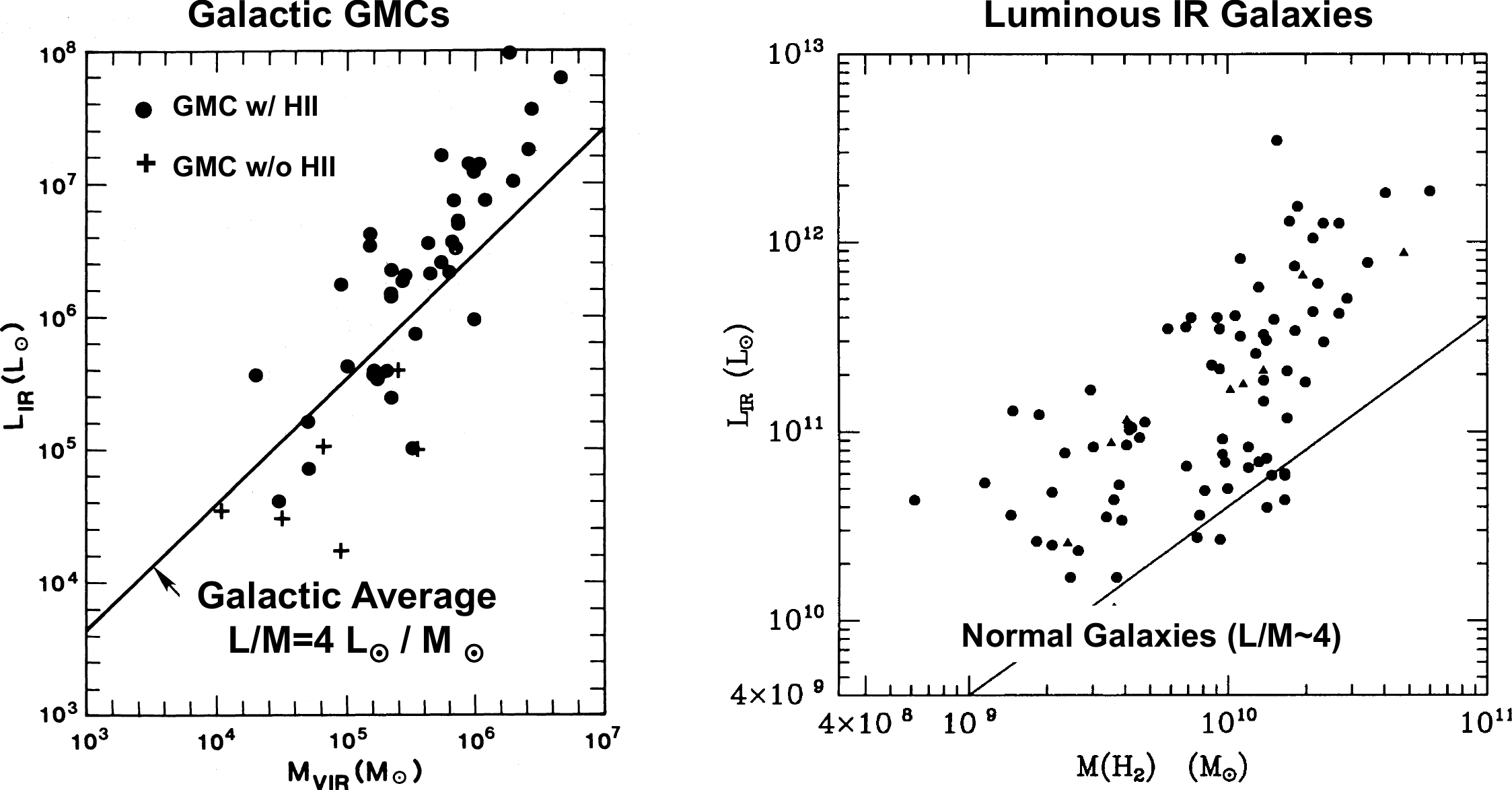}
\caption{Left panel: the relation between IR luminosity and H$_2$ mass is shown
for a large sample of Galactic GMCs with (dots) and without H{\sc ii} regions
(crosses), compared with the overall Galactic average obtained from dividing the
total Milky Way IR luminosity by the total H$_2$ mass (Scoville \& Good 1989).
M\,51 and many other nearby spiral galaxies have mean ratios similar to that of
the Milky Way. The right panel shows a similar plot for a local sample of IR
bright galaxies (Sanders \textit{et al.} 1991). The ULIRGs at $L_{\rm
IR}>10^{12}$\,\lsun\ show much higher luminosity-to-mass ratios, indicating that
they are converting gas to stars 10--20 times faster.}
\label{l_mh2}
\end{figure}

Observations of CO emission (tracing the mass of GMCs) and star formation
tracers (FIR luminosity or H$\alpha$) have generally shown a linear
proportionality between SFR and H$_2$ mass distributions. For example, the
overall Galactic H{\sc ii} region radial distribution shown in
Fig.~\ref{gas_sigma} is similar to that of H$_2$ as traced by CO. On the scale
of individual star-forming clouds, one might expect sizable fluctuations in the
SFR per unit mass since the overall Galactic star formation efficiency (SFE) is
low with a characteristic star formation timescale $\sim$\,$10^9$\,yr, i.e.,
much longer than the GMC internal dynamical timescale $\sim$\,$10^7$\,yr. In
addition, the GMCs may have somewhat variable mean densities which of course
alters their internal dynamical timescale. For a sample of individual Galactic
GMCs (including ones with and without associated H{\sc ii} regions),
Fig.~\ref{l_mh2} (left panel) shows an approximately `linear' correlation
between molecular gas mass $M_{\rm virial}$ and $L_{\rm IR}$ or the SFR
(Scoville \& Good 1989) for GMCs with masses $10^4$ to over $10^6$\,\msun. The
scatter off this relation is a factor $\sim$4. There is also a clear trend for
the most massive clouds which all have H{\sc ii} regions, presumably in the
spiral arms, to have elevated ratios. 

Given the tendency of the spiral arm GMCs to have somewhat elevated
luminosity-to-mass ratios, it is probably wisest to use the mean Galactic
luminosity-to-mass ratio (4\,\lsun/\msun) to define a `normal' mode of star
formation. Such a ratio of $L_{\rm IR}/M_{\rm H_2}$\,$\simeq$\,4\,\lsun/\msun\
implies a characteristic star formation timescale $\sim$\,$10^9$\,yr for
conversion of molecular ISM into stars. The ULIRGs (at $L_{\rm
IR}$\,$>$\,$10^{12}$\,\lsun, most of which are gas-rich galaxy mergers) exhibit
a $\sim$\,10--20 times higher $L_{\rm IR}/M_{\rm H_2}$, hence a 10--20 times
shorter timescale.
 
%-------------------------------------------------------------------------------
%-------------------------------------------------------------------------------

In summary, there is good basis (empirical from the observations and theoretical
intuition from what we know regarding the GMC structures) for an approximately 
linear mode of star formation. From local galaxies such as the Milky Way and M\,51, 
this is quantified at $\simeq$1\,\msun\ per year per 10$^9$\,\msun\ of H$_2$, or
\begin{equation}
\label{sfr} 
{\rm SFR}=\left(M_{\rm H_2}/\msun\right)\cdot10^{-9}\quad{\rm \msun~yr^{-1}}. 
\end{equation}

\subsection{Dynamically-driven starburst mode}

At the same time, there are clearly instances, such as those of the IR-bright
galaxies, galaxy nuclei and the spirals arms and bars of normal galaxies, where
the SFE is significantly enhanced to a level such that they can be classified as
starbursts. Figure~\ref{l_mh2} (right panel) shows a plot of the IR luminosities
and molecular gas masses for IR-luminous galaxies (LIRGs, Sanders \textit{et
al.} 1991) -- all of them have considerably elevated luminosity-to-mass ratios
compared to the average ratios for the Milky Way and nearby spiral galaxies like
M\,51. Both the luminous IR galaxies and the individual spiral arm GMCs have
exceptionally high rates of massive star formation (presumably they are also
forming low-mass stars). In such regions, the SFE (based on the ratio of $L_{\rm
IR}/M_{\rm H_2}$) may be enhanced by a factor ten compared to that in
Equation~\ref{sfr}. 

In both the spiral arms and the ULIRGs, this elevated SFE appears correlated
with non-circular galactic dynamics and high concentrations of dense gas. Galaxy
merging leads to dissipative deposition of gas to the centres of the merging
systems and spiral arm streaming motions cause crowding of the GMC galactic
orbits. As a result of the non-circular motions and concentration of gas clouds,
cloud-cloud collisions will be more frequent. If such collisions occur at
relative velocity greater than the GMC internal velocities (few km/s), the
collisions can compress the internal gas mass and significantly elevate the SFE
(Scoville \& Hersh 1979; Tan 2000). This high-SFE mode might therefore be
referred to as a dynamically-driven starburst where the SFE is 10--50 times that
given in Equation~\ref{sfr} -- occurring in galaxy mergers or localised regions
such as spiral arms and nuclear bars of normal galaxy disks.

\subsection{Star formation `laws'}

Over the years a number of prescriptions have been proposed to describe the
rates of star formation within galaxies. Early on, Scoville \& Good (1989) and
Young \& Scoville (1991) advocated that the SFR varied linearly with the mass of
molecular gas -- based on observations of Galactic GMCs and the similarity of
molecular gas radial distributions in nearby, normal galaxies to the radial
distributions of SFR tracers (e.g., blue light, H$\alpha$ and FIR). Extensive
studies (Kennicutt 1998) which included H{\sc i} in the analysis then led to 
the well-known Kennicutt-Schmidt SFR law with SFR $\propto\Sigma_{\rm
HI+H_2}^{1.4}$ and a threshold cutoff surface density below which the SFR
decreased more steeply. In my opinion, the atomic gas (H{\sc i}) exhibits almost
no correlation with the SFR tracers (recently shown clearly by Bigiel \textit{et
al.} 2008 and Schruba \textit{et al.} 2011) and is not directly relevant to the
star formation process in the inner parts of spiral galaxies. (The non-integral, 
$\sim$1.4, dependence of the SFR in the Kennicutt-Schmidt star formation law is
likely due to including H{\sc i} in the analysis and is without a direct
physical link to star formation except via the formation of molecular clouds and
due to the inclusion of starburst galaxy nuclei. As emphasised above, the GMCs
are long-lived and stars form from them at low efficiency, so the fact that the
H{\sc i} is needed to form molecules does not justify a physical connection to
the star formation process.) 

\begin{figure} 
\includegraphics[width=\linewidth]{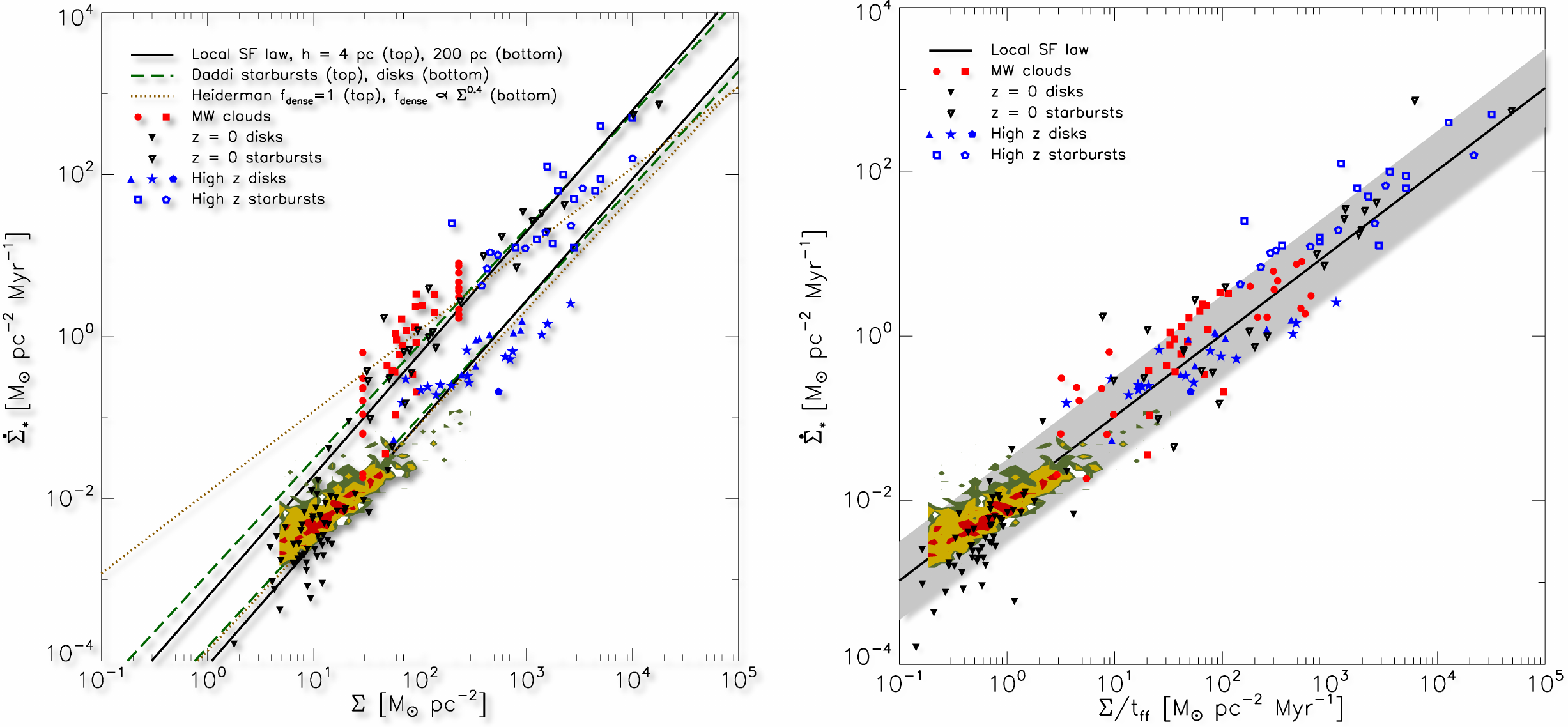}
\caption{Left panel shows the compilation of observational determinations of the
SFR and total gas surface densities by Krumholz \textit{et al.} (2012) together
with several proposed SFR prescriptions. The right panel shows the alignment of
these data improved when the gas surface densities are normalised by a free-fall
timescale, i.e., a scalelength, to provide a fixed `volumetric' law.}
\label{kru}
\end{figure}

The most recent compilations have been done by Bigiel \textit{et al.} (2011) and
Krumholz \textit{et al.} (2012). The former advocate a simple linear dependence
on molecular gas surface density; the latter propose a `volumetric' law
$\dot\Sigma_{*}=f_{\rm H_2}~\epsilon_{\rm ff}\Sigma/\tau_{\rm ff}$ where the
factors entering are the H$_2$ mass fraction, the efficiency of star formation
per free-fall time, the total gas surface density and the free-fall time. Their
compilation of observational data is shown in Fig.~\ref{kru}, together with the
various star formation laws they evaluated. Although their proposed volumetric
law provides a good fit to a large range of data, the evaluation of $\tau_{\rm
ff}$ requires that a scaleheight for the gas be assumed and this is not easily
derived from the observations so the quality of fit is largely dependent on what
is assumed. 

%-------------------------------------------------------------------------------
%-------------------------------------------------------------------------------
      
At this point, I believe the best approach is simply: 1) assume star formation
occurs \emph{only in the molecular gas} (i.e., do not include atomic gas in the
analysis since it is not physically relevant) and 2) assume \textit{two modes of
star formation: a) a quiescent or normal star formation mode which depends
linearly on the mass or surface density of H$_2$; and b) a dynamically-driven
mode (relevant to mergers, spiral arms and bars) which is also linear with
$\Sigma_{\rm H_2}$ but has a rate constant up to 20 times that of the first mode}. 

Clearly, for distant galaxies it would be nice to have a more precise star
formation law with dependence on detailed properties of the clouds (such as
their internal density and the details of the local galactic velocity field
streaming and dispersion) but this will not be observationally accessible for
many many years. Let's keep the star formation prescriptions close to the
relevant observational capabilities!

\subsection{Distinguishing normal star formation and starbursts:\\concentration and timescales}
  
How does one define a starbursting system (as opposed to a normal star-forming
galaxy) and what are the observational discriminants between these two modes? I
think the most physically meaningful definition for a starburst is a galaxy or
area within a galaxy where the rate of conversion of existing ISM into stars is
such that the ISM will be consumed in a time significantly less than the typical
$M_{\rm ISM}/{\rm SFR}\sim1$\,Gyr timescale for local galaxies like the Milky
Way. This SFE can be observationally characterised by the simple ratio $L_{\rm
IR}/M_{\rm H_2}$. 

Alternatively, one might use the specific SFR per unit stellar mass (SSFR =
SFR/$M_*$) to identify galaxies where the characteristic timescale (1/SSFR) is
much shorter than the `likely' age of the galaxy (which is obviously less than the
age of the Universe at the galaxy's redshift). At redshift \hbox{$z$\,=\,1.5--2.5},
Rodighiero \textit{et al.} (2011) argue that there is a tight `main sequence' of
star-forming galaxies with fairly constant SSFR at all stellar masses (see
Fig.~\ref{rod}); they then identify starburst objects with SSFR significantly
above this main sequence. Unfortunately, the SSFRs found for the main sequence
at $z\sim2$ are higher by a factor $\sim$15 than those at low redshift so the
discriminant between the two star formation modes is evolving in time -- without
a clear physical basis, be it a higher gas content, smaller stellar mass at
early epochs or an increasing efficiency/rate in converting existing gas to
stars. And, of course, if the definition of the main sequence is evolving in
time, the fraction of galaxies found to be `bursting' will depend entirely on
how far off the main sequence a galaxy must be to be so classified. If most
galaxies at high redshift were, in fact, bursting, then the `main-sequence' SSFR
would be elevated as observed and the dispersion would simply reflect the range
of burst activity. Clearly, it would be most important now to analyse the
morphologies of the so-called main-sequence galaxies to classify the
merging/interacting versus undisturbed percentages.

Definitions for starburst classification which compare the SSFR to the galaxy
age or better, using the timescale for exhausting the ISM are more physically
based than simply using the dispersion about the main sequence. In the near
future we can anticipate that ALMA will be measuring gas contents for
large samples of high-$z$ galaxies. 
 
Another approach which might be used to identify starburst activity is the
luminosity density or concentration of the activity. In starbursting systems,
one expects that a large fraction of the luminosity from the young stars will be
absorbed by dust and emerge in the FIR and that the luminosity energy density
within the active region will be high. As discussed in Section~\ref{sed_model},
good IR SED sampling allows one to classify the observed SED according to the ratio
$L_{\rm IR}/M_{\rm dust}$ which is proportional to $L_{\rm IR}/M_{\rm ISM}$.
\emph{Sources with high luminosity density (either AGN or intense starbursts)
may also have high `apparent' colour temperatures}, providing they are not
completely enshrouded by dust which is optically thick out to 100\,$\mu$m. One
can reasonably estimate a characteristic source size using Equation~\ref{rthick}
and a luminosity density from $L_{\rm IR}/(4\pi R_{\rm thick}^3/3)$.

\subsection{Starbursts in ULIRGs}

\begin{figure} 
\includegraphics[width=\linewidth]{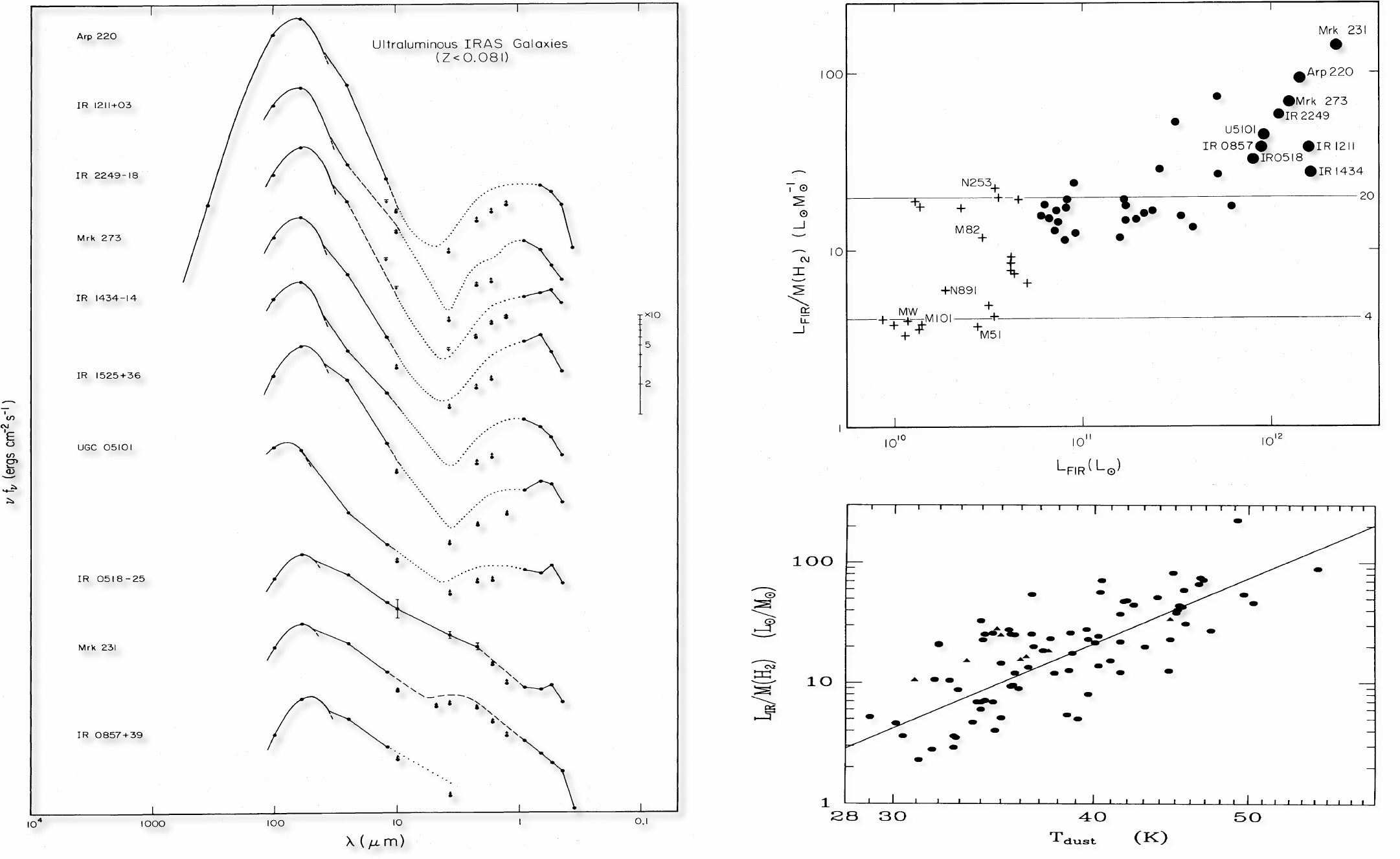}
\caption{The SEDs, $L_{\rm IR}/M_{\rm H_2}$ mass ratios and fitted dust 
temperatures are shown for the complete sample of 20 ULIRGs at $z<0.1$ from
Sanders \textit{et al.} (1988) and Sanders \textit{et al.} (1991).}
\label{ulirgs_fig}
\end{figure}

In Fig.~\ref{ulirgs_fig} the SEDs, $L_{\rm IR}/M_{\rm H_2}$ mass ratios and
fitted dust temperatures are shown for the complete sample of 20 ULIRGs at
$z<0.1$ from Sanders \textit{et al.} (1988) and Sanders \textit{et al.} (1991).
The SEDs peak at $\lambda\gtrsim120\,\mu$m with dust colour temperatures
30--50\,K and the FIR-to-H$_2$ mass ratios are 30--200\,\lsun/\msun. As noted
above, these ratios are a factor 10--20 higher than the luminosity-to-mass
ratios in local spiral galaxies, indicating a 10--20 times shorter timescale for
converting ISM to young stars.

%-------------------------------------------------------------------------------
%-------------------------------------------------------------------------------

\subsection{Arp\,220 -- a prototypical ULIRG}

In the spirit of `one example can provide more understanding than many
generalisations', I think it is very worthwhile to look in detail at current
observations of Arp\,220. Of course not all ULIRGs are the same as Arp\,220, 
but one can readily see the essence of the starburst mode in this object. 

Arp\,220, at 77\,Mpc, is one of the nearest and the best-known ultra-luminous
merging system ($L_{8-1000\,\mu{\rm m}}$\,=\,1.5\,$\times$\,10$^{12}$\,\lsun). To
power the IR output, star formation must be occurring at a rate of
$\sim$10$^2$\,\msun~yr$^{-1}$. Visual wavelength images reveal two faint tidal
tails, indicating a recent tidal interaction (Joseph \& Wright 1985), and
high-resolution ground-based radio and NIR imaging show a double nucleus (Baan
\& Haschick 1995; Graham \textit{et al.} 1990). The radio nuclei are separated
by 0.98\,arcsec (Baan \& Haschick 1995), corresponding to 350\,pc.
Millimetre-wave imaging provides a unique capability to probe such starburst
nuclei in dusty LIRG/ULIRGs since the dust is optically thin at long
wavelengths. Virtually all of the ULIRGs observed so far have massive
concentrations of molecular gas in the central few kiloparsec (Scoville
\textit{et al.} 1997; Downes \& Solomon 1998; Bryant \& Scoville 1999; Tacconi
\textit{et al.} 1999). 

%-------------------------------------------------------------------------------
\begin{figure} 
\includegraphics[width=\linewidth]{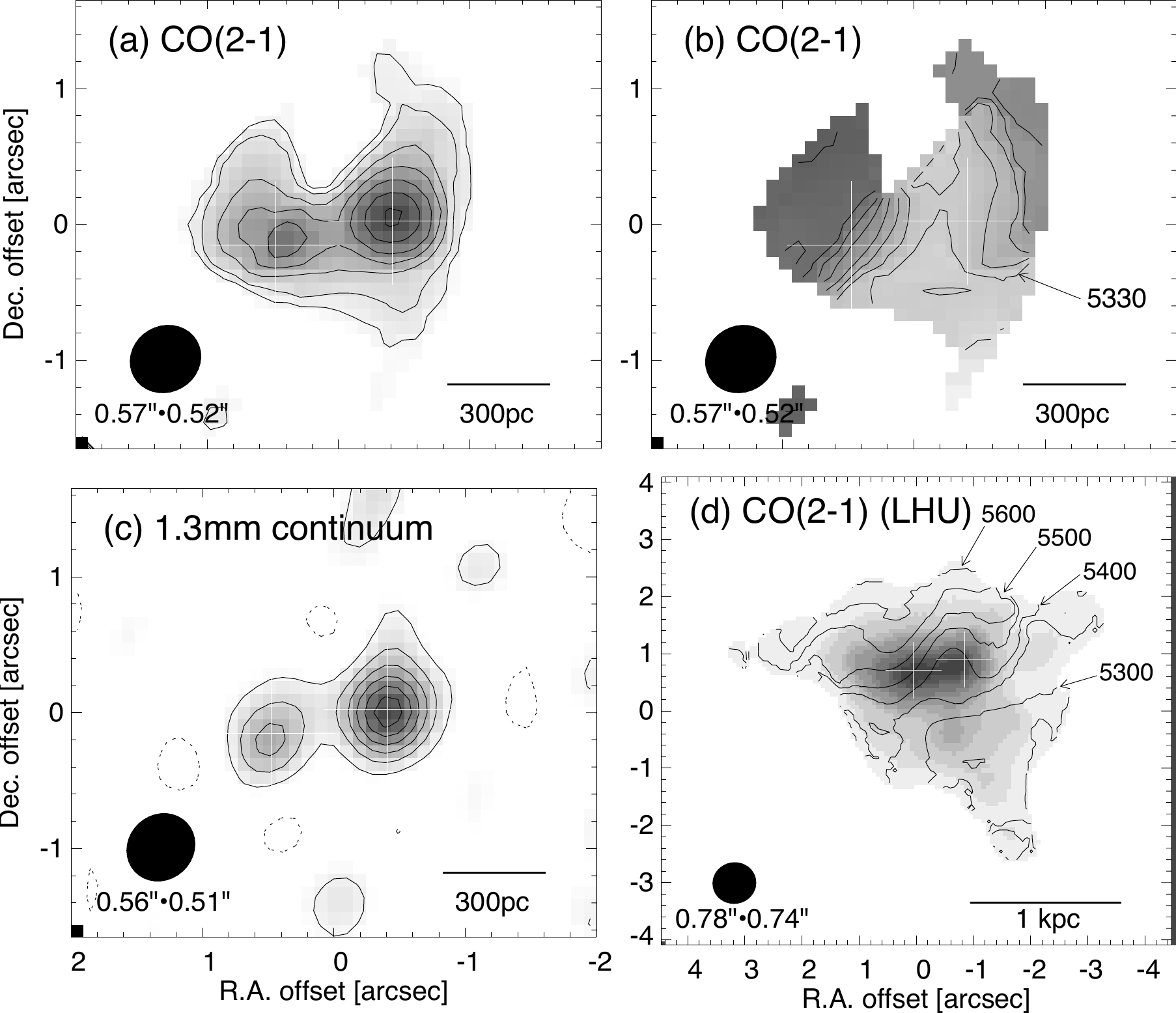}
\caption{The merging galactic nuclei of Arp\,220 are shown in 0.5\,arcsec
resolution imaging of the CO(2--1) and dust continuum emission (Sakamoto
\textit{et al.} 1999). The four panels show: a) continuum-subtracted CO(2--1),
b) the CO mean velocities, c) the 1.3\,mm dust continuum, and d) the total CO
emission including the extended molecular emission. These data resolve the two
nuclei (separated by 1\,arcsec \hbox{E-W} or 350\,pc) and the crosses indicate the
1.3\,mm dust continuum peaks. In the molecular gas emission, the two galaxy
nuclei have counter-rotating disks as seen in the upper-right
image.}
\label{arp220}
\end{figure}
%-------------------------------------------------------------------------------

Arp\,220 has been imaged at high resolution in the 2.6\,mm CO line (Scoville
\textit{et al.} 1991), 3\,mm HCN (Radford \textit{et al.} 1991), 1.3\,mm CO
(Scoville \textit{et al.} 1997; Downes \& Solomon 1998; Tacconi \textit{et al.}
1999) and most recently in the 0.9\,mm CO and HCO$^+$ lines (Sakamoto \textit{et
al.} 2009). The CO(2--1) line emission exhibits two peaks separated by
0.9\,arcsec, and a larger inclined disk of molecular gas (Scoville \textit{et
al.} 1997; Downes \& Solomon 1998; Sakamoto \textit{et al.} 1999; Tacconi
\textit{et al.} 1999). At 0.5\,arcsec resolution (Sakamoto \textit{et al.}
1999), the CO and 1.3\,mm continuum imaging shown in Fig.~\ref{arp220} reveals
\textit{counter-rotating} disks of gas in each of the IR nuclei. The kinematic
data require very high mass concentrations in both of the nuclei. The nuclear
disks are counter-rotating, consistent with the notion that the most complete
and violent merging should be associated with counter-rotating precursor
galaxies in which there is naturally angular momentum cancellation during the
merging. The masses of each nucleus are apparently dominated by the molecular
gas -- a common finding of the ULIRG galaxy studies (Bryant \& Scoville 1999).
The nucleus of Arp\,220 has a total molecular gas mass $M_{\rm
H_2}\sim9\times10^9$\,\msun\ within 300\,pc (Scoville \textit{et al.} 1997;
Downes \& Solomon 1998; Sakamoto \textit{et al.} 1999; Tacconi \textit{et al.}
1999). This H$_2$ mass within 300\,pc is 3--4 times that of the entire Milky Way
(most of which is in the ring at 4--8\,kpc).

Although Arp\,220 has enormous masses of molecular gas, one should not visualise
this gas being contained in self-gravitating GMCs like those in the Galaxy. The
high brightness temperature of the observed CO emission (Scoville \textit{et
al.} 1997) requires that the gas distribution be continuous, i.e., filled
nuclear disks, not a cloudy medium. The mass density within the disks implies a
visual extinction $A_V$\,$\sim$\,2000\,mag perpendicular to the disks! Clearly,
studying such regions can only be superficial in the optical/NIR and must rely
on much longer-wavelength observations.

The western nucleus in Arp\,220 exhibits hard X-ray emission (Iwasawa \textit{et
al.} 2005) and very high-velocity CO emission. The inner dust emission source
seen at $\lambda$\,$\sim$\,1 is extremely compact with 10$^9$\,\msun\ within
$R$\,$<$\,35\,pc (Downes \& Eckart 2007). Downes \& Eckart argue that the
inferred dust temperature of $\sim$175\,K implies a higher radiation energy
density for its heating than could be provided by a compact starburst and
therefore that this nucleus may harbour a luminous and massive black hole. 
  
\subsection{An aside: Sgr\,A$^*$ -- an extraordinary ISM}

\begin{figure} 
\includegraphics[width=\linewidth]{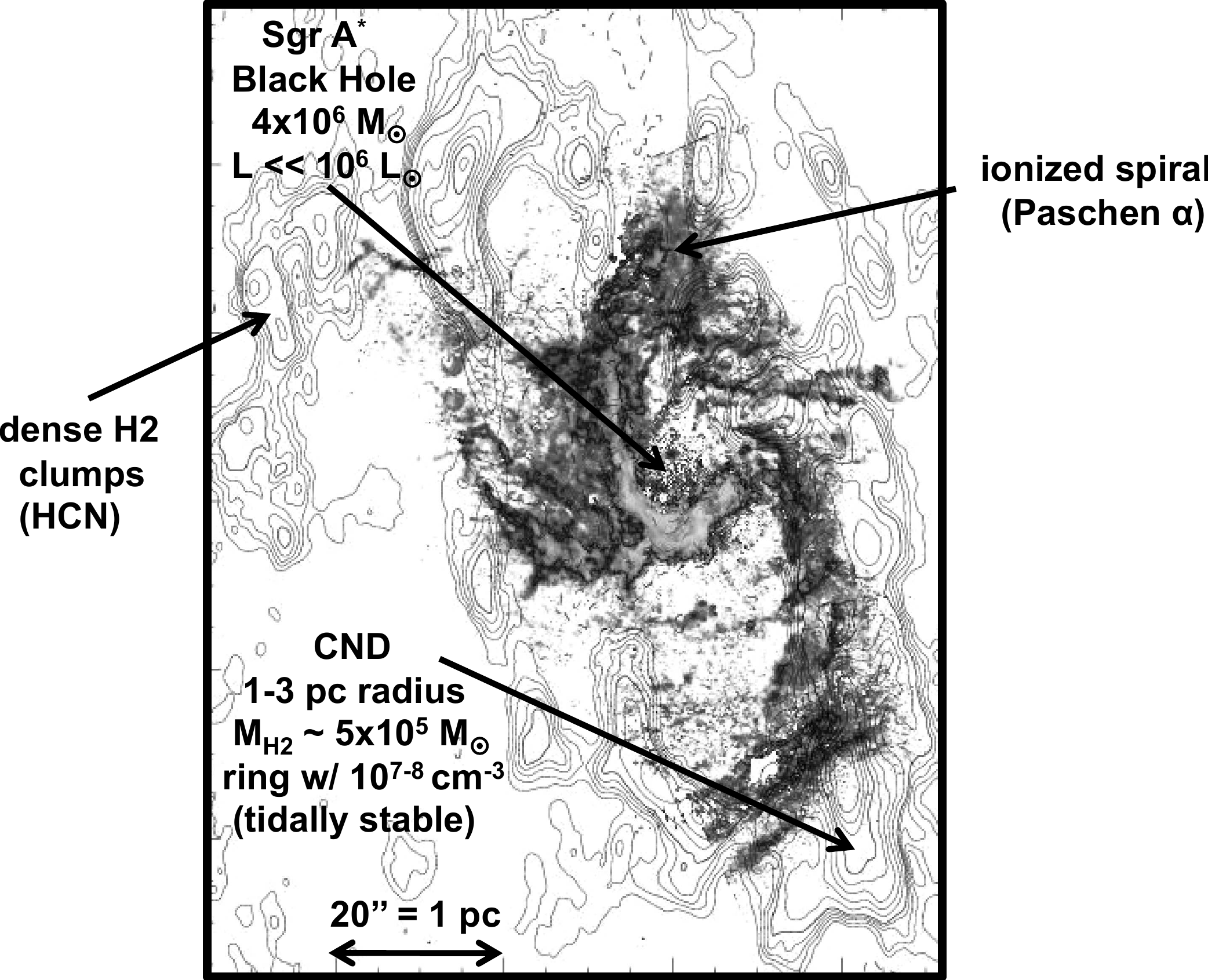}
\caption{The circumnuclear disk (CND) in the Galactic nucleus disk in ionised
gas (greyscale) as imaged in the H{\sc i} P$\alpha$ line at 1.87\,$\mu$m
(Scoville \textit{et al.} 2003) and the dense molecular gas (contours) seen in
HCN emission (Christopher \textit{et al.} 2005). The ionised gas is distributed
in a `spiral' pattern and at the inner surfaces of the neutral molecular clumps.
The densities in the molecular gas are $\sim10^{7-8}$\,cm$^{-3}$(!), orbiting at
1--3\,pc radius.}
\label{sgra}
\end{figure}

Given the very limited spatial resolution in observations of distant nuclei such
as Arp\,220, it is instructive and cautionary to look at the properties of the
ISM in our own Galactic nucleus despite the fact that it is not currently very
active. The massive black hole associated with the radio source Sgr\,A$^*$ has a
mass of 4\,$\times$\,10$^{6}$\,\msun\ -- derived from the motions of stars within the
central 1\,pc (Genzel 2006; Ghez 2007). A circumnuclear disk (CND) at radius out
to $\sim$\,3\,pc is seen with both ionised gas and dense molecular clumps on the
exterior. Figure~\ref{sgra} shows the ionised gas as probed in the H{\sc i}
P$\alpha$ line at 1.87\,$\mu$m (greyscale; Scoville \textit{et al.} 2003) in a
structure termed the mini-spiral and the dense molecular gas as probed in the
3\,mm HCN line (contours; Christopher \textit{et al.} 2005). The molecular gas
is in a clumpy ring at 1--3\,pc radius. The ionised gas in the mini-spiral may
be infalling or outflowing; the ionised gas at the edge of the disk in the
southeastern arm probably arises from photoionisation at the inner edge of the
molecular cloud ring. 

%-------------------------------------------------------------------------------
%-------------------------------------------------------------------------------
 
The total mass of the molecular gas within 3\,pc of the black hole is variously
estimated at 1--5\,$\times$\,10$^5$\,\msun\ and has an orbital time $\sim$\,$10^5$\,yr.
Given that the black hole is only $\sim$\,10 times more massive it is clear that
very little of the nearby ISM will actually make it to the black hole, unless we
are viewing an extremely atypical epoch for Sgr\,A$^*$. In any case, it is
certainly the case that the ISM here is not being accreted on an orbital
timescale since the luminosity of Sgr\,A$^*$ is $\ll$\,10$^6$\,\lsun. 

Just as was the case for Arp\,220, our intuition about the properties of the ISM
in the disk of our galaxy (i.e., self-gravitating GMCs) clearly cannot be
transferred directly to the ISM in the Milky Way nucleus. The molecular gas
clumps in the CND have extraordinary properties! Their sizes are $<$1\,pc but
densities $10^{7-8}$\,cm$^{-3}$ (Christopher \textit{et al.} 2005). (By
contrast, a typical Galactic 10$^5$\,\msun\ GMC has a diameter 40\,pc and mean
density $\sim$300\,cm$^{-3}$, see Section~\ref{prop}). The higher densities are
required near Sgr\,A$^*$ if the clumps are to be stable against tidal
disruption, given their proximity to the central point mass. If the dust-to-gas
ratio in the clumps is similar to the standard Galactic value, then the inferred
column densities of 10$^{25}$\,cm$^{-2}$ imply an extinction of
$A_V\sim10^4$\,mag through each clump. Since the molecular ring/torus is clearly
not appearing edge-on, its angular momentum is not perpendicular to the Galactic
plane. The ring therefore probably resulted from the accretion and disruption of
a single cloud at non-zero height out of the Galactic plane, rather than smooth
accretion over time from the larger nuclear disk. If such a cloud arrived at the
point where it was tidally disrupted with a non-zero height the resulting
fragments might end up as clumps in an orbital plane inclined to the Galactic
plane, as is observed for the CND.

At present there is no evidence of star formation within the molecular clumps
despite their very high densities and extraordinarily short internal dynamical
timescales $\sim10^4$\,yr. However, the very high extinctions suggested above
would preclude detection of embedded young stars and the only means of detecting
them might be via radio detection of ultra-compact H{\sc ii} regions or maser
emission. The unusually high densities of the molecular clumps must reflect
the fact that at lower density they would not be tidally stable and therefore
could not exist for long (see Section~\ref{disks}).

%-------------------------------------------------------------------------------
\begin{figure} 
\includegraphics[width=\linewidth]{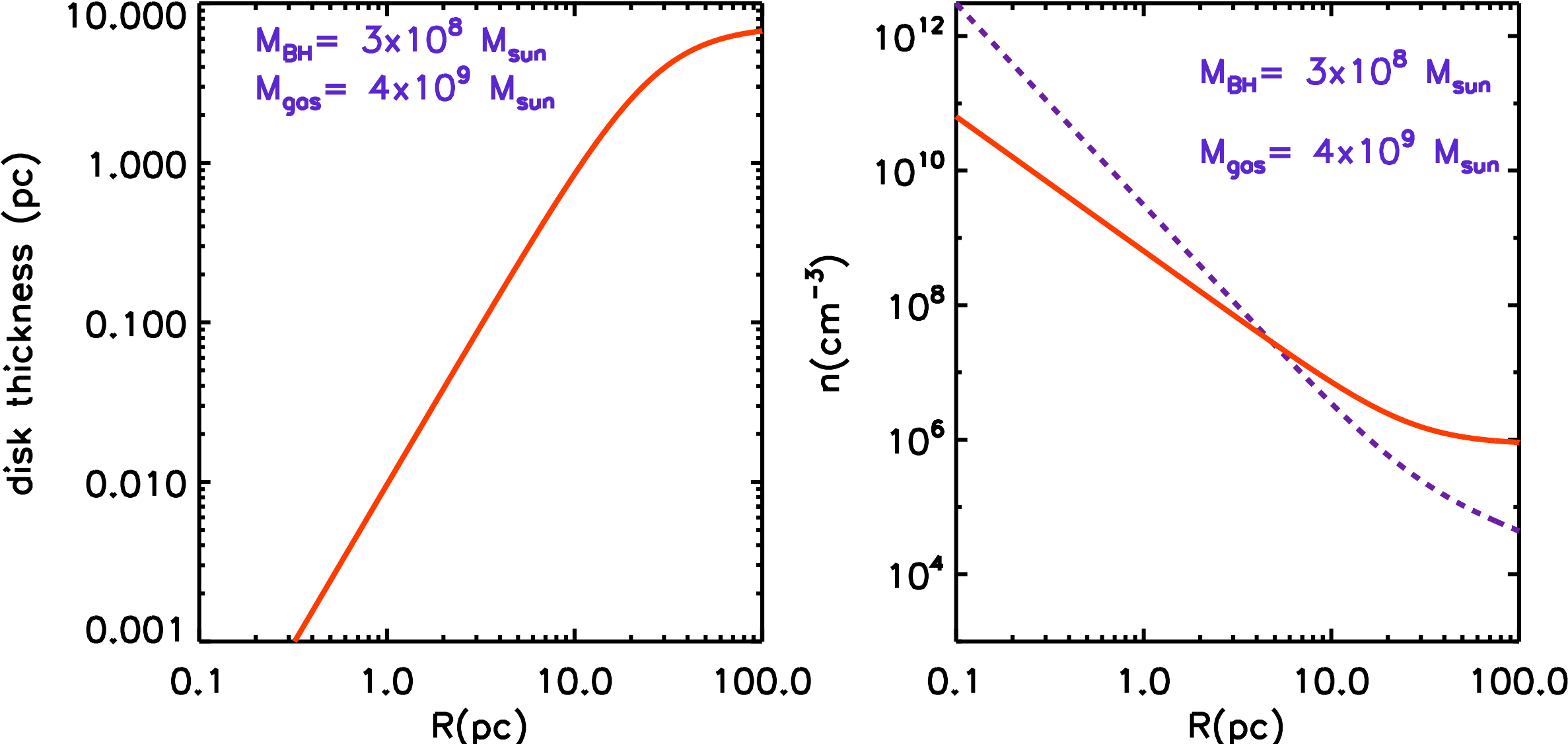}
\caption{A very crude model for the disk in gas-rich merging systems like
Arp\,220 is shown. We assume a central point mass (black hole) and a uniform gas
surface density disk extending to 100 pc radius. The turbulent velocity
dispersion within the disk is taken to be 50 km/s (based on Arp\,220; Scoville
\textit{et al.} 1997). The disk thickness as a function of radius for
hydrostatic equilibrium is shown in the right panel and the mean gas density in
the left panel (solid red line). The critical density required for stability of
a self-gravitating object (e.g., a cloud) in the disk is shown by the dashed
(blue) curve. The right panel shows the extraordinarily high gas densities
expected for these very simple assumptions and underscores the fact that
self-gravitating clouds should not form in the inner disk due to tidal
disruption.}
\label{disk} 
\end{figure}
%-------------------------------------------------------------------------------

\subsection{Nuclear starburst disks}
\label{disks}

As a first step toward understanding the massive gas disks like that seen in
Arp\,220, we consider an extremely simple model with uniform gas surface density
out to radius 100\,pc. At the centre of the disk, we assume a point mass of
4\,$\times$\,10$^8$\,\msun\ -- either a central black hole or a nuclear star cluster.
The disk is assumed to be self-gravitating and in hydrostatic equilibrium with a
sound speed of 50\,km/s, representative of the turbulent velocity dispersion
measured in Arp\,220. Although this is meant only as an illustrative, {\it
gedanken}, experiment, it clearly shows that the conditions are inescapably
different from those of star-forming clouds in the Galactic disk. 

In Fig.~\ref{disk}, the disk thickness and mean gas densities are shown as a
function of radius out to 100\,pc. The thickness of the disk is typically less
than 10\,pc and the mean densities are everywhere $>10^6$\,cm$^{-3}$, getting up
to $10^{10}$\,cm$^{-3}$ near the centre. The right panel of Fig.~\ref{disk}
also shows the gas density required for tidal stability (dashed line) -- clearly
demonstrating that it will be difficult to form stable clouds within the central
10\,pc radius. Within this region, we should expect a more continuous gaseous
disk structure, not a cloudy ISM as found in the centre of our Galaxy.

%-------------------------------------------------------------------------------
\begin{figure} 
\includegraphics[width=\linewidth]{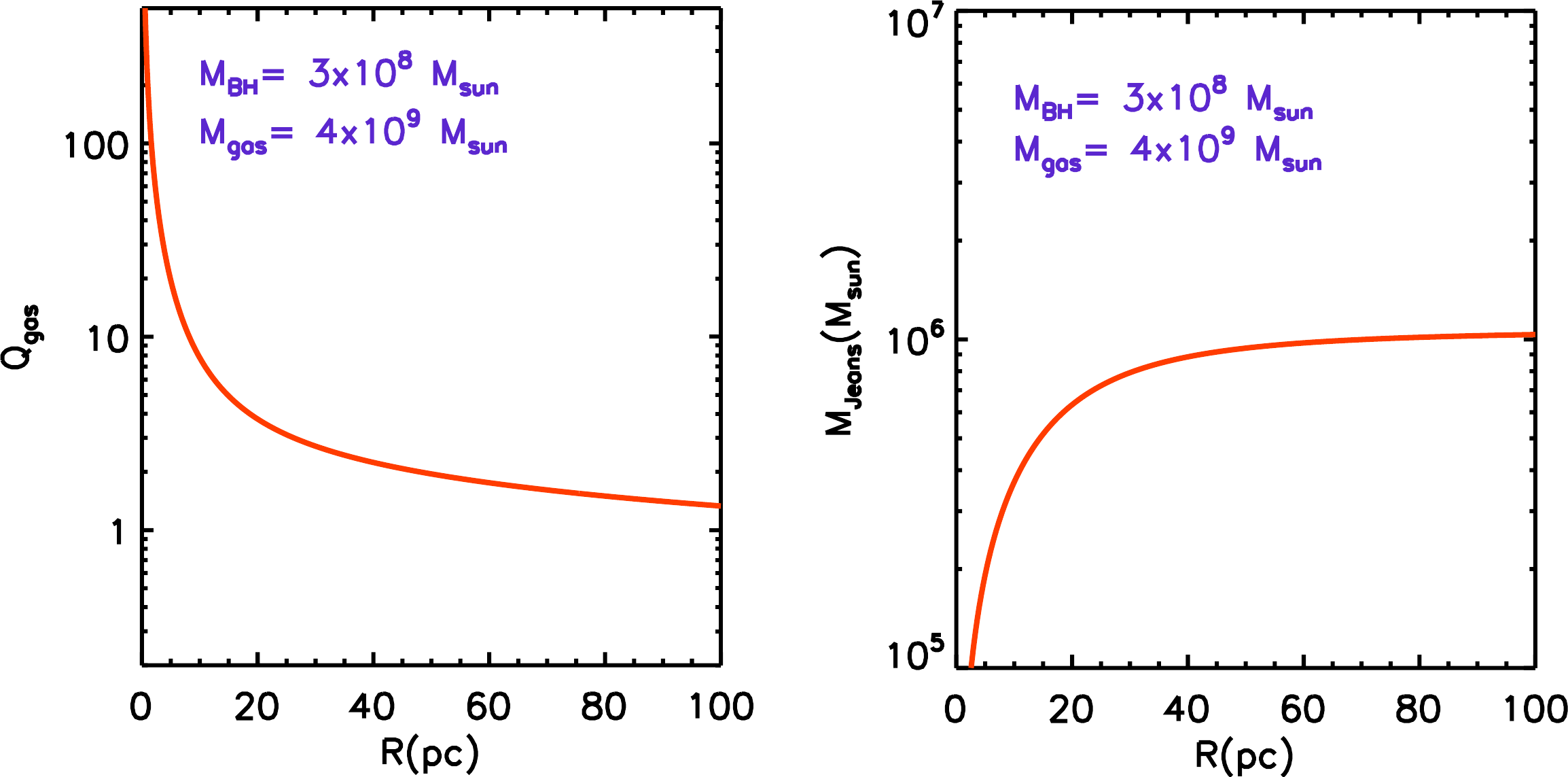}
\caption{The left panel shows the Toomre $Q$ stability parameter ($Q<1$
unstable) for the disk parameters in the previous figure and the right panel
shows the Jeans mass for collapse/self-gravitation  assuming a sound speed of
50\,km/s and neglecting tidal shear.}
\label{stability}
\end{figure}
%-------------------------------------------------------------------------------

The gravitational stability of the disk can also be seen from the Toomre $Q$
parameter shown in Fig.~\ref{stability} (left panel). Within the inner 10\,pc,
$Q\gg$\,1 and it is very hard to form self-gravitating structures in the ISM, such
as clouds. On the other hand, outside 10\,pc radius such clouds might form (if
they are not disrupted by cloud collisions) but their masses are required to be
very large, exceeding $10^6$\,\msun. 

At the inner boundary of the nuclear disk ($R$\,$\sim$\,1--5\,pc), one expects
that the dust will sublimate and any ISM inside this radius will be dust-free
and therefore have much lower radiative opacity, permitting radial accretion
inwards. On the other hand, at the sublimation boundary, there will be a strong
radiation pressure gradient in the outward direction, tending to push the inner
edge of the dust accretion disk to larger radii. The physics of this transition
zone should indeed be interesting -- the material pushed outwards will intercept
orbit disk gas with higher specific angular momentum, and the dust sublimation
interface will also be Rayleigh-Taylor unstable,\linebreak perhaps leading to radial
streams of accretion flow. I leave these details to the students to work out!

\subsection{Maximum-rate starbursts -- the dust Eddington limit}

For Arp\,220, the estimated nuclear SFR of 100--200\,\msun\ per year is at least
a factor of ten less than the maximal rate, obtained by dividing the ISM by the
orbital timescale of $\sim$\,2\,$\times$\,10$^6$\,yr. One very effective means of
restricting the star fromation (or even ejecting the dense ISM) is via the
radiation pressure on dust of the central starburst (or AGN) luminosity. For a
self-gravitating gas and dust mass, the effective `Eddington' limit is
approximately 500\,\lsun/\msun, similar to the overall mass-to-light ratio
measured in the Arp\,220 nucleus ($10^{12}$\,\lsun/$3\times10^9$\,\msun). For
higher luminosity-to-mass ratios, the ISM is blown out of the region. 

The significant role of radiation pressure in high-mass star formation and
starbursts for feedback through radiation pressure on buildup has only recently
been appreciated. The outward radiation pressure in a starbursting cloud core or
galactic nucleus will dominate self-gravity at radius $R$ when 
\begin{equation}
\label{ed}
{L\langle\kappa\rangle\over4\pi R^2 c } > {G M_{\rm R} \over R^2}, 
\end{equation}
where $\langle\kappa\rangle$ is the \textit{effective} radiative absorption
coefficient per unit mass. Although the original stellar radiation is primarily
UV and visible, the dust in the cloud core absorbs these photons and re-radiates
the luminosity in the IR. The `effective' absorption coefficient takes account
of the fact that outside the radius where $A_V$\,$\sim$\,1\,mag, the luminosity is at
longer wavelengths where the dust has a reduced absorption efficiency.  For the
standard ISM dust-to-gas ratio (Draine \& Li 2007), $A_V=1$\,mag corresponds to
a column density N$_{\rm H}$\,=\,2\,$\times$\,10$^{21}$\,cm$^{-2}$ and 
\begin{equation}
\label{kap}
\langle\kappa\rangle=312~{\lambda_V\over\lambda_{\rm eff}(R)}\quad{\rm cm^2 g^{-1}},
\end{equation}
where $\lambda_{\rm eff}(R)$ is the absorption coefficient-weighted mean
wavelength of the radiation field at radius $R$ and for simplicity, we have
adopted a $\lambda^{-1}$ wavelength dependence for the absorption efficiency.

Combining Equations~\ref{ed} and \ref{kap}, we find that the radiation pressure
will exceed the gravity when
\begin{equation}
(L/M)_{\rm cl}\ge42{\lambda_{\rm eff}\over\lambda_{V}}{\lsun\over\msun}. 
\end{equation}
For $\lambda_{\rm eff}$\,$\sim$\,3--10\,$\mu$m, $\lambda_{\rm eff}/\lambda_{V}$
is approximately 10 and \textit{the radiation pressure limit is}
$\sim$\,500\,\lsun/\msun. Thus, for luminosity-to-mass ratios exceeding this
value radiation pressure will halt further accretion. For a forming OB star
cluster, this luminosity-to-mass ratio is reached at about the point where the
upper main sequence is first fully populated, i.e., a cluster with approximately
2000\,\msun\ distributed between 1 and 120\,\msun. A similar `Eddington' limit
applies for any dust-embedded starburst region (see Scoville \textit{et al.}
2001; Scoville 2003; Murray \textit{et al.} 2005; Thompson \textit{et al.}
2005). It turns out that much larger-scale nuclear starbursts such as that in
Arp\,220 have approximately reached the same empirical limit of
500\,\lsun/\msun, suggesting that nuclear starburst activity may also be
regulated by a balance of self-gravity and radiation pressure support. 

\subsection{AGN -- starburst: observational connections}

Evidence has also accumulated for an evolutionary link between merging ULIRGs
and UV/optical quasi-stellar objects (QSOs) as suggested by Sanders \textit{et
al.} (1988): similar local space densities for ULIRGs and QSOs; continuity of
FIR SEDs smoothly transitioning between the two classes; the occurrence of
AGN-like emission lines (Veilleux \textit{et al.} 1999) and significant
point-like nuclei (less than 0.2\,arcsec -- Scoville \textit{et al.} 2000) in
30--40\% of the ULIRGs; and the association of both ULIRGs and some QSOs
(MacKenty \& Stockton 1984; Bahcall \textit{et al.} 1997) with galactic
interactions (see the review article by Sanders \& Mirabel 1996). Whether the
entire QSO population had precursor ULIRGs (implying that galactic merging is
the predominant formation mechanism for AGN) is certainly not yet settled and at
lower AGN luminosities, much of the activity is probably not directly associated
with galaxy merging or interactions.

Two possible scenarios linking the ULIRG and luminous AGN phenomena are: 1) that
the abundant ISM, deposited in galactic nuclei by merging, fuels both the
nuclear starburst and feeds the central black hole accretion disk; or
alternatively, 2) the post-starburst stellar population evolves rapidly with a
high rate of mass return to the ISM in the galactic nucleus -- leading to
sustained fuelling of the black hole (e.g., Norman \& Scoville 1988).

\subsection{AGN -- starburst: theoretical connections}

The largest potential sources of fuel for AGN are the nuclear ISM (until it has
been cleared out) and the mass loss associated with normal stellar evolution of
stars in the galactic nucleus. Most recent discussion has assumed the former
link; so I would like to briefly discuss the latter possibility here --
partially as a stimulus for further investigation by students. For a nuclear
starburst population, approximately 20\% of the initial stellar mass will be
lost via red giant mass-loss winds within the first 2\,$\times$\,10$^8$ years
(Norman \& Scoville 1988). (The mass return associated with supernovae is
probably significantly less.) Scoville \& Norman (1988) and Scoville \& Norman
(1995) examined the fate of this mass-loss material -- specifically to account
for both the broad emission lines (BELs) and the broad absorption lines (BALs)
seen in AGN. Since the stars will not be disrupted by AGN feedback once the
nucleus becomes active, the stars can provide a long-term supply of fuel to
power the AGN.

The physics of the dust shed by the stars in their stellar winds, particularly
its evaporation, is critical to determining whether the mass-loss material
accretes inwards to an accretion disk or is blown outwards by radiation
pressure. This consideration leads naturally to a division of the central
cluster environment into: 1) an inner zone ($r$\,$\leq$\,1\,pc) where the dust is
heated to above the sublimation temperature and the lower opacity of the
dust-free gas allows it to fall inwards, and 2) an outer zone ($r$\,$\geq$\,1\,pc)
where the dust (and gas) survives and is driven outwards at velocities up to
0.1$c$ by radiation pressure.

It is interesting to note that the density required for tidal stability of a BEL
cloud at 1\,pc radius from the black hole requires that the clouds must have
extremely high internal densities. At 1\,pc from a 10$^9$\,\msun\ black hole,
the Roche limit density is 3\,$\times$\,10$^{11}$\,cm$^{-3}$. This provides a strong
argument for the stellar (rather than interstellar) scenario to account for the
BELs, using the stellar masses to bind the emission line gas. 

%
%%%%%%%%%%%%%%%%%%%%%%%%%%%%%%%%%%%%%%%%%%%%%%%%%%%%%%%%%%%%%%%%%%%%%%%%%%%%%%%%
%

\section{Evolution of galaxies at high redshift}

Over the last 15 years extensive surveys of high-redshift galaxy evolution
(e.g., GOODS, UDF, COSMOS, AEGIS and CANDELS) with space- and ground-based
telescopes have dramatically increased our understanding of galaxy evolution in
the early universe. All of these go very deep (typically $>$\,27--30\,mag~AB) but
cover very different size areas -- most also have excellent multi-wavelength,
ancillary data. The largest survey (COSMOS) has over a million galaxies with
photometry and photometric redshifts at $z$\,=\,0.2--6 and the deepest (UDF) now
has detections at $z$\,=\,6--8, probing the first 1\,Gyr of cosmic time. Major
evolution is seen in the galaxy properties: stellar mass, luminosity, size and
SFR.

%-------------------------------------------------------------------------------
\begin{figure} 
\includegraphics[width=0.85\linewidth]{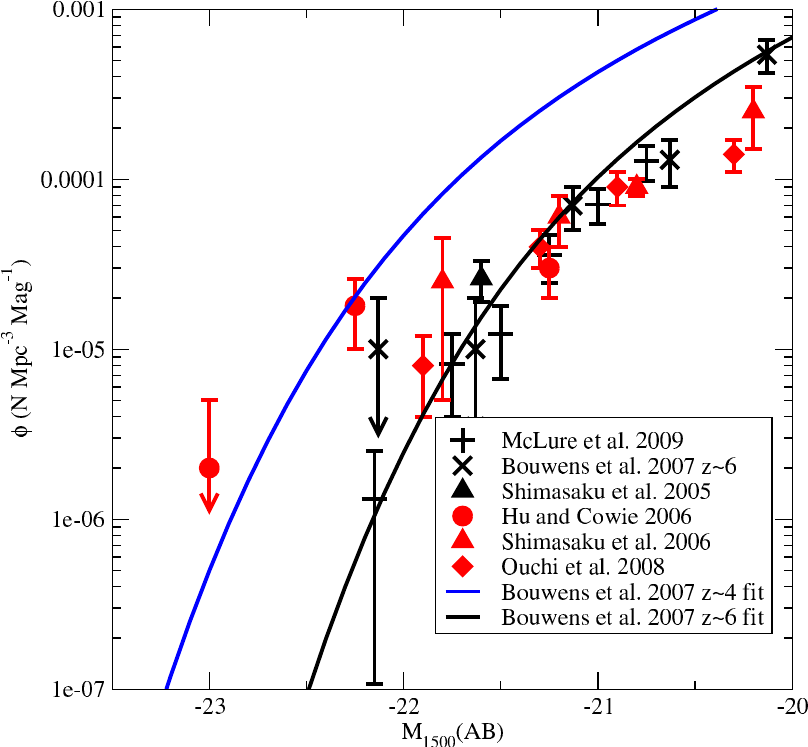}
\caption{A compilation of $z$\,=\,4 and 6 UV LFs for galaxies (Capak \textit{et
al.} 2011). The curves are Schechter function fits by Bouwens \textit{et al.}
(2007) for $z$\,=\,4 (blue) and 6 (black) galaxies and the points are derived
for Lyman alpha emitters (LAEs) and Lyman break galaxies
(LBGs).}
\label{highz_lf}
\end{figure}
%-------------------------------------------------------------------------------

\subsection{Luminosity and Mass Functions}

Figure~\ref{highz_lf} shows a compilation of recent determinations of the
$z$\,=\,4--6 luminosity functions (LFs; Capak \textit{et al.} 2011). Evolution
of the UV LF (probing the distribution of star-forming galaxies) is clearly seen
with the number densities increasing at all luminosities as one goes to lower
redshift and there is apparent steepening of the low-$L$ power law going to
higher redshift, i.e., more low-luminosity galaxies contributing. If this
persists to higher redshift then it can be argued that the reionisation of the
universe at $z$\,$\sim$\,10 must have been produced by relatively low-luminosity
galaxies (Robertson \textit{et al.} 2010). At $z$\,$>$\,6, the samples are very
small, $\sim$\,20--50 objects with virtually no spectroscopic confirmation.  Even
at $z$\,=\,6, one can see large scatter in the different determinations of the
LF, shown as points in Fig.~\ref{highz_lf}. It is noteworthy that the high-$L$
portion of the LF is very poorly constrained (see Fig.~\ref{highz_lf}) and could
even be a power law rather than the Schechter function (exponential fall-off)
used in fitting. 

%-------------------------------------------------------------------------------
\begin{figure} 
\includegraphics[width=0.77\linewidth]{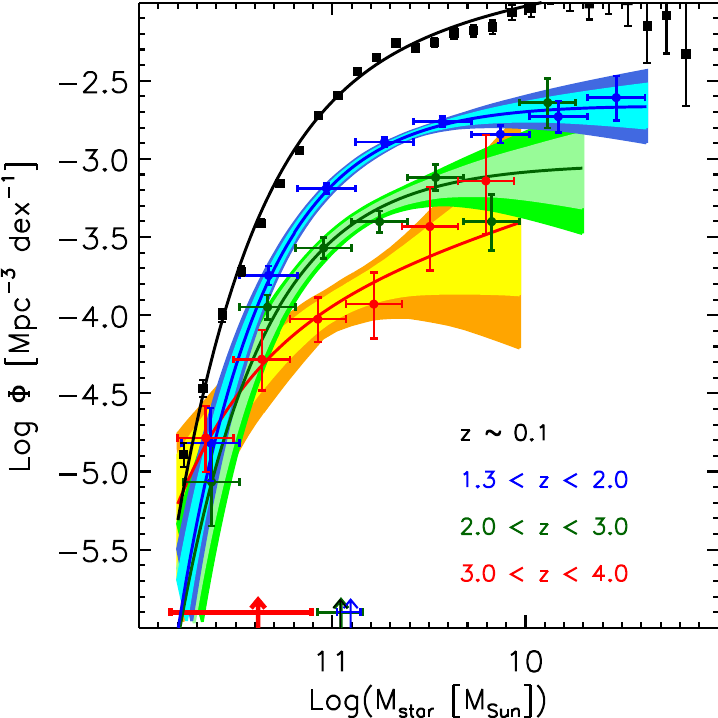}
\caption{The stellar mass function for galaxies at $z$\,=\,0.1--4 from 
Marchesini \textit{et al.} (2009).}
\label{highz_mf}
\end{figure}
%-------------------------------------------------------------------------------
        
At lower redshifts, comparatively good samples exist for the stellar mass
functions of galaxies (see Fig.~\ref{highz_mf}). Strong growth in the number
density of galaxies at all masses is seen but the largest increase at late
epochs \hbox{(low $z$)} occurs in the lower-mass galaxies (see, e.g., Ilbert \textit{et
al.} 2010). This `downsizing' of evolution at later epochs is also seen in
the galaxies with star formation and AGN activity. 

In principle, the buildup of stellar mass in galaxies as a function of time
should be consistent with the evolution of star formation activity within
galaxies in each mass bin. Discrepancies between these two might indicate the
rate of galactic merging (assuming the masses and SFRs are reliable). This has
been explored by Drory \& Alvarez (2008) who find reasonable consistency between the
stellar mass buildup and the SFR evolution -- providing the SFRs drop first in
the highest-mass galaxies and galaxies typically undergo $\sim$\,1 major merger
after $z$\,=\,1.5 (see also Reddy 2011).

%-------------------------------------------------------------------------------
\begin{figure} 
\includegraphics[scale=0.7]{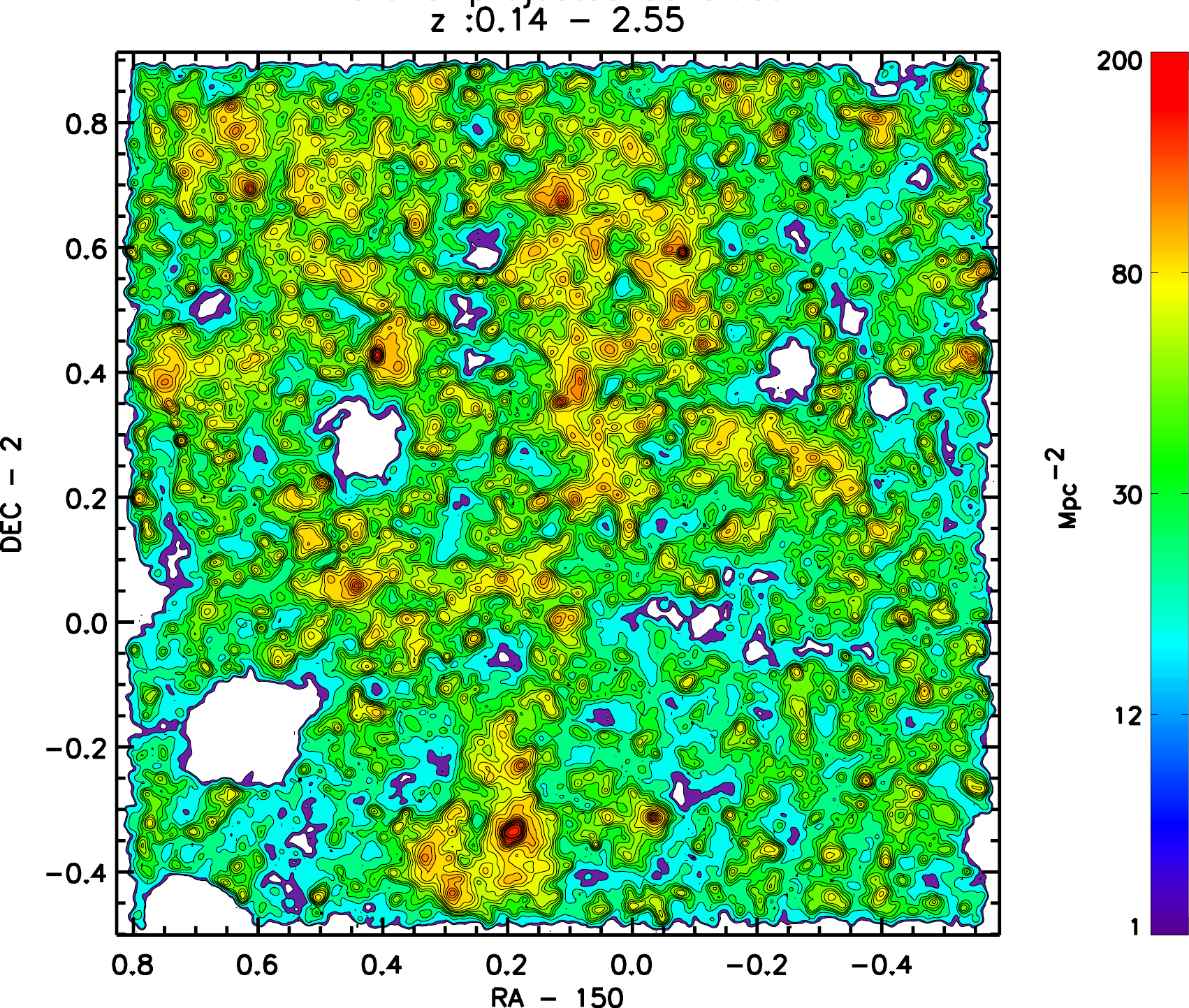}{ht}
\caption{Large-scale structures mapped by the projected density of galaxies in
redshift slices ($\Delta z$\,=\,0.02--0.2) in the COSMOS survey field (Scoville
\textit{et al.} 2007a; Scoville \textit{et al.} 2012). Over 200
significantly overdense regions are seen over the redshift range
$z$\,=\,0.1--2.55.}
\label{lss}
\end{figure}
%-------------------------------------------------------------------------------

\subsection{Environmental correlations}

In $\Lambda$CDM simulations for the early Universe, it is the most massive,
highly biased, structures which form earliest. Such structures will become the
locations of massive galaxy clusters with the most massive galaxies built there
first. The COSMOS survey (Scoville \textit{et al.} 2007b) was specifically
designed to probe a sufficiently large area on the sky (2 square degrees
corresponding to $\gtrsim$\,50--100 comoving Mpc at $z$\,$>$\,0.5) that the full
range of environmental densities could be sampled at all redshifts. This enables
both the mapping of the large-scale structure and the investigation of the
correlation of galaxy evolution with environment. High-accuracy photometric
redshifts and a large sample of spectroscopic redshifts enable the separation of
galaxy structures along the line of sight. Figure~\ref{lss} shows the 200
overdense regions of galaxies seen in COSMOS at $z$\,=\,0.1--2.55 (Scoville
\textit{et al.} 2012).
      
\begin{figure} 
\includegraphics[width=\linewidth]{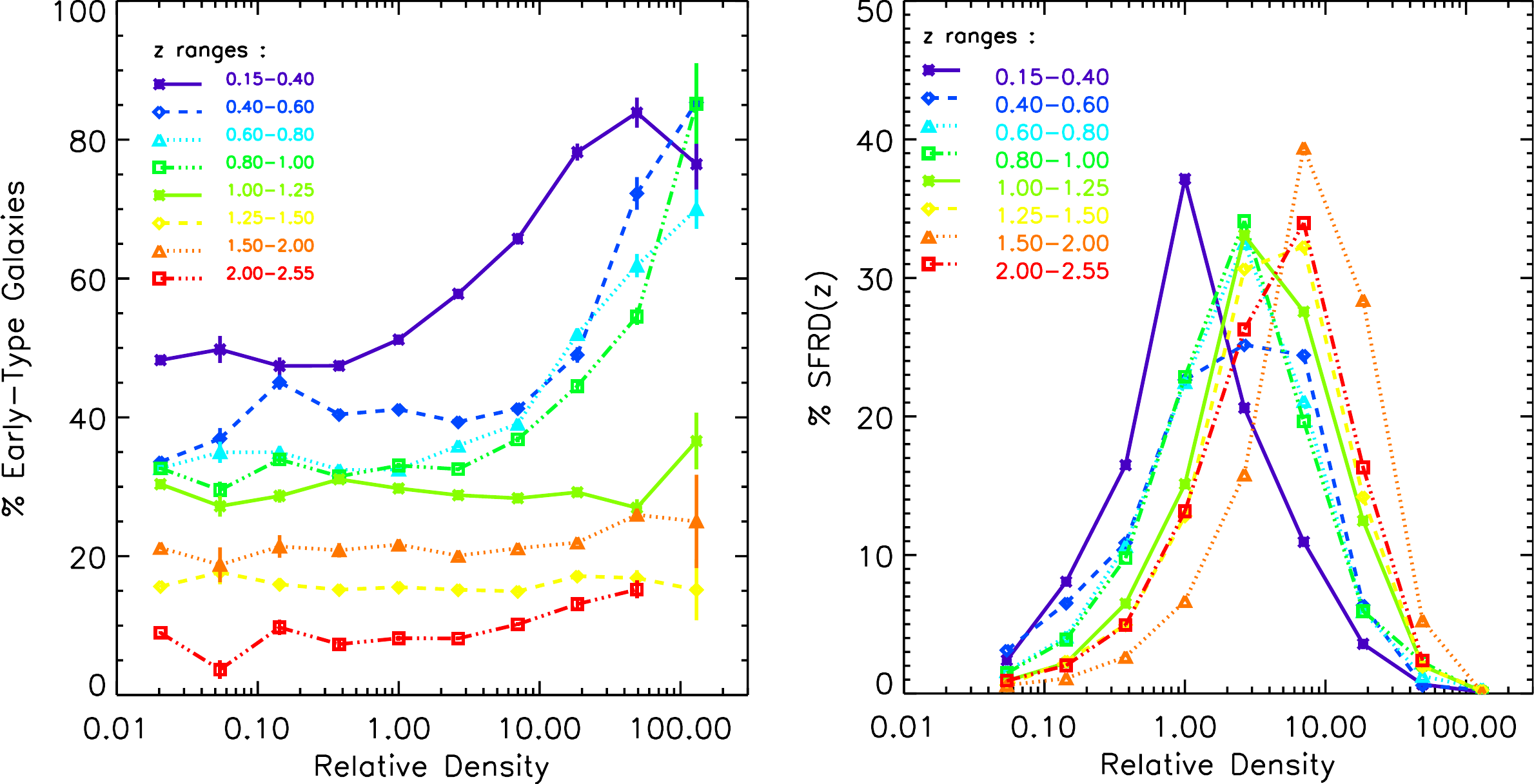}
\caption{Left panel: the early-type (E--Sa) galaxy fraction is clearly
correlated with both redshift and large-scale structure density from
Fig.~\ref{lss} (Scoville \textit{et al.} 2007a; Scoville \textit{et al.} 2012).
The early-type galaxies form first in the highest-density environments. Right
panel: the percentage of the overall SFRD is shown as a function of redshift and
large-scale structure density. The dominant environments for star formation
shift to lower density at later epochs (Scoville \textit{et al.} 2012).}
\label{cosmos}
\end{figure}

Using the environmental densities shown in Fig.~\ref{lss}, I show the percentage
of early-type galaxies (with SED corresponding to E--Sa galaxies) as a function
of density and redshift in Fig.~\ref{cosmos} (left panel). As is well known from
many studies, the early-type galaxy fraction increases systematically to a
fraction $\sim$\,50\% at $z$\,=\,0. However, Fig.~\ref{cosmos} (left panel) also
clearly isolates the differential evolution associated with large-scale
structure density -- the early-type galaxies form first in the highest-density
regions of the large-scale structure. 

The evolution of the SFR density (SFRD\,=\,SFR per comoving volume) rises a
factor 20 from $z$\,=\,0 to a peak at $z$\,=\,2--3. An important question is: in
which environments is the star formation occurring at each epoch?
Figure~\ref{cosmos} (right panel) shows that the SFRD systematically shifts from
dense to less dense regions of the large-scale structure as time progresses.
Peng \textit{et al.} (2010), using both COSMOS and SDSS data, show that the
quenching of star formation activity in galaxies has two \emph{separable} terms:
one dependent on galaxy mass and the other on environmental density.

%-------------------------------------------------------------------------------
%-------------------------------------------------------------------------------

%-------------------------------------------------------------------------------
\begin{figure} 
\includegraphics[width=\linewidth]{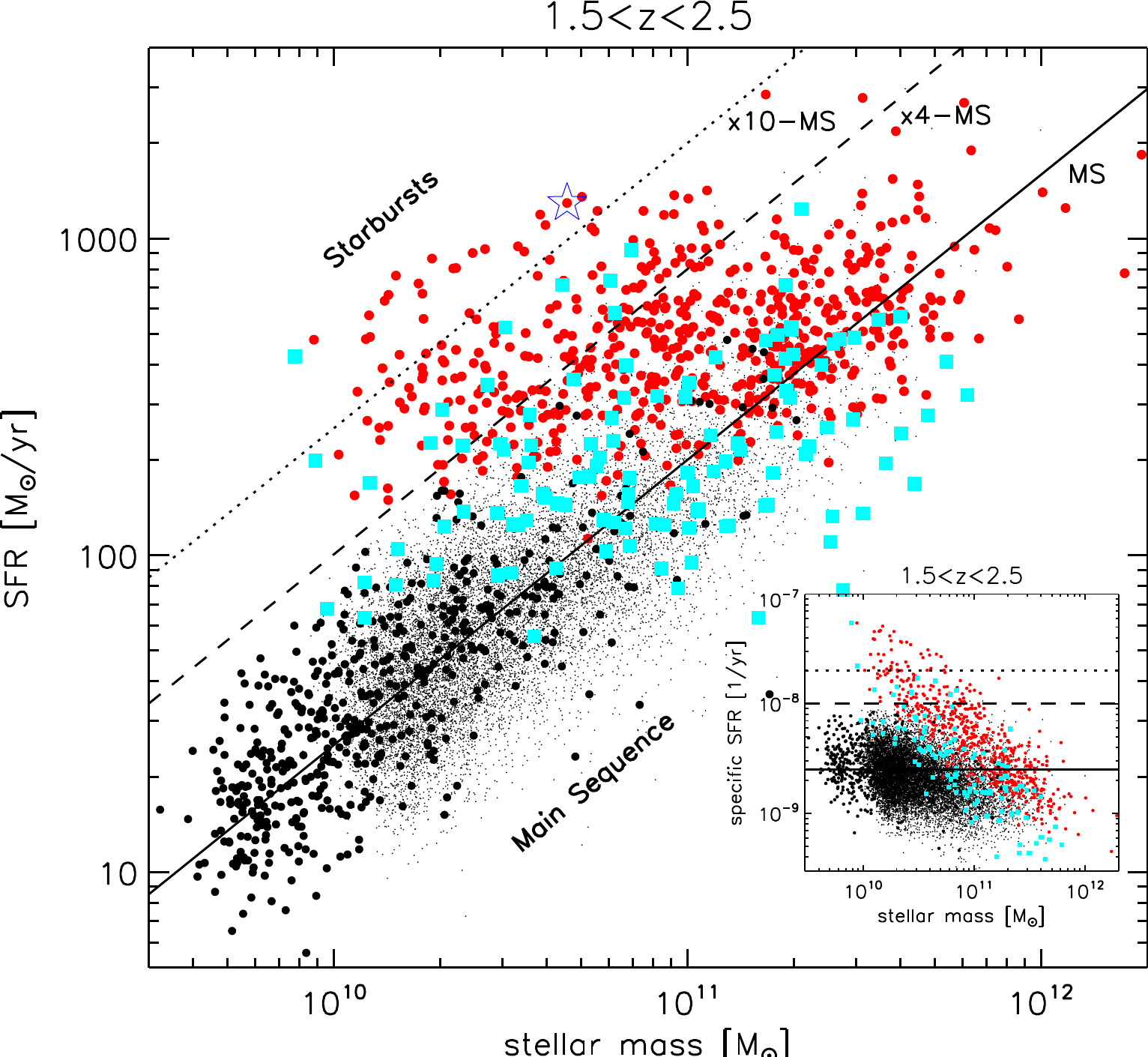}
\caption{The SFRs of three samples of galaxies at $z$\,=\,1.5--2.5 and stellar
masses are shown (Rodighiero \textit{et al.} 2011). The samples include
$BZK$-selected galaxies (black dots), GOODS (cyan) data and COSMOS (red)
galaxies with \textit{Hershel} PACS FIR detection. They use the $BZK$ galaxies
to define a `main sequence' of star-forming galaxies at this epoch and then
classify the outlier (mostly IR-detected) galaxies at $>$\,4 times the SSFR on the
main sequence as starburst-mode galaxies.}
\label{rod}
\end{figure}
%-------------------------------------------------------------------------------

Recently, Rodighiero \textit{et al.} (2011) have attempted to quantify the
fraction of star formation occurring in starbursts versus normal-rate star
formation at $z$\,=\,1.5--2.5. They make use of a $BZK$ colour-selected sample
of galaxies with low SFR and samples from GOODS and COSMOS, probing with
\textit{Herschel} and \textit{Spitzer} the moderate- and high-FIR luminosity
galaxies (see Fig.~\ref{rod}). The $BZK$ galaxies are used to define a `main
sequence' of star-forming galaxies at this epoch. The outliers above the main
sequence (mostly IR-detected) with more than four times the SSFR are then taken to
be galaxies undergoing burst-mode star formation. Rodighiero \textit{et al.}
(2011) suggest that only 2\% of the galaxies by number are undergoing starbursts
and this population contributes $\sim$\,10\% of the total SFRD at this epoch.
This is an interesting approach to the issue, but it should be clear that the
adopted definition of the main sequence and its spread will entirely determine
the starburst percentages. In addition, some of the galaxies within the main
sequence could well be merger-starbursts of originally lower-SSFR galaxies. As
noted earlier, it would be advantageous to have a physically-based definition of
bursting rather than simply picking outliers as bursts. To take an extreme
example, the latter definition would clearly be wrong if essentially a large
fraction of the population were, in fact, bursting.

%
%%%%%%%%%%%%%%%%%%%%%%%%%%%%%%%%%%%%%%%%%%%%%%%%%%%%%%%%%%%%%%%%%%%%%%%%%%%%%%%%
%

\section{Modelling star formation at high redshift:\\ same modes but different frequency}

It is now well established that the activity associated with both star formation
and AGN increases dramatically ($20\times$) out to a peak at $z$\,$\sim$\,2--3
using SFRs determined by the UV continuum, FIR and radio continuum and that
beyond $z$\,=\,3 there may be a gradual decline in the SFR density, although the
measurements become increasingly difficult at these redshifts.

At high redshift, one might expect that the same two modes pertain, yet their
relative importance could be quite altered. On the one hand, high-redshift
galaxies should have higher gas mass fractions than the typical 5--10\%  of
low-redshift spiral galaxies like the Milky Way -- leading to higher rates of
star formation associated with the quiescent linear mode of star formation. But
at the same time, one expects a greatly elevated rate of galactic
merging/interaction, increasing the frequency of the starburst mode. Which of
these dominates is not at all clear without a numerical simulation to track
their relative importance, keeping track of the rare and brief mergers.

\subsection{Cosmic evolution:\\ \textbf{{\rm M}$_{*}$} and \textbf{{\rm M}$_{\rm ISM}$} and star formation luminosities}

In order to test the framework developed above against the observed cosmic
evolution of galaxies, I developed a simple Monte Carlo simulation -- including:
star formation in the quiescent and merger-driven burst modes, merging of dark
matter haloes and their contained galaxies, and accretion of fresh gas from the
external large-scale structure environment. The simulation starts at $z$\,=\,6
with a population of 10 million dark matter haloes having a mass function
approximating that seen in simulations for $z$\,=\,6 (e.g., Heitmann \textit{et
al.} 2010). Each halo is started at $z$\,=\,6 with a galaxy of baryonic mass
equal to the dark matter mass times the universal baryon fraction, with 90\% of
the baryons being gaseous ISM and 10\% stellar mass. I then let the galaxy and
dark matter halo population evolve with 50\,Myr timesteps down to $z$\,=\,1. At each
timestep, gas is converted to stellar mass at the quiescent SFR given by
Equation~\ref{sfr}. At each timestep, haloes are also randomly sampled for having
undergone a merger with a probability weighted as $(1+z)^{2.5}M_{\rm
halo}^{0.125}$ (Fakhouri \& Ma 2010). This merger rate is normalised such that
2\% of the haloes merge per 50\,Myr at $z$\,=\,6. The secondary galaxy for each
merger was also selected randomly from the galaxy population weighted by $({\rm
mass~ratio})^{-2.1}$ (Fakhouri \& Ma 2010). For those galaxies selected to
merge, the SFE was increased by a factor of 10--50 (but only for one timestep).
Gas accretion to the galaxy halo was taken as $M=6.6(M_{\rm
halo}/10^{12})^{1.15} (1+z)^{2.25}\times0.165$ (Dekel \textit{et al.} 2009) for
$M_{\rm halo}$\,$<$\,10$^{12}$\,\msun. For larger-mass haloes, we assumed simply
that the accretion was cut off -- either by the accretion shocks or by AGN
feedback -- the former is a departure from Dekel \textit{et al.} but some
reduction of accretion is required in order to have the massive objects become
gas-poor ellipticals at modest redshifts as shown observationally.
  
Minor non-critical details which were included were that: 1) the effective
accumulation of stellar mass was taken to be 70\% of the integrated star
formation (i.e., assume 30\% of the stellar mass is recycled eventually in mass
loss), 2) the accretion of external gas to star-forming galaxy was delayed by
1\,Gyr after it accreted to the halo boundary (to account for the infall time)
and 3) during starbursts, ISM mass was shed from the galaxy at a rate equal to
the SFR (only for the burst mode). The star formation luminosity (motivated to
model the IR luminosity function) was taken {\it very crudely} as the total
luminosity from stars formed in the last 50\,Myr plus that from young stars in
earlier timesteps reduced by a factor of two in each timestep. Specifically, the
luminosity associated with recent star formation was taken to be
$10^{10}$\,\lsun\ per \msun\ per year of star formation, based on observations of
local galaxies.

\subsection{Need ISM replenishment by accretion}

\begin{figure}
\includegraphics[width=\linewidth]{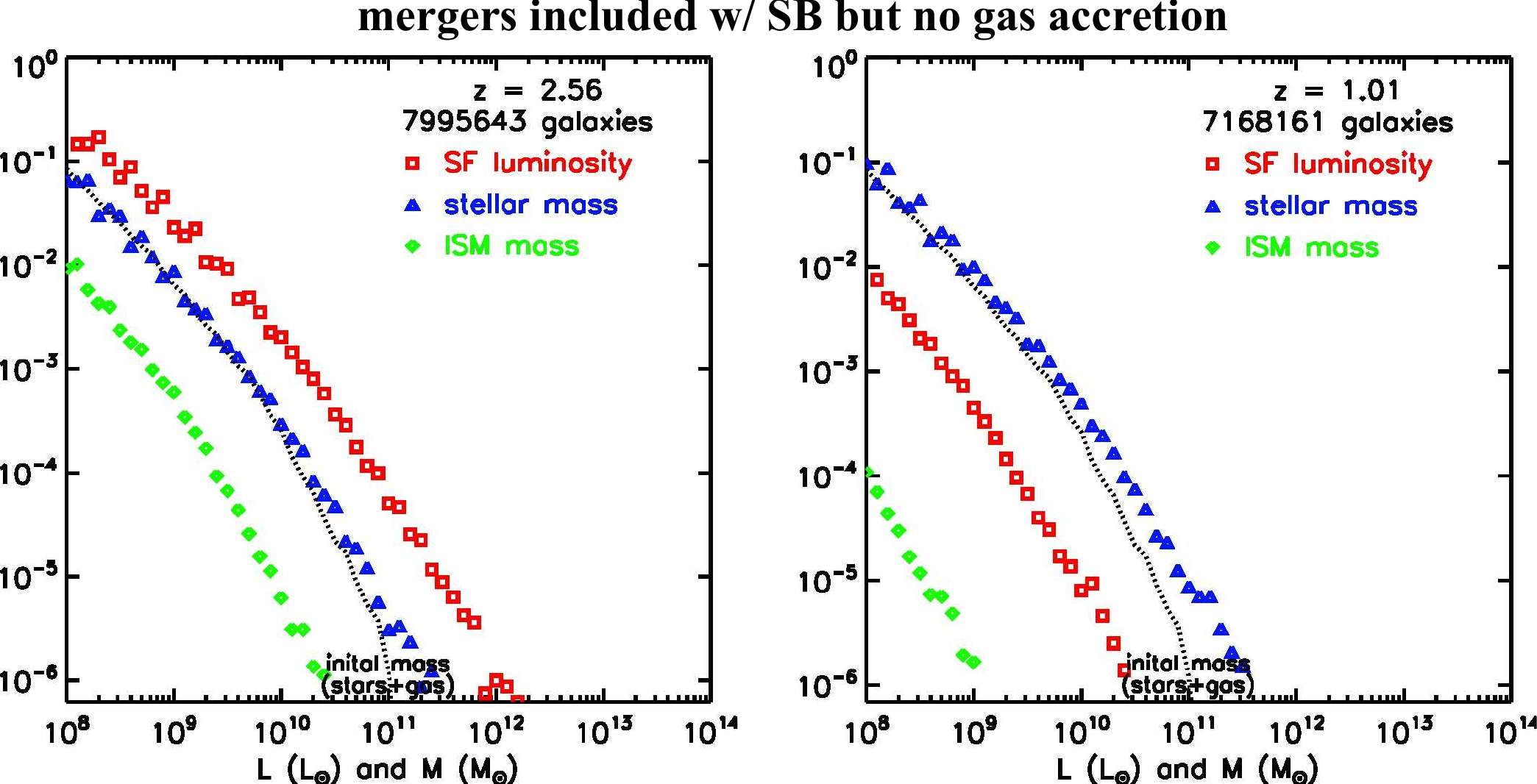}
\caption{The mass and luminosity functions for galaxies at $z$\,=\,2.5 (left) and $z$\,=\,1 
(right) derived from a Monte Carlo simulation including galaxy merging and
associated starburst activity but no gas replenishment via accretion from the
external environment. The dashed line is the original ($z$\,=\,6) galaxy mass
distribution (gas+stars). The original gas supply is exhausted far too quickly,
even by $z\sim2$.}
\label{no_gas}        
\end{figure}

This simplistic model was remarkably useful to explore critical aspects of the
evolutionary scheme involving quiescent and burst-mode star formation with a
reasonable gas accretion hypothesis. Figure~\ref{no_gas} shows the evolved mass
functions of stars and ISM gas and the star formation luminosity at $z$\,=\,2.5 and
1. For this figure, the simulation included merging galaxies and their
associated starbursts but was without accretion of gas from the environment. In
this case, the ISM runs down at a rate given by local Universe star formation
laws and the original gas content of the haloes is exhausted by $z$\,=\,2.5 to a
level much less than that seen either at $z$\,=\,2 or in present-epoch galaxies.

%-------------------------------------------------------------------------------
%-------------------------------------------------------------------------------
%-------------------------------------------------------------------------------
\begin{figure}[ht]
\includegraphics[width=\linewidth]{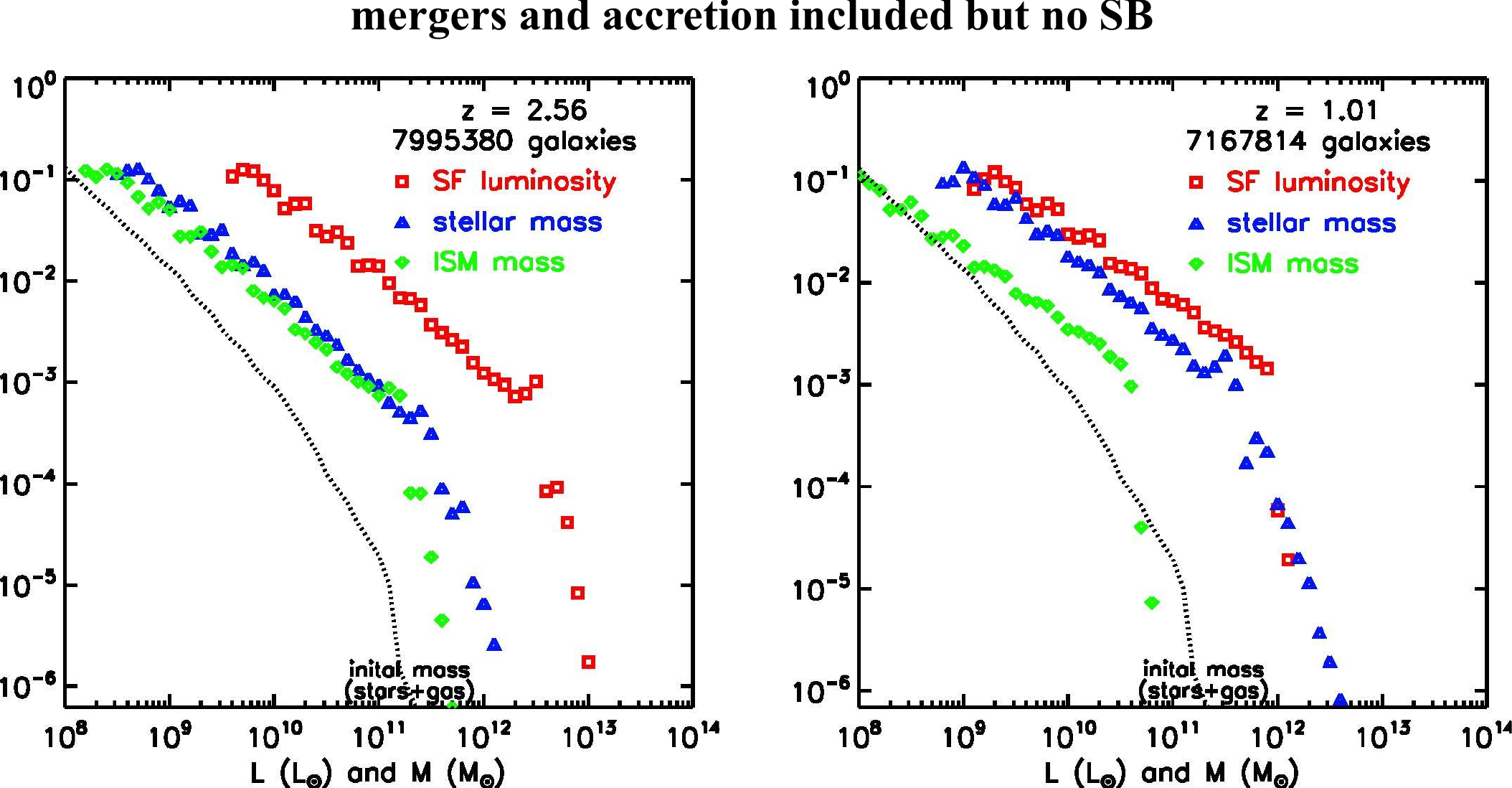}
\caption{Similar to Fig.~\ref{no_gas} except that gas accretion at halo masses
below $10^{12}$\,\msun\ is included, following the prescription of (Dekel
\textit{et al.} 2009). In this simulation, the starburst activity associated
with merging has been removed and the high end of the luminosity function
naturally must follow the exponential form of the mass distribution.}
\label{no_sb}
\end{figure}
%-------------------------------------------------------------------------------
%-------------------------------------------------------------------------------
\begin{figure}[!h]
\includegraphics[width=\linewidth]{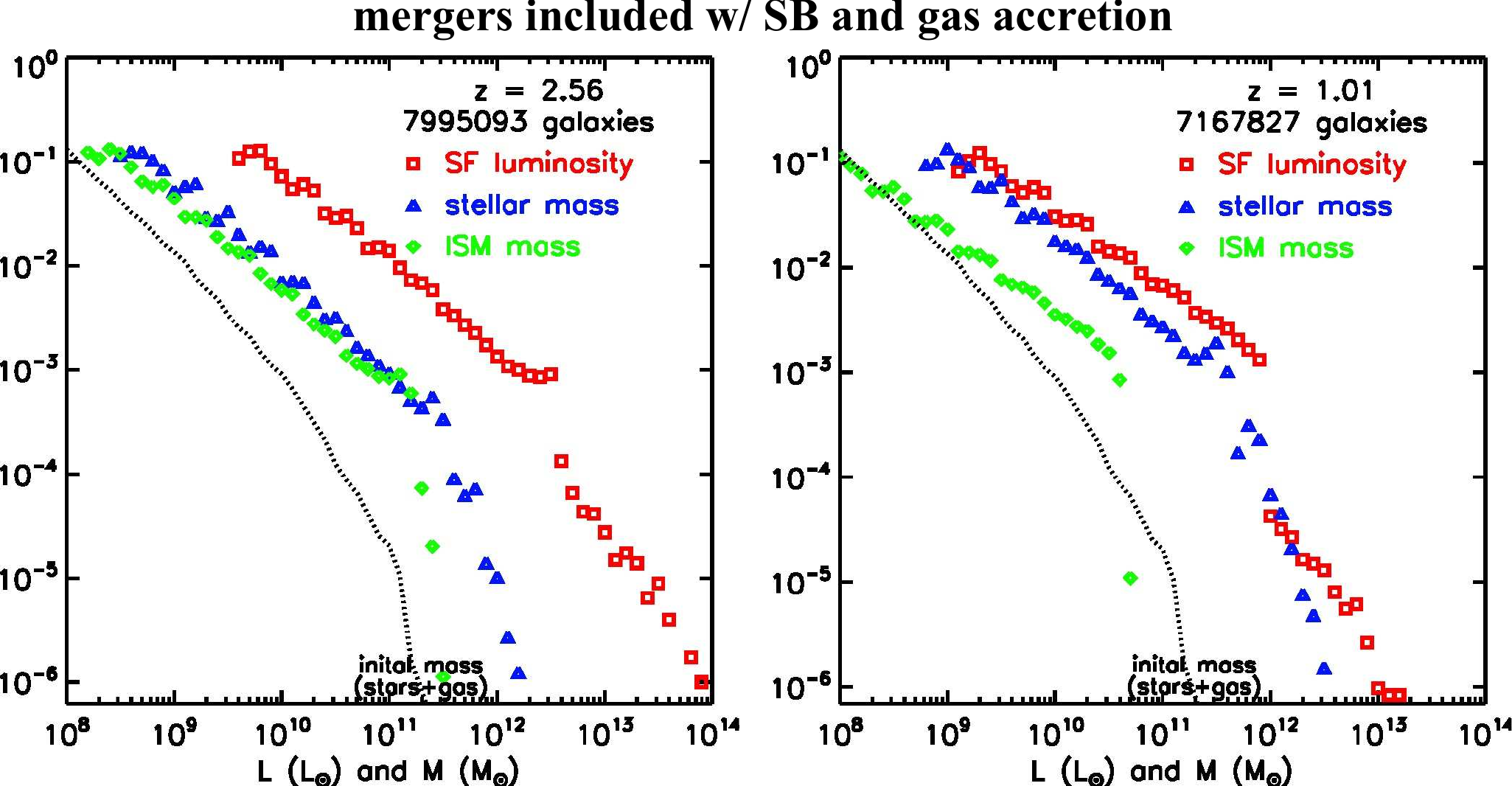}
\caption{Here both accretion and starburst activity during galaxy merging are
included, giving a reasonable qualitative match to the observed high-end mass
and luminosity functions.}
\label{all}
\end{figure}
%-------------------------------------------------------------------------------

\subsection{ULIRG starbursts account for high-\textit{L} tail}

The simulation shown in Fig.~\ref{no_sb} includes gas accretion as formulated
above but without the starburst activity associated with galaxy merging. Now,
the ISM and stellar mass functions exhibit characteristics similar to those
observed at $z$\,=\,2.5 and 1, i.e., $\lesssim50$\% gas mass and 10\%, respectively.
However, even at $z$\,=\,2.5, the star formation luminosity function is lacking the
power-law tail at the high-luminosity end. To reproduce the power-law tail, the
starburst activity associated with galaxy merging is needed, as shown in
Fig.~\ref{all}. (The low-mass and low-luminosity ends of the distribution
functions rise more steeply than is observed since no effort was made to model
the behaviour there -- this shallow slope is often attributed to star-formation
winds driving galactic mass-loss at velocities above the escape velocity of
lower-mass galaxies.)

\section{Conclusions}

In these lectures, I have attempted to summarise a physically intuitive approach
toward understanding the characteristics of star formation in high-redshift
galaxies making use of our intuition gained from the low-$z$ Universe. I have
devoted considerable effort towards developing an understanding of the molecular
clouds (GMCs) in which most low-redshift star formation occurs in normal
galaxies. 

I have also provided a summary of the physics of IR emission from optically
thick dust clouds. Since there is very little systematic treatment from a
theoretical point of view of the observed IR SEDs, I spent considerable effort
to develop a systematic approach. Most of these IR sources are centrally heated
by a starburst or AGN, so the theoretical framework for the dust heating and
radiative transfer can be greatly simplified and distilled. Interpreting the
observed SEDs is particularly prone to mistakenly evaluating a dust temperature
from the wavelength of the SED peak when in reality a significant range of dust
temperatures is contributing to the emergent flux and the peak is equally
determined by the overall dust opacity (since obviously the radiation at a given
wavelength cannot escape if the opacity is too high at that wavelength). I
underscore also the point that the long-wavelength R-J tail fluxes
should be used to derive dust (and ISM) masses -- not total IR luminosities (as
is commonly done for submm galaxies). 

Given the known internal structure of the star-forming GMCs, it is reasonable to
expect that star formation activity will have two modes: 1) a quiescent mode in
which stars form in GMCs at a rate roughly proportional to the mass of H$_2$ gas
(accounting for the linear correlations observed between the CO emission and
star formation tracers such as FIR over the disks of local galaxies including
our own) and 2) a dynamically-triggered starburst mode in which the rate per
unit gas mass is elevated by factors of 10--50 accounting for the
preferential formation of H{\sc ii} regions in spiral arms, despite the presence
of GMCs throughout the disk, and the high luminosity-to-gas ratios seen in local
ULIRGs. For the starburst mode, collisions of GMCs may significantly
increase the internal structure/density of the H$_2$ gas and stimulated star
formation due to supernovae and expanding H{\sc ii} regions.

At high redshift, the relative importance of these modes will be shifted, albeit
in different directions, by two changes: 1) higher gas-mass fractions
(increasing the quiescent mode) and 2) higher rates of galactic merging
(increasing the starburst mode). Some observational determinations of the galaxy
merger rates at high redshifts have shown very discrepant results for the
evolution. However, my recent analysis of evolution in the frequency of galaxy
close pairs from COSMOS data suggests merger evolution as $(1+z)^{2.3}$, in good
agreement with the dark matter halo merger rates seen in simulations. 

At high redshift the conditions in the gas may be quite different from those in the
star-forming clouds at low redshift. It has been speculated that the stellar IMF
might be top-heavy in starbursts or low-metallicity environments. This is not an
insurmountable problem -- observations of the 4000\,\AA ~break may be used to
constrain the mass of low-mass stars. Until such observations are done, my
conservative opinion would be to avoid the `last refuge of scoundrels' since
very little hard evidence is found in the local Universe to support such
variations.

To test whether the ingredients discussed above provide a reasonable basis for
understanding high-redshift galaxy evolution, I show the results of a Monte
Carlo simulation starting at $z$\,=\,6 with 10 million haloes distributed in mass and
with merging rates as found in $\Lambda$CDM simulations. Starting with galaxies
for which the gas content is 90\% of the baryonic mass, the galaxies are evolved
including both star formation modes and gas accretion. The simulations clearly
show the need for gas replenishment through accretion from the external
environment; otherwise, the $z$\,=\,2 and present-epoch gas contents are far too low. 

This need for accretion is simply a reflection at high-$z$ of the well-known
requirement of gas replenishment for the Milky Way. Specifically, for the Milky
Way the SFR\,$\sim$\,3\,\msun\ per year and the present gas content is
$\sim$\,3\,$\times$\,10$^9$\,\msun, implying an exhaustion timescale of only 1\,Gyr. In
the case of local galaxies, it does not appear that the accreting gas is H{\sc
i} since the high velocity H{\sc i} clouds do not constitute a sufficient
influx. More likely the inflow is in the form of diffuse ionised hydrogen (H{\sc
ii}) for which imaging with high sensitivity to low surface brightness emission
is required. Clearly, this is an important direction for future observations.

Another conclusion from the simulations is that the starburst mode triggered by
galactic merging is necessary to account for the high-luminosity power-law tail
of the IR luminosity functions at high redshift. The quiescent mode cannot do
this since the luminosity generated would simply reflect a scaled version of the
galaxy mass function which is exponentially falling at the high mass end.

\section*{Acknowledgments} 
I would like to thank Zara Scoville for help in editing this manuscript and
Andreas Schruba and Kevin Xu for a careful proofreading. I would also like to
thank some of my close colleagues over the years who contributed much to this
enjoyable research: Herve Aussel, Peter Capak, Peter Goldreich,\linebreak Jeyhan
Kartaltepe, Jin Koda, Colin Norman, Brant Robertson, Dave Sanders, Kartik Sheth,
Phil Solomon and Min Yun.

\end{document}